\pdfoutput=1
\documentclass[pre,aps,twocolumn,showpacs,floatfix,nofootinbib,
superscriptaddress]{revtex4-1}

\usepackage{subfigure}
\usepackage{amssymb}
\usepackage{amsfonts}
\usepackage{amsmath}
\usepackage{amsthm}
\usepackage{epsfig}
\usepackage{graphicx}

\usepackage[usenames,dvipsnames]{color}
\usepackage[latin1]{inputenc}

\usepackage[hidelinks,unicode=true]{hyperref}
\hypersetup{colorlinks=true,
 	linkcolor=blue,
 	urlcolor=blue,
 	citecolor=blue,
 	pdfhighlight=/N
 }
 \usepackage{comment}
 \usepackage[normalem]{ulem}

\begin{document}

\title{Field emitter electrostatics: a review with special emphasis on modern high-precision finite--element modelling}

\author{Thiago A. de Assis}
\email{thiagoaa@ufba.br}
\address{Instituto de F\'{\i}sica, Universidade Federal da Bahia,
  Campus Universit\'{a}rio da Federa\c c\~ao,
  Rua Bar\~{a}o de Jeremoabo s/n, 40170-115, Salvador, BA, Brazil}

\author{Fernando F. Dall'Agnol}
\email{fernando.dallagnol@ufsc.br}
\address{Department of Exact Sciences and Education (CEE), Universidade Federal de Santa Catarina,
 Campus Blumenau, Rua Jo\~{a}o Pessoa, 2514, Velha, Blumenau 89036-004, SC, Brazil}

\author{Richard G. Forbes}
\email{Author to whom correspondence should be addressed: r.forbes@trinity.cantab.net}
\address{Advanced Technology Institute \& Department of Electrical and Electronic Engineering, University of Surrey, Guildford, Surrey
GU2 7XH, UK}

\begin{abstract}

This review of the quantitative electrostatics of field emitters, covering analytical, numerical and ``fitted formula" approaches, is thought the first of its kind in the 100 years of the subject. The review relates chiefly to situations where emitters operate in an electronically ideal manner, and zero-current electrostatics is applicable.  Terminology is carefully described and is ``polarity independent", so that the review applies to both field electron and field ion emitters. It also applies more generally to charged, pointed electron-conductors---which exhibit the ``electrostatic lightning-rod effect", but are poorly discussed in general electricity and magnetism literature. Modern electron-conductor electrostatics is an application of the chemical thermodynamics and statistical mechanics of electrons. In related theory, the primary role of classical electrostatic potentials (rather than fields) becomes apparent. Space and time limitations have meant that the review cannot be comprehensive in both detail and scope. Rather, it focuses chiefly on the electrostatics of two common basic emitter forms: the needle-shaped emitters used in traditional projection technologies; and the post-shaped emitters often used in modelling large-area multi-emitter electron sources. In the post-on-plane  context, we consider in detail both the electrostatics of the single post and the interaction between two identical posts that occurs as a result of electrostatic depolarization (often called ``screening" or ``shielding"). Core to the review are discussions of the ``minimum domain dimensions" method for implementing effective finite-element-method electrostatic simulations, and of the variant of this that leads to very precise estimates of dimensionless field enhancement factors (error typically less than 0.001 \% in simple situations where analytical comparisons exist). Brief outline discussions, and some core references, are given for each of many ``related considerations" that are relevant to the electrostatic situations, methods and results described. Many areas of field emitter electrostatics are suggested where further research and/or separate mini-reviews would probably be useful.


\end{abstract}

\maketitle

\tableofcontents

\section{General Introduction}
\label{GI}
Field electron emission (FE) \cite{EHYMP15,Lia16,ES17,Jen18,For20a} involves electron tunnelling through an energy barrier lowered by a high negative electrostatic field, of magnitude typically of order $4$ V/nm. The various types of field ion emission, notably field evaporation (FEV) \cite{MF14} and gas field ionization (GFI) \cite{For10a} require high positive electrostatic fields, typically many tens of V/nm.
These high-field-magnitude emission processes play an important role in many scientific and technological contexts, particularly in theories of electrical breakdown in air and in vacuum, in theories of the optically bright electron and ion sources used in many machines of modern nanotechnology, in the theory of large-area field electron sources, and in the theory of some experimental techniques, for example atom probe microscopy.

With field ion emission, the field magnitudes are so high that, for most applications, they have to be created at the apex of a specially fabricated individual needle-shaped structure, referred to as a field (ion) emitter, or often (colloquially) as a ``tip". There are also situations where a sharp point forms naturally, as result of electroformation or (with liquids) electrohydrodynamic effects. Specially fabricated individual needle-like structures are also used in electron emission contexts, for example the field electron sources used in electron microscopes and in field electron microscopy and related techniques.

In the 1970s, developments in lithography allowed the fabrication of large arrays of miniaturised conical field electron emitters, in the form of the so-called Spindt array \cite{Spindt,SBHS10}. This was an early form of so-called \textit{large-area field electron emitter (LAFE)}, which has many or very many individual emitters or emission sites fabricated onto a substrate typically of area 1 to 25 mm$^2$ (but not limited to this size).

Driven by the technological interest in large-area high-current-density electron sources, many different emitting materials and LAFE geometrical configurations have been investigated in the last twenty years or so (e.g., \cite{Sai10,LAFE1,LAFE2,LAFE3}). LAFEs based on carbon nanotubes (CNTs) have emerged as one of the most promising options \cite{Sai10}. Emitters based on emission from graphene edges are a future prospect \cite{Sai22}.

In all these cases, there are theoretical issues of how to relate the electrostatic field at the emitter surface (in particular at the emitter apex) to the remainder of the system geometry and to the applied voltage (or applied voltages, in the case of non-diode-like systems). In some cases the interest has mainly been directed towards information that will guide the practical operation of machines and techniques, or potentially provide information about an electrical breakdown process. But, particularly with LAFEs, there has been interest in how to choose/optimise emitter and system geometry in order to get the highest apex field (or, with LAFEs, the highest ``macroscopic" (i.e., LAFE-average) emission current density.

Consequently, particularly with LAFEs, there has been interest in both analytical and numerical methods of analysing the electrostatics of field emitters, and in the underlying theory. Some of this underlying theory is not widely familiar, because the electrostatics of good electrical conductors is to some extent part of the theory of electrical effects in thermodynamics, rather than part of the theory of classical electrostatics alone. The ideal goal is to establish some relatively simple formula for a parameter that characterises the electrostatic situation, often (but not always) a dimensionless field enhancement factor as defined below.

With LAFEs, because of the enormous complexity of the general situation, much attention has been focussed on the development and analysis of simple models involving a ``post on a plane" and ``arrays of posts on a plane" (mostly, regular arrays of identical posts). Because of the particular interest in the electrostatics of carbon nanotubes, the so-called ``hemisphere-on-a-cylindrical-post" (HCP) shape model has attracted particular attention. Usually, the post is modelled as standing on one of a pair of well-separated parallel planar plates, of lateral extent large compared with the plate separation: this situation is called here \textit{PPP geometry}.

The main topic of this review is the use and verification of finite-element methods for high-precision ``pure" electrostatic modelling of post-like shapes, mainly in the context of classical-conductor models and mainly in the context of field electron emission as it occurs in PPP geometry. By ``pure", we mean the modelling of the essentially geometric electrostatic effects that occur in the absence of free electric charge in the vacuum space (such as would be generated as field emitted vacuum space-charge) and in the absence of current flow in surrounding conductors. These ``pure" effects are described in some detail, and an important role of this modelling is to check on the accuracy of approximate formulae derived by other means. However, it has seemed useful to set this material in the context of a wider account of the electrostatics of field ion and electron emitters. Since, at a classical level, electrostatics is symmetrical as between positively charged and negatively charged structures, the wider aspects of this review will apply to both polarities. Parts of the review material may also be relevant to other situations where pointed electrical probes are used, in particular the scanning-probe techniques;  parts may be relevant more widely, in discussion of the ``electrostatic lighting-rod effect".

Formally, we call the effect under discussion \textit{electrostatic field enhancement}. There is a separate physical effect, which we call \textit{electromagnetic (EM) field enhancement}, that occurs when an EM wave is directed onto a needle-like structure, with the EM electric field ($E$-field) parallel to the needle axis: this effect leads to enhancement of the $E$-field at the needle apex, but the physics of the two effects is different. The theory in this review does \textit{not} apply to EM field enhancement.

\section{Physical Introduction}
\label{PI}

\subsection{Background Issues}
\label{BI}
The electrostatics of field electron emission (FE) is most commonly (and most easily) discussed by using classical-conductor models in which the emitter and relevant parts of the enclosing experimental system are represented by smooth geometrical surfaces (the ``conductor surfaces") on which the classical electrostatic potential $\mathit{\Phi}$ or (in some contexts) the conventional classical electrostatic field $\textbf{\textit{E}}$ $\left[=-\mathbf{grad} \mathit{\Phi}\right]$ is specified. Inside the conductor surfaces (i.e., on the conductor side) the field is taken as zero; outside, the classical electrostatic potential and field are taken to obey Maxwell's equations in the form of Laplace's equation. Classical-conductor models work best for metals, but are commonly applied to FE from non-metals as a first approximation. 

 We now prefer to call $\textbf{\textit{E}}$ the ``electrostatic field" rather than the ``electric field", because the latter term can be understood to also cover the optical $E$-fields generated by lasers. Lasers are increasingly used in FE contexts. However, as indicated above, applied electrostatic fields and applied optical $E$-fields can generate different physical effects. In particular, the processes of field enhancement are physically different in the two cases.
 
In most FE situations, it is customary to disregard work-function variations across the emitter and system surfaces (and the resulting patch fields). Thus, in the absence of current flow along an emitter, the potential $\mathit{\Phi}$ in the part of the conductor surface that defines an emitter is treated as constant. Further, if the surfaces of all system electrodes are treated as having the same work function as the emitter, as is usually done for simplicity, then the classical potential difference $\Delta \mathit{\Phi}$ between a counter-electrode and the emitter eletrode  is equal to the voltage $V$ applied between them. In this case, the technical distinction between $\Delta \mathit{\Phi}$ and $V$, which only rarely has any significant quantitative consequences, can be disregarded when formulating basic theory.

The above remarks apply to classical-conductor models. However, with both field ion and field electron emission there are experimentally observable atomic-level phenomena. There exist both an older body of atomic-level ``electrostatic" theory (developed in conjunction with the theory of how the field ion microscope works), and an increasing modern body of atomic-level electrostatic theory--mainly in the context of FE (and particularly in the context of FE from carbon nanotubes), and mainly implemented via density functional theory (DFT). Deep issues arise about how to properly connect classical-conductor and modern atomic-level theories of electrostatics. These issues, and atomic-level theories of electrostatics in general, are outside the scope of this review, which focuses on the accuracy of classical-conductor electrostatic models. However, the material in this review should assist in later discussion of atomic-level electrostatic models.

\subsection{Equation conventions (including electrostatic conventions)}
\label{EC}

All quantities and equations in this article are defined using the modern international system of equations that has the vacuum electric permittivity $\epsilon_{0}$ in Coulomb's Law. Since 2009, this system has been called the ``International System of Quantities" (ISQ). Thus, voltages and classical electrostatic potentials are measured in volts, and fields are measured in V/m (or a related unit with same dimensions, such as V/nm). Most published FE articles now use the ISQ, though some older FE textbooks use Gaussian system equations to formulate aspects of FE theory.

In field emitter physics, it is customary to take the direction away from the emitter into the vacuum as the positive direction. This convention implies that the electrostatic fields that induce field electron emission are negative in sign. However, the ``fields" that go into the related equations for \textit{local emission current density (LECD)} ($J_{\rm{L}}$) and current ($I$) need, by convention, to be positive in sign. Customary practice in FE literature is to use the symbol $E$ or $F$ to denote a positive quantity that is the absolute magnitude of electrostatic field. We think it clearer here to denote this positive quantity by $|E|$. However, we use the customary convention that the symbols $J$ and $I$ denote positive quantities that are the absolute magnitudes of current density and current, respectively.

\subsection{Relevant elements of field electron emission theory}
\label{RE}

In FE theory, the main role of an electrostatic field is as a parameter in an emission equation. A convenient starting point is the zero-temperature version of Murphy-Good (MG) FE theory \cite{MG}, which was developed in 1956 in order to correct serious errors in the original 1928 theory of Fowler and Nordheim and/or in directly related 1928 and 1929 papers. (For a modern derivation of MG FE theory, see \cite{Forbes2007RFN}.)

Notwithstanding this, many modern FE technological papers use a simplified version of the original 1928/29 FN FE theory, rather than MG FE theory. It has been argued \cite{JAP2019RG,MMA} that this is weak scientific practice, and that using MG FE theory would lead to better science and to better value for research funders.

Both FN FE theory and MG FE theory are based on a simple underlying physical model, originally introduced in the 1920s, that (a) disregards the existence of atomic structure and the role of atomic-level wave-functions on the emission process, and (b) models a field electron emitter as a Sommerfeld-type free-electron conductor with a smooth planar surface of large lateral extent. This has been called \textit{smooth planar metal-like emitter (SPME) methodology}.

In MG theory, FE takes place by tunnelling from emitter electron states near the emitter Fermi level, through a barrier modelled as a planar image-rounded barrier--often now called a \textit{Schottky-Nordheim (SN) barrier} \cite{Forbes2007RFN}. In the zero-temperature theory, the resulting local emission current density $J_{\rm{L}}^{\rm{MG0}}$ is a function of the local work function $\phi$ and of the absolute magnitude $|E_{\rm{L}}|$ of the constant electrostatic field above the planar emitter surface.

For present purposes it is convenient to write the zero-temperature 1956 MG FE equation in the ``linked" form
\begin{equation}
J_{\rm{L}}^{\rm{MG0}} = {\rm{t}_{\rm{F}}^{-2}} J_{\rm{kL}}^{\rm{SN}},
\label{MG0a}    
\end{equation}
\begin{equation}
J_{\rm{kL}}^{\rm{SN}} \equiv a {\phi}^{-1}|E_{\rm{L}}|^{2} {\exp{[-\rm{v}_{\rm{F}}} b {\phi}^{3/2}/|E_{\rm{L}}|]},
\label{MG0b}    
\end{equation}
where $a$ and $b$ are the Fowler-Nordheim constants \cite{ForA146}, and $\rm{v}_{\rm{F}}$ and $\rm{t}_{\rm{F}}$ are particular values (appropriate to a SN barrier defined by $\phi$ and $|E_{\rm{L}}|$) of the field emission special mathematical functions ${\rm{v}}(x)$ and ${\rm{t}}(x)$ \cite{For20a}, where $x$ is the \textit{Gauss variable}, i.e., the independent variable in the Euler/Gauss Hypergeometric Differential Equation. The function $J_{\rm{kL}}^{\rm{SN}}$ is defined by eq. (\ref{MG0b}) and is called the \textit{local kernel current density for the SN barrier}.

Obviously, many real emitters do not have large flat emitting surfaces, but have the shape of a pointed needle or a rounded post, or are otherwise ``pointy". We call these \textit{pointform emitters}. For pointform emitters that are ``not too sharp", Murphy-Good FE theory can be applied by using the so-called ``planar emission approximation".  

Consider a location ``L" in the emitter surface, and assume the kernel current density at ``L" is given by eq. (\ref{MG0b}). Integration over the emitter surface yields the \textit{notional emission current} (for the SN barrier)
\begin{equation}
I_{\rm{n}}^{\rm{SN}} = \int_{\Sigma} J_{\rm{kL}}^{\rm{SN}} \rm{d} \Sigma .
\label{FEcurrent-notional}    
\end{equation}
$I_{\rm{n}}^{\rm{SN}}$ also depends on the emitter shape and the assumed common work-function value, but these dependences are not explicitly shown.

The planar emission approximation is poor if the emitter apex radius is ``too small". This is because the correct curved-surface expressions for the barrier transmission probability and the electron supply then diverge significantly from those applicable to planar geometry. ``Too small" is often taken to mean ``radius less than 10 to 20 nm"---though recent work \cite{BR19a} suggests that ``less than 100 nm", or even ``less than 1 $\mu$m", might be a better criterion, depending on what level of LECD precision is required (the 1 $\mu$m criterion  corresponds to 5\%). Nevertheless, the planar emission approximation has been very widely used in practice (possibly often without recognition).

After this integration, it is convenient to choose a \textit{characteristic location} ``C" on the emitter surface where the local field magnitude (and hence the LECD) are particularly high, and then define a \textit{notional emission area, using ``C" and the SN barrier}, $(A_{\rm{nC}}^{\rm{SN}})$ by
\begin{equation}
A_{\rm{nC}}^{\rm{SN}} \equiv I_{\rm{n}}^{\rm{SN}}/J_{\rm{kC}}^{\rm{SN}}.
\label{An defintion}    
\end{equation}
In modelling, where cylindrical symmetry and uniform work function are nearly always assumed, ``C" is usually taken at the emitter apex.

To the best of our knowledge, at no stage in the last 90 years or so since the 1928 Fowler and Nordheim paper \cite{FowlerN} have there been any comparisons of theoretical and experimental LECD values that were reliable, precise and decisive. Further, it is clear that theories that disregard the existence of atoms (which SPME methodology does) are not physically complete. Hence, it can be argued that FE theory is still in a pre-scientific state, and that it must be assumed that $I_{\rm{n}}^{\rm{SN}}$ is not an accurate estimate of  experimentally measured current---even for so-called ``\textit{electronically ideal}" FE systems where the measured current is determined by the emission physics and the system geometry alone, and is not influenced by any so-called ``system complications" (see Section \ref{FVR}).

To deal formally with this situation, for electronically ideal systems, a \textit{prediction uncertainty factor} (previously called a ``knowledge uncertainty factor") has been introduced (e.g., see \cite{For20a}). The treatment here aligns with that in \cite{MMA}, is more complete than given previously, and supersedes earlier discussions.

We denote the real physical LECD at a given surface location ``L" by $J_{\rm{real,L}}$ and take it as given by an equation of the form
\begin{equation}
J_{\rm{real,L}} = {\lambda_{\rm{kL}}^{\rm{SN}}} J_{\rm{kL}}^{\rm{SN}},
\label{lam-kl}    
\end{equation}
where the \textit{LECD prediction uncertainty factor} $\lambda_{\rm{kL}}^{\rm{SN}}$ is defined by this equation. For simplicity, the pre-factor ${\rm{t}}_{\rm{F}}^{-2}$ in eq. (\ref{MG0a}), and the temperature-dependent correction factor that appears in the finite-temperature version of MG FE theory, are ``swept up" into $\lambda_{\rm{kL}}^{\rm{SN}}$.   

Values and functional dependences of $\lambda_{\rm{kL}}^{\rm{SN}}$ are not known.  At present, the best guess of Forbes \cite{For-U535} is that it is a function of field and lies in the range $0.005 < \lambda_{\rm{kL}}^{\rm{SN}} < 14$.

When modified in this way, MG theory has been called \cite{For19a} \textit{Extended Murphy-Good (EMG) FE theory}.

For an emitter of given ``\textit{configuration}" (i.e., of given shape and of given work-function distribution over its surface $\Sigma$), a \textit{true model emission current} $I_{\rm{tm}}^{\rm{SN}}$ would be given by integration, and could be put in the form shown:

\begin{equation}
{I_{\rm{tm}}^{\rm{SN}} = \int_{\Sigma} J_{\rm{real,L}} \rm{d} \mathit{\Sigma}} = \lambda_J I_{\rm{n}}^{\rm{SN}} = \lambda_J A_{\rm{nC}}^{\rm{SN}} J_{\rm{kC}}^{\rm{SN}}.
\label{I-tn}    
\end{equation}
Here, $\lambda_J$ is an LECD--related prediction uncertainty factor of the same general kind as $\lambda_{\rm{kL}}^{\rm{SN}}$. $\lambda_J$ will depend on many things, and we do not think it helpful here to introduce complicated notation about this: $\lambda_J$ is basically a ``place-holder" indicating that uncertainty exists. 

We now take into account that the true configuration of a real emitter may be different that assumed during modelling. This introduces a second uncertainty factor ($\lambda_{\rm{EM}}$), called here the \textit{emitter-model prediction uncertainty factor}, and results in a formula for the \textit{predicted emission current} $I_{\rm{p}}^{\rm{SN}}$:
\begin{equation}
{I_{\rm{p}}^{\rm{SN}} =  \lambda_{\rm{EM}} I_{\rm{tm}}^{\rm{SN}} = \lambda_{\rm{EM}} \lambda_J A_{\rm{nC}}^{\rm{SN}} J_{\rm{kC}}^{\rm{SN}}}.
\label{Ip1}    
\end{equation}
If a leakage current exists, in a FE system that is otherwise electronically ideal, this can be treated as equivalent to a small change in $\lambda_{\rm{EM}}$.

This formula can be simplified by defining a new parameter, $A_{\rm{fC}}^{\rm{SN}}$, the \textit{characteristic formal emission area as defined using ``C" and the SN barrier}, by
\begin{equation}
A_{\rm{fC}}^{\rm{SN}} = \lambda_{\rm{EC}} \lambda_J A_{\rm{nC}}^{\rm{SN}}.
\label{AfCSN}    
\end{equation}
Equation (\ref{Ip1}) can then be written
\begin{equation}
I_{\rm{p}}^{\rm{SN}} = A_{\rm{fC}}^{\rm{SN}} J_{\rm{kC}}^{\rm{SN}}
\label{Ip2} .    
\end{equation}

We now assert that, in EMG FE theory, for an electronically ideal emitter, a formula for the actual experimentally measured current $I_{\rm{m}}^{\rm{EMG}}$ is
\begin{equation}
I_{\rm{m}}^{\rm{EMG}} = A_{\rm{fC}}^{\rm{SN}} J_{\rm{kC}}^{\rm{SN}}  =  A_{\rm{fC}}^{\rm{SN}} a {\phi}^{-1}|E_{\rm{C}}|^{2} {\exp{[-\rm{v}_{\rm{F}}} b {\phi}^{3/2}/|E_{\rm{C}}|]}  .
\label{Im-2}    
\end{equation}

In this equation, both the measured current and the kernel LECD are well defined in principle, so if (a) the work-function value $\phi$ is well known, and (b) the relationship between measured voltage and characteristic local field is well known (or can be determined experimentally), then a value for formal emission area can be extracted from experimental current-voltage measurements.

Note that, for a given emitter of presumed known configuration, the \textit{notional} emission area is the area that would be predicted by theory, but the \textit{formal} emission area is the parameter extracted from experiments. As things stand at present, even for a hypothetical emitter of known well-defined configuration, it would not be possible for theoreticians to accurately predict the ``area" value that would be measured by experimentalists, and it would not be possible for experimentalists to accurately measure an ``area" value predicted by theoreticians.

To ameliorate this problem and put FE theory onto a proper scientific basis, one needs much better knowledge (than currently exists) of the behaviour of all four of $A_{\rm{fC}}^{\rm{SN}}$, $\lambda_{\rm{kL}}^{\rm{SN}}$, $A_{\rm{nC}}^{\rm{SN}}$ and (ideally) $\lambda_{\rm{EM}}$. In particular, it would be helpful to know how the notional emission area depends on the assumed work-function value and on emitter shape, particularly on the apex radius, and (using a slightly modified version of the theory) on temperature. An important new role for electrostatic simulations using finite element methods is to help explore questions of this kind. Some useful progress has been made \cite{Jen18,Mauro2018,Forbes_preexponential}, but much more remains to be done.

\subsection{Field-voltage relationships}
\label{FVR}

A \textit{FE system} is defined to include all aspects of the experimental system that can affect the measured current-voltage [$I_{\rm{m}}(V_{\rm{m}})$] relationship, including: (a) emitter composition, geometry and surface condition; (b) the mechanical, geometrical and electrical arrangements in the vacuum system; (c) all aspects of the electronic circuitry and all electronic measurement instruments; (d) the emission physics; and (e) all relevant physical processes that might be taking place (for example, the generation of field emitted vacuum space-charge, Maxwell-stress-induced reversible changes in emitter geometry, and adsorbate-atom dynamics).

The \textit{measured voltage} $V_{\rm{m}}$ (normally measured at the high voltage generator) is defined as the voltage applied between the counter-electrode side of the system and the emitter side of the system. Thus, in field electron emission the voltage $V_{\rm{m}}$ is positive in sign.

In general, the relationship between the electrostatic field $E_{\rm{L}}$ at location ``L" in the emitter's electrical surface and the \textit{\textbf{measured}} voltage $V_{\rm{m}}$ can be written in the form
\begin{equation}
E_{\rm{L}} = -V_{\rm{m}}/\zeta_{\rm{mL}},
\label{zeta-mL}    
\end{equation}
where $\zeta_{\rm{mL}}$ is the ``\textit{\textbf{measured}" local voltage conversion length (LVCL)}. It is usually convenient to give LVCLs in nm.

For the analysis of measured $I_{\rm{m}}(V_{\rm{m}})$ characteristics, the relationship between $V_{\rm{m}}$ and the emitter apex field $E_{\rm{a}}$ is of particular interest, since this can be combined with eq. (\ref{Im-2}) (with ``C" = ``a") to yield a formal expression for the $I_{\rm{m}}(V_{\rm{m}})$ characteristic.

An FE system is described as \textit{electronically ideal} if, for all surface locations ``L",  the parameter $\zeta_{\rm{mL}}$ is constant, independent of the measured current or voltage, or of time. For ideal systems, the symbol $\zeta_{\rm{mL}}$ can be replaced by $\zeta_{\rm{L}}$, and this ``ideal LCVL" is simply called the ``local VCL". (Earlier treatments did not make this distinction, and the symbol $\zeta_{\rm{L}}$ was used for both meanings.)  

For these ideal systems, it is better to replace eq. (\ref{zeta-mL}) by the ``ideal" or ``purely electrostatic" equation
\begin{equation}
E_{\rm{L}} = -V_{\rm{M}}/\zeta_{\rm{L}},
\label{zeta-L}    
\end{equation}
where $V_{\rm{M}}$ is the voltage between the counter-electrode and the emitter electrode; $V_{\rm{M}}$ has sometimes been called the ``macroscopic voltage" but is called here the \textit{inter-electrode voltage}.  

Strictly, electrostatic discussions should be formulated in terms of fields and electrostatic potentials, but it is often more convenient---and closer to actual practice in the literature---to formulate discussions as a relationship between fields and voltages, and ignore the distinction between voltage and classical electrostatic potential difference. (If the simplifying assumption of uniform local work function is made, there is no numerical distintion, anyway.)  

In some real systems the inter-electrode voltage $V_{\rm{M}}$ is not equal to the measured voltage $V_{\rm{m}}$, because there is significant series resistance in the current path between the high-voltage generator and the \textit{front} (vacuum facing) surface of the emitter electrode. In particular, this can happen if an emitting protrusion stands on a poorly conducting semiconductor substrate. Thus, it is essential to make the distinction between $\zeta_{\rm{mL}}$ and $\zeta_{\rm{L}}$.

More generally, many real FE systems are not electronically ideal, due to the occurrence of one or more ``system complications".  Possible complications include, amongst others: (1) significant series resistance in the conducting path between the protrusion and the high-voltage generator, as just discussed; (2) work-function changes, due to current-related heating (Joule and/or Nottingham heating) and resulting desorption of surface adsorbates; (3) geometrical emitter-shape changes due to Maxwell stress; (4) shape changes due to electroformation or emitter erosion; (5) effects due to field emitted vacuum space-charge (FEVSC); (6) current dependence in LVCLs, and in the related field enhancement factors defined below, due to voltage loss along the emitter; and (7) (for semiconductors) field penetration into the emitter surface.

A practical validity check, called an \textit{orthodoxy test} \cite{Forbes2013}, has been created that can indicate whether or not it is very probable that a given set of measured current-voltage characteristics relate to a FE system that is electronically ideal.  

In FE literature, the apex (``a") version of equation (\ref{zeta-L}) has been written in several equivalent forms, namely (using the notation preferred here)
\begin{equation}
(-E_{\rm{a}}) = \beta_{\rm{a}} V_{\rm{M}} = V_{\rm{M}}/{\zeta_{\rm{a}}} = V_{\rm{M}}/{k_{\rm{a}} r_{\rm{a}}}.
\label{E-VM}    
\end{equation}

The parameter $\beta_{\rm{a}}$ is defined as $1/\zeta_{\rm{a}}$, and is also written ``$\beta$" and ``$\beta_V$" in the literature. These ``betas" are conveniently measured in m$^{-1}$. The symbol ``$\beta$" has been widely used in older literature (e.g. \cite{DykeTrolan}) but clashes with the widespread modern use of $\beta$ to denote the dimensionless field enhancement factor denoted below by $\gamma$.

Confusingly, the parameter $\beta_{\rm{a}}$ in the equation $|E_{\rm{a}}| = \beta_{\rm{a}} V_{\rm{m}}$ is also sometimes called a ``field enhancement factor". This dual use of the symbol $\beta$ and the term ``field enhancement factor" can and sometimes does give rise to misunderstandings; hence, we now prefer to use $\zeta_{\rm{a}}$. More generally, our view is that a parameter measured in nm is easier to interpret than its reciprocal.

When eq. (\ref{zeta-mL}) is rewritten in the form $V_{\rm{m}}=\zeta_{\rm{ma}} |E_{\rm{a}}|$, it is readily seen that $\zeta_{\rm{ma}}$ is a measure of how easy it is to ``turn an emitter on" in a given FE system. If, say, the emission onset value for $|E_{\rm{a}}|$ is 2 V/nm, then a $\zeta_{\rm{ma}}$-value of 10 nm implies a turn-on voltage of 20 V, whereas a $\zeta_{\rm{ma}}$-value of 100 nm implies a turn-on voltage of 200 V, etc. Note that $\zeta_{\rm{ma}}$ and $\zeta_{\rm{a}}$ are characterization parameters, not physical distances.

In eq. (\ref{E-VM}), $r_{\rm{a}}$ denotes the emitter apex radius, and the dimensionless parameter $k_{\rm{a}} =\zeta_{\rm{a}}/r_{\rm{a}}$ is called a \textit{shape factor}, or sometimes a ``field factor". The form of eq. (\ref{E-VM}) involving $k_{\rm{a}}$ is more commonly used in field ion emission contexts, but is also useful for the needle-like emitters used in electron microscopes and traditional field electron microscopes.

Equation (\ref{E-VM}) is also sometimes seen in the form
\begin{equation}
E_{\rm{a}} = E_{\rm{free}} / k_{\rm{a}} ,
\label{Efree}    
\end{equation}
where $E_{\rm{free}}$ is the field at the surface of a ``free" sphere charged to the same potential as the needle-like emitter.

For electronically ideal FE systems, empirical values of a characteristic LVCL (usually interpreted as an apex LVCL) can be extracted from current-voltage characteristics as analysed with Fowler-Nordheim plots (e.g., \cite{ForbesJordan}) or (better) with Murphy-Good plots \cite{For19a}. However, applying these procedures to non-ideal systems can generate spurious LVCL values \cite{Forbes2013}. (Also see \cite{MMA}.)

This article is primarily about the ``pure" electrostatics of electronically ideal FE systems.

\subsection{Basic geometrical and electrical definitions}

\textit{Emitter geometry.} The main emitter geometry considered in this review is a cylindrically symmetric protrusion, with its apex labelled ``a", having total height $h$, apex radius of curvature $r_{\rm{a}}$, and defined shape. This protrusion stands on an \textit{emitter substrate}, with the protrusion symmetry axis passing through a point ``b" on the substrate surface.

The ratio $(h/r_{\rm{a}})$ has an important theoretical role: we call it the \textit{apex sharpness ratio} and denote it by $\sigma_{\rm{a}}$. Thus
\begin{equation}
\sigma_{\rm{a}} \equiv h / r_{\rm{a}} .
\label{Apexsigma}    
\end{equation}

The substrate plus the protrusion forms the \textit{emitter electrode}. In the diode geometries discussed in this review, the emitter faces a single counter-electrode, called the \textit{collector}. In many cases of theoretical interest, the emitter substrate will be one of a pair of well-separated parallel planar plates, of large lateral extent compared to their separation $d_{\rm{sep}}$: as already indicated, this will be called \textit{PPP geometry.} Usually, it will be assumed that the protrusion height $h \ll d_{\rm{sep}}$.

However, a later Section formally extends analysis to the situation where the protrusion stands at the apex ``b" of a substrate of defined non-planar geometry, with apex radius of curvature much larger than $r_{\rm{a}}$. In this case, the substrate electrostatics can sometimes be the electrostatics of the needle-shaped emitters used in electron microscopes and in field electron and field ion microscopes, i.e. so-called \textit{needle geometry}. It will be useful to consider this substrate electrostatics first, in Section \ref{ClasNeedle}.

In principle, field emitter electrostatics may need to consider other geometries---for example those involving concentric spheres, cylindrical wires, blade-shaped emitters or parallel plates of limited lateral extent. However, this review focuses mainly on PPP and needle geometries.
\bigskip

\textit{Macroscopic fields.} As discussed in more detail below, when a protrusion is present, then so-called \textit{macroscopic-field enhancement factors (MFEFs)} can be defined. A local MFEF $\gamma_{\rm{ML}}$ applies to some specific location ``L" on the protrusion surface and is defined by an equation of the form
\begin{equation}
\gamma_{\rm{ML}} = E_{\rm{L}} / E_{\rm{M}}.
\label{gam-ML1}    
\end{equation}
Here, $E_{\rm{L}}$ is the local electrostatic field at location ``L", and the so-called \textit{macroscopic field} $E_{\rm{M}}$ is an electrostatic field (different in different circumstances) that characterises the FE system geometry more generally.

For the electronically ideal FE systems under consideration, the value of $E_{\rm{M}}$ depends on the interelectrode voltage $V_{\rm{M}}$ and the system geometry, and is given formally by   
\begin{equation}
E_{\rm{M}} = -V_{\rm{M}} / \zeta_{\rm{M}},
\label{zeta-b}    
\end{equation}
where $\zeta_{\rm{M}}$ is the \textit{(zero-current) macroscopic voltage conversion length (MVCL)}. $V_{\rm{M}}$ is considered positive if the collector is positive relative to the emitter electrode.

For electronically ideal FE systems, all parameters are defined in the zero-current approximation (i.e., it is assumed there are no significant voltage differences in the system other than that across the high voltage generator). (However, in-system voltage differences can occur in non-ideal systems---see Sections \ref{SerResSect} and \ref{voltloss}.)

From the above, it follows formally that
\begin{equation}
\gamma_{\rm{ML}} = \zeta_{\rm{M}} / \zeta_{\rm{L}}.
\label{gam-ML2}    
\end{equation}


Macroscopic fields fall into two broad types: base-fields ($E_{\rm{B}}$ or $E_{\rm{b}}$); and gap-fields ($E_{\rm{G}}$).

A \textit{base-field} is the field-value that exists at substrate point ``b" as defined earlier, in the absence of the protrusion and of certain effects that would be induced by its presence. (More careful discussions are given later.)

For a given interelectrode voltage, the value of $E_{\rm{B}}$ will depend on the functional form and value of the corresponding VCL $\zeta_{\rm{B}}$, which in turn will depend on the system geometry. For PPP geometry, $\zeta_{\rm{B}}$ becomes given by the \textit{plate separation} $d_{\rm{sep}}$, and $E_{\rm{B}}$ becomes the \textit{inter-plate field} $E_{\rm{P}}$ (also called the ``plate field") given by
\begin{equation}
E_{\rm{P}} = -V_{\rm{M}} / d_{\rm{sep}} \;\; or \; = -V_{\rm{P}} / d_{\rm{sep}} .
\label{E-P1}    
\end{equation}
For notational consistency, in PPP geometry the symbol ``$V_{\rm{M}}$" can be replaced by the symbol ``$V_{\rm{P}}$", although the meanings are the same.

Other geometries involving base-fields are discussed below.

Note a change in terminology as compared with some earlier papers. What was previously called a ``macroscopic field" is now being called an ``(inter-)plate field" and is being denoted by $E_{\rm{P}}$ rather than $E_{\rm{M}}$ (or $|E_{\rm{P}}|$ would be used instead of $F_{\rm{M}}$), and the term ``macroscopic field" is being used in a more general sense.

A \textit{gap-field} $E_{\rm{G}}$ is defined as the mean field between the \textit{apex} of the protrusion and an adjacent counter-electrode, between which a voltage $V_{\rm{m}}$ exists. $E_{\rm{G}}$ is given by 
\begin{equation}
E_{\rm{G}} = -V_{\rm{M}} / d_{\rm{gap}} .
\label{E-P2}    
\end{equation}
where $d_{\rm{gap}}$ is the gap length.

Gap-fields  are used in various geometrical situations, including close collector adjacency in PPP-type geometry, and more generally in point-point and point-plane geometries.
\bigskip

\textit{Macroscopic field enhancement factors.} Associated with the different types of macroscopic field there are different types of \textit{macroscopic field enhancement factor (MFEF)}. In some situations (e.g., PPP geometry) both types of MFEF can be defined. At large plate separations the two types have similar values, but as the plate separation diminishes to the point where the gap length is comparable with the protrusion height the behaviours of the two MFEF types diverge (see Section \ref{adjacency}).

More generally, it is often not electrostatically legitimate to compare the numerical values of MFEFs of different kinds, as is sometimes done in the literature.

With field enhancement factors, some considerations and formulae apply to all MFEF types: these common formulae retain the label ``M" in the subscript. Some considerations and formulae apply only to the particular type of MFEF under discussion: in these cases the label ``M" is replaced by an appropriate alternative (e.g., ``P", in the case of base-FEFs in PPP geometry). Some generally applicable formulae are now discussed. Many detailed formulae are given in later Sections.

When the characteristic location ``C" is of interest, eqns (\ref{gam-ML1}) and (\ref{gam-ML2}) are replaced by
\begin{equation}
\gamma_{\rm{MC}} = E_{\rm{C}} / E_{\rm{M}}  =  \zeta_{\rm{M}} / \zeta_{\rm{C}} .  
\label{gam-MC}    
\end{equation}

When the characteristic location is taken at the emitter apex ``a", then this is replaced by
\begin{equation}
\gamma_{\rm{Ma}} = E_{\rm{a}} / E_{\rm{M}}  =  \zeta_{\rm{M}} / \zeta_{\rm{a}} .  
\label{gam-Ma}    
\end{equation}

For electronically ideal FE systems, these last two formulae are used for deriving FEF-values from the VCL-values extracted from data-analysis plots, but will yield correct results only if the parameter $\zeta_{\rm{M}}$ has been determined correctly for the actual geometry of the FE system used.

 In general, the value of $\zeta_{\rm{M}}$ may depend on the whole geometry of the FE system and thus may be difficult to calculate accurately. Most basic modelling and data-analysis use some sort of simple ``special-case" geometry. 

In the special case of PPP geometry, the formula for the ``apex plate-FEF (PFEF)" is
\begin{equation}
\gamma_{\rm{Pa}} = E_{\rm{a}} / E_{\rm{P}}  =  d_{\rm{sep}} / \zeta_{\rm{a}} .  
\label{gam-Pa}    
\end{equation}

In PPP geometry (and also in some other geometries), the formula for the apex gap-FEF (GFEF) is
\begin{equation}
\gamma_{\rm{Ga}} = E_{\rm{a}} / E_{\rm{G}}  =  d_{\rm{gap}} / \zeta_{\rm{a}} .  
\label{gam-Ga}    
\end{equation}

A merit of PPP geometry is that, provided $d_{\rm{sep}}$ is much greater than the protrusion height $h$ (say, $d_{\rm{sep}}/h > 5$) then the values of $\gamma_{\rm{ML}}$ (and hence $\gamma_{\rm{Ma}}$) are effectively independent of $d_{\rm{sep}}$, and depend only on the protrusion shape. The value of $\gamma_{\rm{Pa}}$ thus characterises the sharpness of the emitter.  This basic theoretical situation, and related situations where arrays of identical emitters on the emitter plate are involved, has attracted much theoretical attention, and is a primary focus of this review. 

A related situation of practical interest (e.g., \cite{Lauritsen_PhD}) is the concentric cylinders case, where (like PPP geometry) an exact analytical expression can be given for $\zeta_{\rm{M}}$.

Usually, it will be difficult to relate a gap FEF to the true electrostatics of the real geometry, and (except for very small gaps) the value of the gap FEF will vary with the gap length (see Section \ref{adjacency}). Thus, gap FEFs do not depend \textit{only} on the emitter shape, and can have only limited value as emitter characterization parameters. 

We stress that formulae (\ref{gam-MC}) to (\ref{gam-Ga}) are ``purely electrostatic" formulae that have been formulated for electronically ideal FE systems, and should not be applied blindly to systems that have not been subjected to some form of validity check that the related FE system is ideal.

\begin{figure}
\includegraphics [scale=0.06] {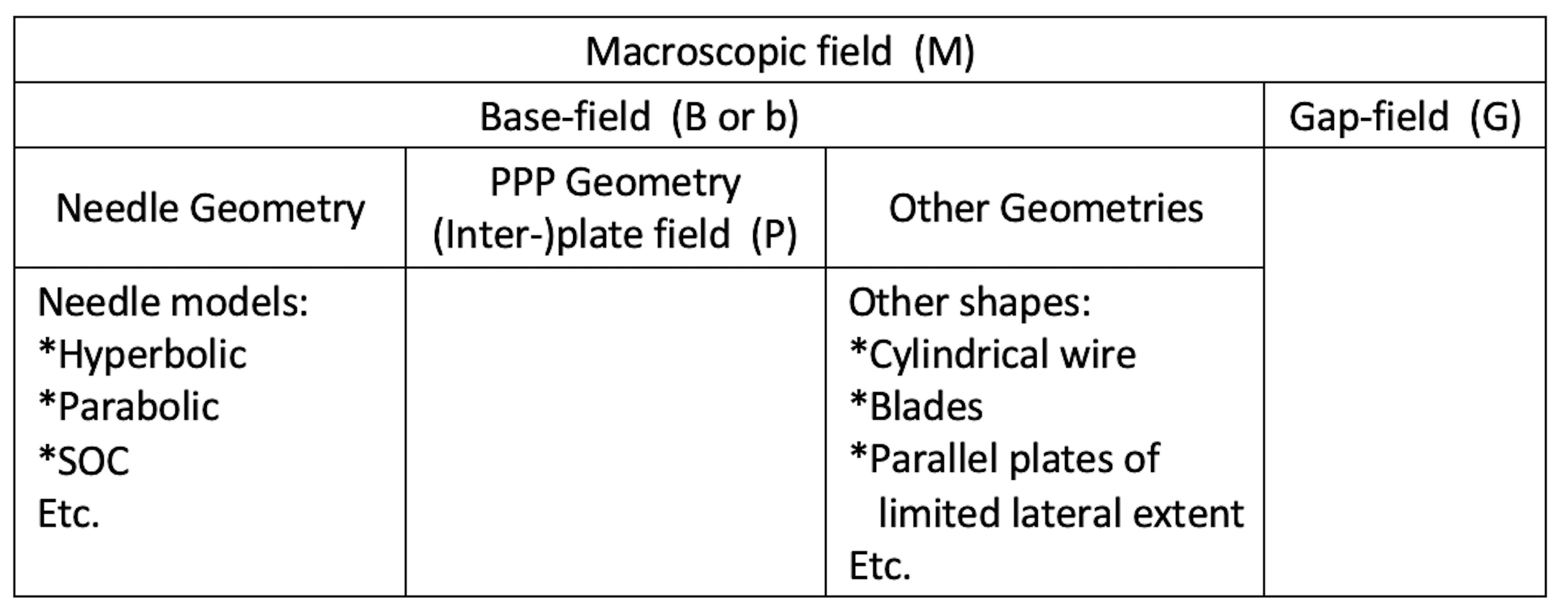}
\caption{The hierarchy of terms being used to label specialised types of macroscopic field and the related field enhancement factors.}
\label{FieldTerms}
\end{figure}

Figure \ref{FieldTerms} provides a visual reminder of the hierarchy of terms currently being used to label fields and field enhancement factors.

\subsection{Basic principles of the physics of charged metal-like conductors}

The classical theory of electricity is built on work first comprehensively summarised by Maxwell \cite{Max1873}, and on many related later classical developments. However, in the early decades of the 20th Century, it became clear that the theory of the behaviour of charged metal and metal-like conductors is (at a more basic level) the theory of electron behaviour in these conductors, and is thus governed by the principles of chemical thermodynamics, quantum mechanics and quantum statistical mechanics.

In particular, as shown by Fowler \cite{Fow36}, an appropriately defined electron Fermi level acts as the chemical potential for electrons. In a connected system of electrical conductors in which there are no current-driving elements such as batteries, the condition for static electrical equilibrium is that the Fermi level be constant throughout the system. This is the situation in which the relevant Gibbs function (that includes electron energy and work terms) has its minimum value. Constancy of the Fermi level throughout the system is the \color{black}atomic-level equivalent of the classical condition that voltage be constant throughout the system. 

It has long been known empirically that, when a system is in static electrical equilibrium, the electrostatic field is highest over the most sharply pointed features of the system, with the field being particularly high at the apex of any spike-like feature of the system surface. The fundamental underlying reasons for this are poorly explained both in the electrostatics literature as a whole and in most electricity and magnetism textbooks. At the classical level, the most careful discussion we know of is the 1886 discussion of Robin \cite{Rob1886}, who (in effect) constructed and minimised an electrostatic energy functional involving the charge distribution across the surface of an irregularly shaped body. A modified version of this approach (using point charges at surface atom positions) has been developed by Rolland \textit{et al.} \cite{Rolland15} in the context of atom probe microscopy, in order to model the process of field evaporation.

In an atomic-level approach, the requirement (as just indicated) is that the relevant Gibbs function have its minimum value and hence that the Fermi level be constant throughout the emitter electrode. When the simplifying assumption is made that the local work function has the same uniform value for all conducting surfaces in the system, then the condition for static electrical equilibrium of the emitter electrode becomes that the electrostatic potential is uniform across the surface of the emitter electrode. Across the surface of the counter-electrode a different constant electrostatic potential can be assumed. These conditions allow the field distribution across the emitter electrode (or relevant parts of it) to be determined using classical electrostatics.

Although there are a few configurations that can be exactly solved analytically (and hence have an important role in validating approximate and numerical solutions), most geometrical configurations cannot be exactly solved analytically. At a classical level, there are two main ways of proceeding: (a) charge-distribution methods; and (b) numerical solution of Laplace's equation.

In a charge-distribution method, one determines an arrangement (positions and magnitudes) of point charges and/or dipoles that yields the correct electrostatic-potential value(s) at a relevant number of chosen points on the active conductor surface (in practice, on the protrusion surface). Then, using these charges and dipoles, and Coulomb-law arguments, one can determine the electrostatic field(s) at a relevant point or relevant points on the protrusion surface. In practice, it is often only the electrostatic field at a protrusion apex that is of significant  interest. The simpler forms of this method can be easy to apply and can lead to simple formulae that are physically instructive, although not accurate. The more sophisticated forms lead (in the cylindrically symmetric geometries of interest here) to so-called ``line-charge models". 

Several different techniques exist for solving Laplace's equation numerically. These include finite element methods, finite difference methods and boundary element methods. For numerical work, this paper uses finite element methods exclusively. Relevant basic principles and issues are outlined in Section \ref{FEM}.

\section{Structure of remaining discussion}

The structure of the remainder of this review is as follows. We first review the existing simple emitter models, mainly (but not exclusively) those related to needle geometry (Section \ref{ClasNeedle}) and PPP geometry (Section \ref{AFEFSCM}). Models relating to physical emitter shape and simplified electrostatic models are discussed.  The next two Sections then deal with the theory of finite-element electrostatic simulations, including (a) a brief outline of the basic theory (Section \ref{FEM}) and (b) discussion of the theories of our ``Minimum Domain Dimensions" (MDD) methodology (Section \ref{MDD}) and of the variant of it that enables high-precision estimates of dimensionless inter-plate field enhancement factors (Sub-section \ref{extrap}).

For electronically ideal FE systems, these methods are then applied: (a) to single-emitter shapes that have no easy analytic solution (Section \ref{OtherSingle}); and (b) to pairs of identical post-like emitters  where electrostatic depolarization is an important effect (Section \ref{DepolSect}).  In both cases there is particular emphasis on the hemisphere-on cylindrical-post (HCP) emitter model. Section \ref{AdvSingle} provides brief discussion of (and some core references for) some additional effects that may be important in practical situations, including some effects that cause FE systems to be electronically non-ideal.

Finally, Section \ref{Sum&Tasks} provides a summary and an indication of some areas of field emitter electrostatics where additional research and/or mini-reviews would probably be useful. An Appendix discusses units for electric dipole moment.

\section{Apex shape-factors for some simple classical models}
\label{ClasNeedle}

The Subsections below relate to the pure electrostatics of the substrate. Some good analytical results for various well-defined shapes, and then some approximate formulae, are given. In these formulae, $\zeta_{\rm{b}}$ is the substrate-apex VCL, as defined by eq.(\ref{zeta-b}), and $d$ represents the distance between the emitter electrode and the counter-electrode, along a radius or the system axis, as appropriate.

\subsection{Some good analytical results}
\label{Class-Good-Anal}

\noindent (1) For \textit{coaxial cylinders} an exact formula is
\begin{equation}
\zeta_{\rm{b}}  = R_{\rm{cyl}} \ln(1 + d / R_{\rm{cyl}}).
\label{cylinders}    
\end{equation}
where $R_{\rm{cyl}}$ is the cylinder radius.

\noindent (2) For \textit{concentric spheres} an exact formula is 
\begin{equation}
\zeta_{\rm{b}} = R_{\rm{b}} d / (R_{\rm{b}} + d).
\label{concsph}    
\end{equation}
This formula has two limiting cases, as follows.

\noindent (3) When $R_{\rm{b}}  \gg d$ then
\begin{equation}
\zeta_{\rm{b}} \approx d.
\label{planar}    
\end{equation}
In the limit of very large $R_{\rm{b}}$ this becomes PPP geometry.

\noindent (4) When $d \gg R_{\rm{b}}$ then
\begin{equation}
\zeta_{\rm{b}} \approx R_{\rm{b}} .
\label{freesph}    
\end{equation}
In the limit of very large $d$ this becomes \textit{free-sphere geometry}.

As already noted, the practical situation where---as in a traditional field electron or field ion microscope---a needle-shaped emitter faces a distant counter-electrode is called here ``needle geometry". With needle geometry, it is more convenient to use the form $\zeta_{\rm{b}}=k_{\rm{b}}R_{\rm{b}}$, with $k_{\rm{b}}>1$ and to specify a value for $k_{\rm{b}}$. The electrostatic argument is that, for a given contribution to apex potential, the contribution to apex field is less when a charge element is distant from the apex. Expressions and values for $k_{\rm{b}}$ depend on the needle-model used. The commonly used models are as follows.

\noindent (5) \textit{Hyperboloid model}. This model appears to have been introduced by Eyring et al. \cite{Eyring28} in 1928 but the Coelho and Debeau 1971 discussion \cite{Coelho71} is mathematically clearer. For a \textit{hyperboloid} facing a distant plane, a distance $d$ from its apex we have
\begin{equation}
{\zeta_{\rm{b}}(\rm{hyperb})} = k_{\rm{b}} R_{\rm{b}} \approx R_{\rm{b}} \times \tfrac{1}{2} \ln(4d / R_{\rm{b}}).
\label{hyper}    
\end{equation}
This is a satisfactory approximation if $d \gg R_{\rm{b}}$ \cite{Coelho71}.

\noindent (6) \textit{Paraboloid model}. This model appears to have been introduced by Becker \cite{Becker51} in 1951, but the Rose 1956 discussion (see Appendix I in \cite{Rose56}) is mathematically clearer. The \textit{focal length} $z_{\rm{f}}$ is the distance from the paraboloid \textit{focus} to its \textit{vertex} (i.e., the paraboloid apex), and the apex radius of curvature $R_{\rm{b}} = 2 z_{\rm{f}}$. The distance $d$ between the apexes of two confocal paraboloids with focal lengths $z_{\rm{fe}}$ (modelling the emitter) and $z_{\rm{fc}}$ (modelling the counter-electrode) thus is $d=z_{\rm{fc}}-z_{\rm{fe}}$. From eq. (15) in \cite{Rose56} it follows that
\begin{equation}
{\zeta_{\rm{b}} = z_{\rm{fe}} \ln{(z_{\rm{fc}} / z_{\rm{fe}})}}  =  {R_{\rm{b}} \times \tfrac{1}{2} \ln{[(d/z_{\rm{fe}})+1]}} , 
\label{parab1}    
\end{equation}
\begin{equation}
{\zeta_{\rm{b}}(\rm{parab})} = k_{\rm{b}} R_{\rm{b}} \approx R_{\rm{b}} \times \tfrac{1}{2} \ln(2d / R_{\rm{b}}).
\label{parab2}    
\end{equation}
This paraboloid formula is sometimes given incorrectly in the literature.

There is a question of what constitutes a good model for a field emitter needle. Probably, a good model should ideally provide adequate representations of all of the following:

\noindent (a)  The local shape of the needle close to its apex.

\noindent (b)  The shape of the needle distant from its apex.

\noindent (c)  The shape (and location) of the counter-electrode in front of the needle-like apex.

\noindent (d)  The shape and location of the counter-electrode ``around the sides of" the needle, both above and below the needle apex.

\noindent (7) Thus, possibly the best analytical model developed so far for fitting the shapes of real needle-like emitters is the \textit{sphere-on-orthogonal-cone} (SOC) model, introduced by Dyke and colleagues \cite{DykeSOC}. The slightly complicated details are well described in the Dyke et al. paper and will not be reproduced here. This model provides a mathematical description of how (the reciprocal of) the LVCL varies across the emitter surface and down the shank.

\subsection{Some approximate formulae for shape factor}
\label{Approx.}

Notwithstanding the above analytical formulae, simpler (approximate) numerical formulae have been developed for use in needle geometry, and have found much empirical use.

\noindent (8) The simplest is the well-known ``rough" formula
\begin{equation}
k_{\rm{b}} \approx 5 .
\label{k-five}    
\end{equation}
This is stated by Gomer in a 1955 review article \cite{Gom55} (see p. 97), but without detailed justification.

Using the hyperboloid formula (\ref{hyper}), Coelho and Debeau \cite{Coelho71} show that $k_{\rm{b}}$ varies from 3, for $d/R_{\rm{b}}=10^2$, to 5 for $d/R_{\rm{b}}=10^4$, to 6.5 for $d/R_{\rm{b}}=10^5$. For a field electron microscope, one might have $d \approx 10^{-2}$ m, $R_{\rm{b}} \approx 10^{-6}$ m, i.e. $d/R_{\rm{b}}=10^4$; thus, $k_{\rm{b}} \approx 5$ looks a plausible value.

\noindent (9) However, other researchers have suggested that their emitters conform better to shape-factor values that lie somewhere in the range
\begin{equation}
3 < k_{\rm{b}} < 8 .
\label{k-3to8}    
\end{equation}

\noindent (10) An empirical formula found by Charbonnier (\cite{Gom61}, see p. 45), apparently by fitting SOC-model profiles to electron micrographs of field electron emitters, is
\begin{equation}
k_{\rm{b}} \approx 0.59 \alpha_{\rm{c}}^{1/3} (d/R_{\rm{b}})^{0.13} ,
\label{k-0.59}    
\end{equation}
where $\alpha_{\rm{c}}$ is the half-angle of the cone that defines the shape of the needle shank. This approach is said to generate shape-factor values close to 5 in many practical cases.

\noindent (11) Biswas \cite{BiswasFEF,SarBisEnd} has presented a line-charge analysis of needle geometries similar to those discussed above.

We emphasize that the simple formulae discussed above do \textbf{not} apply well to the post-on-plane shapes discussed next.

\section{Apex field enhancement factors for some simple post-on-plane models}
\label{AFEFSCM}

This Section presents the pure  electrostatics of some simple classical post-on-plane models, either because they are analytically exact (and thus useful as reference situations for comparison with approximate or analytical results) or because they are used widely. The parameter of interest is nearly always the apex plate-FEF $\gamma_{\rm{Pa}}$ defined by eq. (\ref{gam-Pa}), using the inter-plate field $E_{\rm{P}}$.

As part of this review, we wish to introduce standard acronymns/labels for the shape models most commonly used. These labels are used in the subsection titles and are also listed in Table \ref{tab-models} , together with any older labels now considered obsolete. The label ``HECP" was proposed in \cite{SarBisEnd}.  
\begin{table*}
\caption{Proposed labels for some simple emitter-shape and electrostatic (ES) models.}
\bigskip
\begin{tabular}{|c|c|c|c|}
\hline
Model type  & Model label & Model name & Older labels\\ & & & (if any)  \\
\hline
\hline
Shape & HSP & Hemisphere-on-plane  & HP  \\ \hline
Shape & HEP & Hemiellipsoid-on-plane & Elli \\ \hline
Shape & HCP & Hemisphere-on-cylindrical-post &   \\ \hline
Shape & HECP & Hemiellipsoid-on-cylindrical-post & \\ \hline
Shape & SOC & Sphere-on-orthogonal-cone &  \\ \hline
Shape & SRC & Spherically rounded cone & hSoC  \\ \hline
ES & FSEPP & Floating sphere at emitter plate potential & \\ \hline
ES & BFSSP & Basic floating sphere at substrate potential & \\ \hline
ES & TCS & Two-connected-spheres & \\ \hline
\end{tabular}
\label{tab-models}
\end{table*}

\subsection{Hemisphere-on-plane (HSP)}
\label{AFEFHOP}
A particularly simple physical situation is the hemisphere of radius $r_{\rm{H}}$ on an infinite plane, in a uniform applied field $E_{\rm{P}}$. This is modelled by placing, at the centre of the hemisphere, a dipole of ISQ polarisability $\alpha_{\rm{dip}} = 4\pi \epsilon_0 r_{\rm{H}}^3$. At the hemisphere apex, the field due to the dipole is $E_{\rm{dip}}=2E_{\rm{P}}$. This adds to the applied field, and results in the apex-FEF value $\gamma_{\rm{Pa}}=3$. Fields and potentials at other points on the hemisphere can be obtained by combining results due to the usual dipole formulae with those related to the applied field. Note that the total field at any point where the hemisphere intersects the plane is zero, and that at all points on its surface the electrostatic potential of the hemisphere is (necessarily) equal to that of the plane.

\subsection{Hemi-ellipsoid on a plane (HEP)}
\label{AFEFHEP}

Another analytically solvable situation is the cylindrically symmetric hemi-ellipsoid on an infinite plane, in a uniform applied field. Let the lengths of the semi-major axis (or ``height") be $h$, and of the semi-minor axis (or ``base-radius") be $\rho$, with $h>\rho$. An expression for $\gamma_{\rm{Pa}}$, in the form used by Forbes \textit{et al.} \cite{Forbes2003}, is as follows. The axis ratio (or \textit{aspect ratio}) $\nu$ of the ellipsoid, a parameter $\xi$, the apex radius of curvature $r_{\rm{a}}$, and the apex plate-FEF $\gamma_{\rm{Pa}}$ are given by
\begin{equation}
 \nu \equiv \frac{h}{\rho},
 \label{nu}
 \end{equation}
\begin{equation}
 \xi \equiv  \left(\nu^2 -1 \right)^{1/2},
 \label{zetadef}
 \end{equation}
\begin{equation}
r_{\rm{a}} = \frac{\rho^2}{h} = \frac{\rho}{\nu},
 \label{rarb}
 \end{equation}
\begin{equation}
\gamma_{\rm{Pa}} = \frac{\xi^3}{\left\{\nu\ln{\left[\nu + \xi\right]} - \xi\right\} }.
 \label{gama1ellip}
 \end{equation}
It also follows from (\ref{nu}) and (\ref{rarb}) that

\begin{equation}
\nu \equiv \frac{h}{\rho} = \left(\frac{h}{r_{\rm{a}}}\right)^{1/2} =  \sigma_{\rm{a}}^{1/2}.
 \label{arrarb}
 \end{equation}

These results are based on the form given by Latham \cite{Lat81} and attributed to Rohrback \cite{Roh71}.  A more complicated (but equivalent) form of the same result is proved by Komsahl, eq. (10) in \cite{Kom91}, starting from the electrostatics of ellipsoids in uniform fields, as given in standard textbooks.

Obviously, in the limit that $\nu \gg 1$ (i.e., for a long thin ellipsoid), then $\zeta \approx \nu$ and eq. (\ref{gama1ellip}) reduces to the simpler form
\begin{equation}
\gamma_{\rm{Pa}} \approx \frac{\nu^2}{\left[\ln{\left(2\nu \right)} - 1\right] }.
\label{gama1ellipapprox}
\end{equation}
Further, since $\nu^2 = h/r_{\rm{a}}$, result (\ref{gama1ellipapprox}) can also be put in the instructive form
\begin{equation}
\gamma_{\rm{Pa}} \approx c_{\rm{HEP}} \times  ( h / r_{\rm{a}} ),
\label{gama1ellipif}
\end{equation}
where the correction factor $c_{\rm{HEP}}$ is given by
\begin{equation}
c_{\rm{HEP}} = \frac{1}{ \tfrac{1}{2} \ln{ \left[ 4\left( h/r_{\rm{a}}\right) \right]} - 1},
\label{cHEP}
\end{equation}

For example, for $h/r_{\rm{a}}=100$, $c_{\rm{HEP}} \approx 0.501$. An alternative approach, based on the use of a non-uniform axial line-charge, has been developed by Pogerolov and colleagues \cite{Pog1, Pog2}. This may not be an exact general treatment, but leads again to formulae (\ref{gama1ellipif}) and (\ref{cHEP}) in the case that $\nu \gg 1$.

\subsection{The concept of driving potential difference}
\label{CDPD}

On moving away from physical models with exact analytical solutions, one may need to distinguish between: (a) exact physical models; (b) approximate physical models; and (c) related electrostatic models. For the hemisphere-on-cylindrical-post (HCP) physical model discussed below, a useful approximate physical model is the so-called ``floating sphere at emitter-plane potential" (FSEPP) model (e.g. \cite{RFJAP2016}); a related basic electrostatic model uses only a point charge $q$ placed at the sphere centre.

A slightly generalised version of this basic model is more useful, and is presented here. This generalised model does not require the substrate to be a plane and can be called the \textit{basic floating sphere at substrate potential (BFSSP) model}.

Consider the situation that exists in the absence of the protrusion (or a model for it), and in the absence of effects due to ``vertically related" electrode images or equivalent (i.e., images in the substrate or in the collector, or equivalent effects), but in the presence of any other electrostatic effects would be induced by the protrusion if present (such as depolarization of neighbouring protrusions in a cluster or array, or ``lateral" mirror images in the sides of the simulation box).

Let $(\Delta \mathit{\Phi})_{\rm{XD}}$ [$\equiv  (\mathit{\Phi}_{\rm{a}} - \mathit{\Phi}_{\rm{b}})_{\rm{XD}}$] be the potential difference (between the protrusion apex ``a" and base ``b") that is generated by ``external" electrostatic sources (i.e., sources other than those directly associated with the protrusion model and its ``vertical" electrode images). Further, let $E_{\rm{Xa}}$ be the field generated at ``a" by these external electrostatic sources. We refer to $(\Delta \mathit{\Phi})_{\rm{XD}}$ as the \textit{driving potential difference}, and to $E_{\rm{Xa}}$ as the \textit{external field contribution} to the apex electrostatic field.

When the protrusion is present, the point charge in the basic electrostatic model generates a potential difference $ (\Delta \mathit{\Phi})_{\rm{model}}$ between ``a" and ``b". Thus, the condition that the protrusion/post apex (and hence the sphere apex) be at substrate potential requires that 
\begin{equation}
(\Delta \mathit{\Phi})_{\rm{model}} + (\Delta \mathit{\Phi})_{\rm{XD}} = 0 ,
\label{Phi}
\end{equation}

The solution of this equation generates a value for $q$ and hence a \textit{primary contribution} $E_q$ to the total electrostatic field $(E_{\rm{a}})_{\rm{tot}}$ at the model apex. Thus, in this BFSSP model, the total apex field is
\begin{equation}
(E_{\rm{a}})_{\rm{tot}} = E_q + E_{\rm{Xa}} , 
\label{Field}
\end{equation}
and the apex value $(\gamma_{\rm{Ba}})_{\rm{model}}$ of the model-predicted base-field enhancement factor (BFEF) is given by
\begin{equation}
(\gamma_{\rm{Ba}})_{\rm{model}}  =  (E_{\rm{a}})_{\rm{tot}} / E_{\rm{B}} , 
\label{PFEF}
\end{equation}
where $E_{\rm{B}}$ is the base-field, defined here (as before) as given by the field at point ``b", in the absence of the protrusion and of vertically-related electrode images (or equivalent).

In the definition of $(\Delta \mathit{\Phi})_{\rm{model}}$, the related potentials $\mathit{\Phi}_{\rm{a}}$ and $\mathit{\Phi}_{\rm{b}}$ can be taken as measured relative to infinity. In circumstances where $|\mathit{\Phi}_{\rm{b}}| \ll |\mathit{\Phi}_{\rm{a}}|$ (which requires $(h/r_{\rm{a}}) \gg 1$ (say, greater than 10), the above theory can be developed further. In this case it follows from Coulomb's Law and eq. (\ref{Phi}) that $E_q  \approx  - (\Delta \mathit{\Phi})_{\rm{XD}} / r_{\rm{a}}$, and hence from eq. (\ref{Field}) that
\begin{equation}
(E_{\rm{a}})_{\rm{tot}} \approx  - (\Delta \mathit{\Phi})_{\rm{XD}} / r_{\rm{a}}  +  E_{\rm{Xa}} .    
\label{Field2}
\end{equation}

In many practical circumstances (a possible exception being when emitters in a cluster are close together), the second term on the r.h.s. of eq. (\ref{Field2}) is much smaller than the first term, and can be neglected. The formula for the model BFEF then simplifies to
\begin{equation}
(\gamma_{\rm{Ba}})_{\rm{model}} \approx  - (\Delta \mathit{\Phi})_{\rm{XD}} / E_{\rm{B}} r_{\rm{a}} . 
\label{BFEF2}
\end{equation}
As shown below, this relation between the apex BFEF and the driving PD is a key formula that can be applied in various contexts. It is called here the \textit{driving potential-difference formula}, and applies to both planar and non-planar substrates.

Formula (\ref{BFEF2}) is not an accurate formula, but is considered to display the physics clearly and to provide a ``reasonable first estimate".

\subsection{Floating Sphere at Emitter Plane Potential (FSEPP) model}

The well-known basic FSEPP-model formula is obtained from eq. (\ref{BFEF2}) by replacing $E_{\rm{B}}$ by $E_{\rm{P}}$ and setting $(\Delta \mathit{\Phi})_{\rm{XD}} = - E_{\rm{P}} h$, which yields
\begin{equation}
{\gamma_{\rm{Pa}}(\rm{FSEPP}}) \approx  h/r_{\rm{a}}. 
\label{gam-FSEPP}
\end{equation}

The FSEPP model was first applied to field emission by Vibrans \cite{Vibrans1964a, Vibrans1964}. Subsequently, many researchers \cite{Miller1,Roh71,JVSTB1993,Forbes2003,ZPCL11,RFJAP2016}  have explored both the basic model and more sophisticated versions that involve placing a dipole at the sphere centre (which makes the sphere surface an approximate equipotential), and/or various image configurations. Review-type discussions are presented in \cite{Forbes2003,ZPCL11,RFJAP2016}.

In earlier stages of the development of FE electrostatics, there was a tendency to treat the FSEPP-derived formulae as useful (albeit approximate) working formulae, and attention was given to getting the small correction terms right. With the development of the more accurate approaches discussed later, this role has fallen away, but formulae derived from eq. (\ref{BFEF2}) remain highly useful as models that give insight into the basic physics of sharply pointed conductors---which is not widely understood in the general scientific community.

\subsection{The Two-Connected-Spheres (TCS) model}
\label{gam-TCS-Sec}

The \textit{Two-Connected-Spheres (TCS) model} is another  more general version of the FSEPP model and allows basic physical discussion of the electrostatics of protrusions on spherical and on needle-geometry substrates. The substrate is treated as a sphere of large radius $R_{\rm{b}}$ that can be modelled electrostatically by placing a large point-charge $Q$ at its centre. From the electrostatics of spheres, it is readily shown that the driving PD now becomes
\begin{equation}
{(\Delta \mathit{\Phi})_{\rm{XD}}(\rm{TCS})} =  k_{\rm{C}} Q [1/(R_{\rm{b}}+h) - 1/R_{\rm{b}}]
\label{DrPD-TCS2}
\end{equation}
\begin{equation}
\approx (k_{\rm{C}} Q / R_{\rm{b}}) \left[ (1+h/R_{\rm{b}})^{-1} - 1 \right] 
\label{DrPD-TCS3}
\end{equation}
where $k_{\rm{C}} \equiv 1/4 \pi \epsilon_0$. On using binomial expansion, and then noting that (in this mode) $(k_{\rm{C}} Q / R_{\rm{b}}^2) = E_{\rm{B}}$ the above equations yield
\begin{equation}
{(\Delta \mathit{\Phi})_{\rm{XD}}(\rm{TCS})} \approx - E_{\rm{B}} h ,
\label{DrPD-TCS4} 
\end{equation}
\begin{equation}
{\gamma_{\rm{Ba}}(\rm{TCS}}) \approx  h/r_{\rm{a}}. 
\label{DrPD-TCS5}
\end{equation}
That is, the basic TCS result is the same as the basic FSEPP result. (However, as will be shown later, results are different if second-order terms are included.)

In the context of FE from a mercury whisker grown on what appears to be a nearly flat tungsten substrate \cite{Gom57,Gom58}, a version of the TCS model was used by Gomer \cite{Gom57} in 1957 to derive a formula for the shape factor $k_{\rm{a}}$ that appears in eq. (\ref{Efree}) above. In our notation, Gomer's eq. (1) becomes
\begin{equation}
E_{\rm{a}}/E_{\rm{free}} = 1/k_{\rm{a}} = h/(R_{\rm{b}}+h) \approx  h/R_{\rm{b}}. 
\label{Efree-Gomer}
\end{equation}
Since, for two spheres at the same potential, $E_{\rm{free}} / E_{\rm{b}} = R_{\rm{b}} / r_{\rm{a}}$, eq. (\ref{Efree-Gomer}) is equivalent to eq. (\ref{DrPD-TCS5}).

A different approach to the mathematical analysis of the two-connected-sphere situation has been presented by Feynman \cite{Fey63} and also by Loudin \cite{Lou21}. However, their formula predicts the apex FEF for the smaller sphere to become infinite as $R_{\rm{b}}$ becomes large; thus, we regard their analyses as either physically inappropriate or physically incorrect \cite{For21a}. This point has previously been made by Fricker \cite{Fricker}. 

A feature learnt by comparing the Feynman treatment with the treatment here is the following. For a given charge under analysis, best practice is to write down expressions for the potential difference between two well-defined points in the system model (even if this results in a small term that is later discarded); it can be poor practice to write expressions for the ``potential relative to infinity". This is because the latter practice can lead to the accidental omission of apparently small terms that turn out to be important in some contexts.

\subsection{Conducting post formula and related formulae}

It is clear that the \textit{apex sharpness ratio} $h/r_{\rm{a}}$ is the most important factor in the electrostatics of sharp points and posts. Since eqns. (\ref{gam-FSEPP}) and (\ref{DrPD-TCS5}) are not accurate, it is convenient to introduce a correction factor $c_{\rm{a}}$ and write
\begin{equation}
{\gamma_{\rm{Pa}}(\rm{CP})} = c_{\rm{a}} (h/r_{\rm{a}}). 
\label{gam-Fcondpost}
\end{equation}
This has been called the \textit{conducting post formula}.

When a conducting but electrically isolated ``floating cylindrical rod" of length $L_{\rm{rod}}$ is situated in an uniform applied field  $E_{\rm{P}}$ and aligned parallel to it, then a planar equipotential surface passes through the centre of the rod, normal to its length. The top half of this configuration is equivalent to conducting post geometry. Hence, the apex FEF $\gamma_{\rm{Pa}}(\rm{FR})$ for the floating rod is given by
\begin{equation}
{\gamma_{\rm{Pa}}(\rm{FR})} = \tfrac{1}{2} c_{\rm{a}} (L_{\rm{rod}}/r_{\rm{a}}). 
\label{gam-Fcondpost2}
\end{equation}
This has been called \textit{floating rod formula}.

For reasons that are not entirely clear, these formulae do not normally appear in electricity and magnetism textbooks, and appear not to be widely known. They promise to have significant applications outside FE when discussing electrical phenomena (particularly breakdown phenomena) associated with pointed conductors, even in the approximation $c_{\rm{a}}=1$.  

From about 2000, serious attempts were made to develop more accurate analyses of the HCP model. These included a complicated charge-distribution model devised by Xanthakis and colleagues \cite{KokXan02}, and FEM treatments by Edgcombe and Valdr\`e \cite{Edgcombe2001,Edgcombe3,Edgcombe2}. For a review in 2003 and comparison of detailed formulae, see \cite{Forbes2003}. An useful conclusion \cite{Edgcombe2001,Forbes2003} was that the approximation
\begin{equation}
c_{\rm{a}} \approx 0.7 
\label{ca-0.7}
\end{equation}
was valid to within about 25 \% in the parameter range $ 30 < (h/r_{\rm{a}}) < 2000$.

In PPP geometry, a related formula for the apex shape factor $k_{\rm{a}}$ can be obtained from the conducting post formula, by making use of eq. (\ref{gam-Pa}) in the form
\begin{equation}
k_{\rm{a}}= \frac{d_{\rm{sep}}}{\gamma_{\rm{Ma}} r_{\rm{a}} }.
\label{ka-gen}
\end{equation}
The result is
\begin{equation}
{k_{\rm{a}}(\rm{PPP})} \approx \frac{d_{\rm{sep}}}{c_{\rm{a}} h }.
\label{ka-needle}
\end{equation}
As illustrative values, we can take $d_{\rm{sep}}=25$ $\mu$m, $h=1$ $\mu$m, $c_{\rm{a}}=0.7$, which yields  
$k_{\rm{a}}(\rm{PPP}) \approx 36$. This shows that the needle-substrate approximation $k_{\rm{b}} \approx 5$ does \textbf{not} work for a post in PPP geometry. The failure of this approximation to describe a complex tip geometry was previously shown by Edgcombe and Valdr\`e \cite{Edgcombe98}.

\section{The Finite Element Method}
\label{FEM}

The electrostatic simulations in this paper are performed using a finite element method (FEM), specifically the Galerkin approach \cite{Galerkin}. There are now several software packages that implement finite-element methods; specifically we have used COMSOL, version 5.3a. Mathematical and technical details of FEM methodology are well outside the scope of this review, but we have thought it useful to provide a brief qualitative introduction.

For simplicity, the discussion here is restricted to so-called \textit{zero-current electrostatics}. That is, it is assumed that the flow of current through the emitter (and through any contact resistance at its base) does not cause any significant voltage difference between the emitter apex and the substrate on which it stands. (This is not necessarily true for all practical emitters---see Section \ref{voltloss}.)

In our work, the objective is to solve Laplace's equation in a geometry where there are no analytical solutions. In general, the geometry is 3-dimensional. However, many of the problems of interest to this review have cylindrical symmetry. In this case the solution methodology becomes ``essentially 2-dimensional";  we describe this simpler case here.

The cylindrically symmetric emitter-shape model is placed on the axis of a cylindrically symmetric \textit{simulation box} with planar top and bottom, and one considers a vertical plane that passes through the axis. The enclosed area between the emitter profile and the box outline (see later figures for examples) constitutes the \textit{solution domain}. The aim is to numerically solve Laplace's equation at a finite number of points within this domain. The solution at any other point within the domain or on its boundary is obtainable from an appropriate interpolation formula.

The positions of the solution points are decided by (a) covering the solution domain with a ``mesh" or ``net" of triangles (or tetrahedrons in a 3-dimensional domain), and (b) taking the chosen solution points to be the mesh ``nodes" common to the vertices of two or more triangles. The triangles need to be smaller in regions where the electrostatic field is higher. Hence the solution domain will usually contain several regions, each with triangles of a different size.  The meshing is done semi-automatically by the software package, in accordance with higher-level instructions provided by the user. The user also has to specify the \textit{boundary conditions} that apply at the edges of the simulation domain.

As part of the meshing procedure, a large set of approximate equations is generated that link the values of classical electrostatic potential at the various nodes. A ``starting set" of values of the node potentials is also determined.

A numerical algorithm is then used that runs repeatedly and (in effect) minimises the errors in the node potentials until some ``stopping criterion" is met. In principle, this error minimisation can be done in various detailed ways. Details of the mathematical principles and procedures used are too complicated to summarize effectively here, but can be found in standard FEM textbooks (e.g. \cite{Jin14}). The Wikipedia entry on ``Finite element methods" may also be helpful.

\section{Minimum Domain Dimensions and related topics}
\label{MDD}

For accuracy, FEM simulations that use finite-element methods require a minimum simulation domain surrounding the shape of interest. For a given system geometry, the minimum dimensions of this domain depend on the desired solution precision. If the simulation-box boundaries are inadvertently close to the shape of interest, then this will affect the calculated potentials, due to ``image-type" effects (or equivalent) resulting from the mathematical conditions applied at the box boundaries. As a result, local field enhancement factors may not be evaluated with the desired precision.

To avoid this, it is tempting to overestimate the \textit{minimum domain dimensions (MDD)} needed. However, the resulting calculations can then become seriously time and memory consuming, as the simulations become more demanding. (This applies mainly to three-dimensional and/or time-dependent models). Thus, finding the MDD for particular system geometries is of significant practical interest. 

\begin{figure}
\includegraphics [scale=0.4] {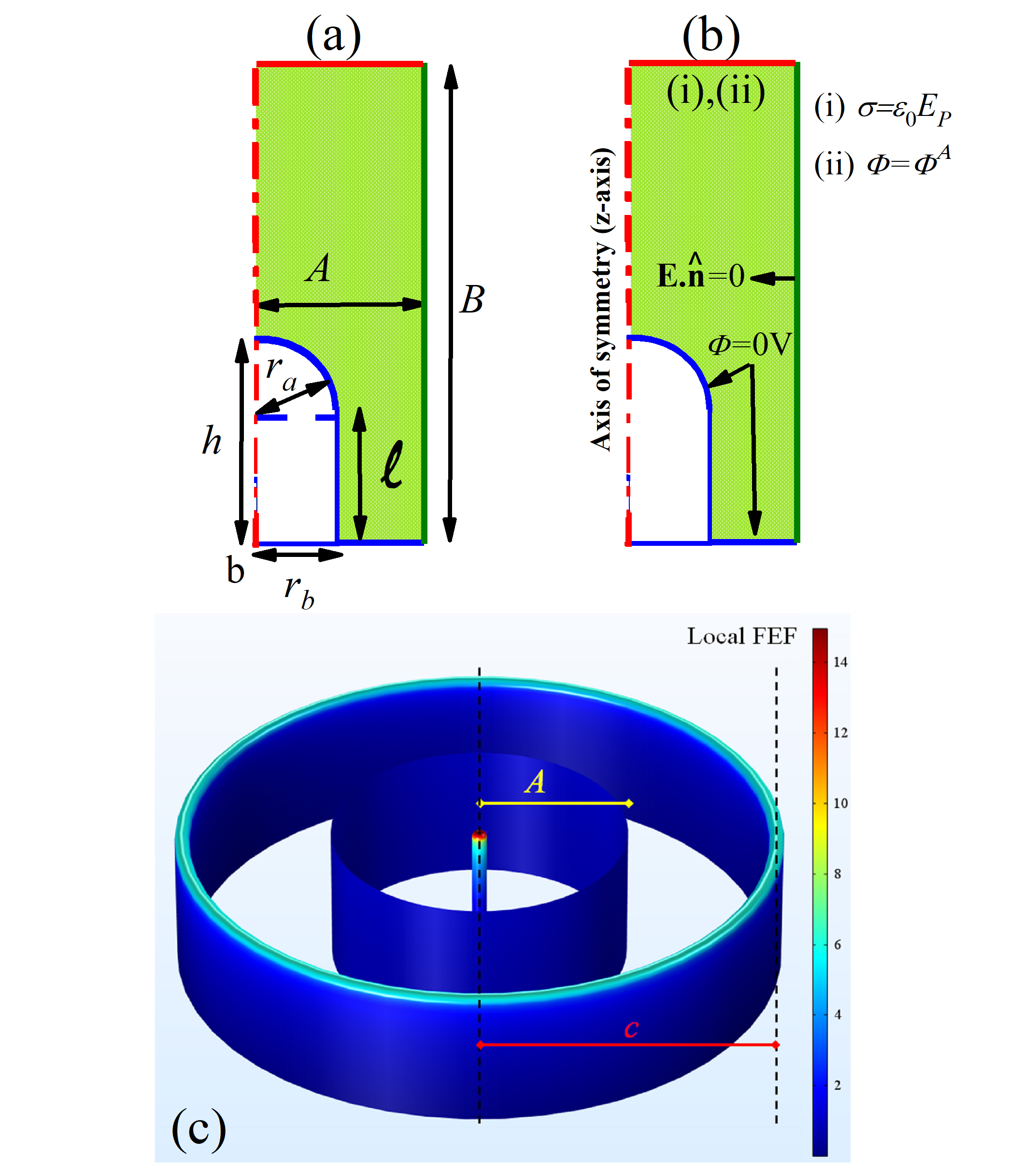}
\caption{(color online) (a) Two-dimensional profile of a hemisphere-on-cylindrical-post (HCP) model, showing defining parameters. (Note that the apex radius, $r_{\rm{a}}$, is equal to the cylinder base radius, $r_{\rm{b}}$). (b) Alternative cell-top boundary conditions (BCs): (i) a Neumann BC imposes a uniform vertically aligned electrostatic field on the boundary; (ii) a Dirichlet BC imposes an uniform electrostatic potential ${\mathit{\Phi}}_{\rm{A}}$. (c) Corresponding rotationally symmetric three-dimensional system.  The on-axis HCP model has apex sharpness ratio $h/r_{\rm{a}} = 20$, with its ring-like image with respect to the rotated right-hand boundary in (a) shown. The parameter $c$ is the distance between the emitter axis and the ring image. In the color map, a red (blue) colour indicates a higher (lower) local value of local field enhancement factor; the value $c/h= 2$ has been used.}
\label{figmdd}
\end{figure}
In 2019, Ref. \cite{JVSTB2019} investigated how the chosen domain dimensions influenced systematic errors in FEM-determined FEF-values. This was done for some simple post shapes, assuming PPP geometry. In that work the term ``minimum domain size (MDS)" was used: it is now thought that the term ``minimum domain dimensions" is clearer. This Section summarises the findings of \cite{JVSTB2019}.

Minimum domain dimensions were established by using versions of the system shown in Fig. \ref{figmdd}. This has the following characteristics. (i) The domain analyzed is rotationally symmetric. (ii) The shapes initially considered have an analytical solution for the local FEF, thereby enabling comparison between numerical and exact analytical results. (iii) The post and the bottom boundary of the simulation box have their electrostatic potentials set equal to zero (this corresponds to Dirichlet conditions).  (iv) The right-hand side boundary is set as a symmetry boundary, i.e., the boundary condition (BC) requires the electrostatic field in the boundary to be normal to the system base-plate.

(v) The post and bottom boundary are taken to be negatively charged, as in a field electron emission situation. The bottom boundary creates a positively charged electrical image of the shape, and this joins with the negatively charged original shape to create a \textit{distributed-dipole} charge distribution with its centre at point ``b" in Fig. \ref{figmdd}. The side-wall boundary condition then works as a cylindrical mirror and generates a cylindrical mirror-image of this distributed-dipole charge distribution. We refer to mirror images of this general kind as \textit{mirror dipoles}, in order to distinguish them from electrical images.

(vi) Many authors define the top boundary as the anode with given positive electrostatic potential (Dirichlet BC); in this case, a positively charged electrical image is generated by the top boundary. However, based on earlier work, this review uses a Neumann BC and imposes a vertically aligned uniform electric field $E_{\rm{P}}$ at the top boundary. It will be noted later that using this Neumann BC leads to smaller minimum domain height.

With the lateral boundary, the systematic errors (although caused by the boundary) can be conceptualised as resulting from electrostatic depolarization effects due to the electrostatic interactions between the emitter shape and its boundary-created mirror-dipole charge distribution. These effects are often called ``mutual screening" or ``field shielding", but we consider these terms to be poor physical descriptions of the underlying origin of the effect. With the top boundary, qualitatively similar systematic errors occur, but these cannot be exactly conceptualised as an image effect (see below).

\subsection{Hemisphere-on-Plane (HSP) emitters}
\label{HEP1MDS}
\begin{figure}
\includegraphics [scale=0.32] {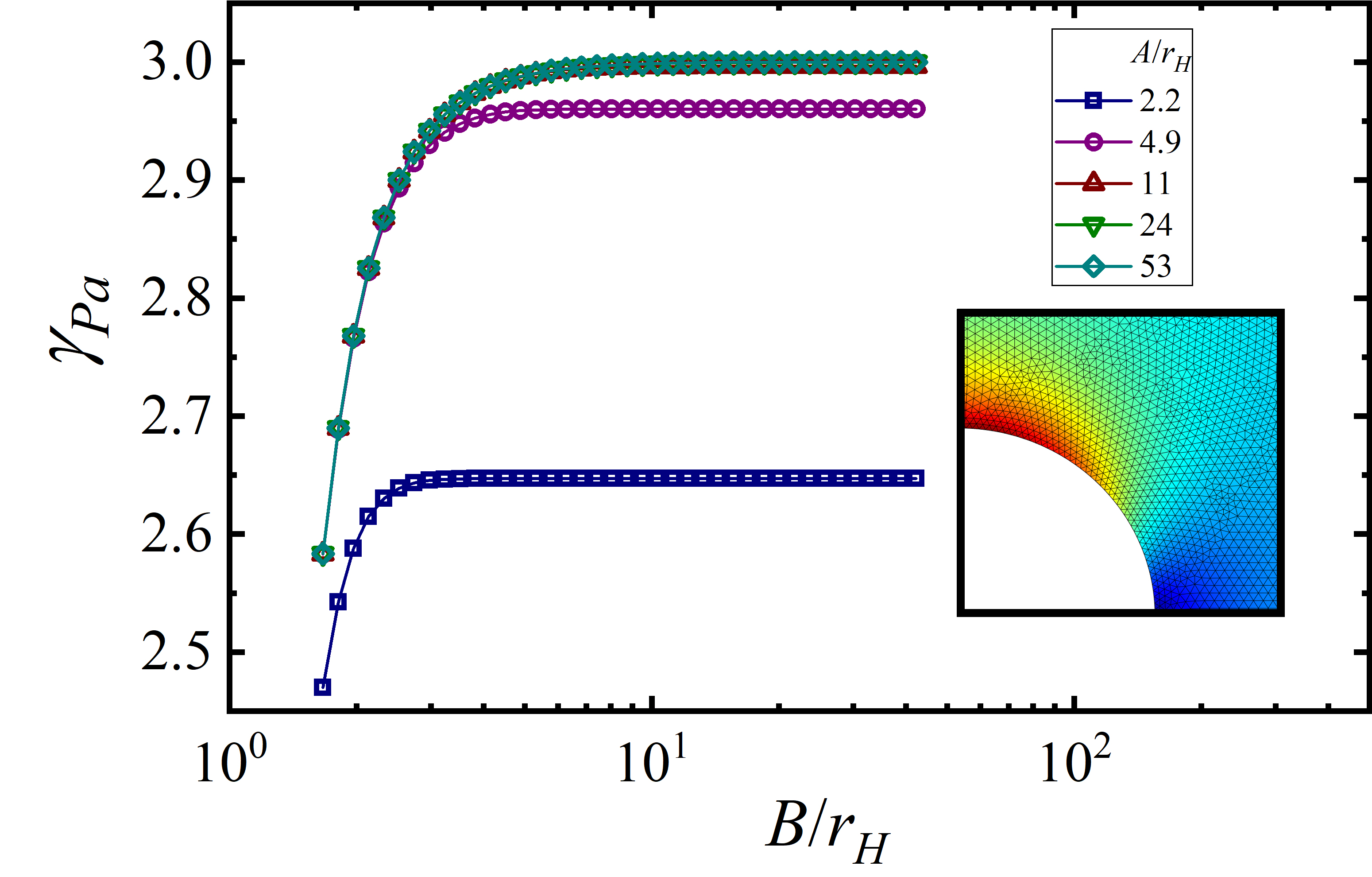}
\caption{(color online) To show, for the hemisphere-on-plane (HSP) model, with hemisphere radius $r_{\rm{H}}$, how the apex-PFEF $\gamma_{\rm{Pa}}$ depends on the height $B$ of the simulation domain, for several different values of the domain radius $A$. For the values $A/r_{\rm{H}}=6$ and $B/r_{\rm{H}}=5$, the inset shows the mesh superposed onto the solution of the local electrostatic field distribution in the vicinity of the emitter tip. In the emission region near the apex, using elements of size $r_{\rm{H}}/30$ is small enough to provide the precision goal. In the color map, red (blue) indicates  higher (lower) local electrostatic field.} 
\label{figmdsstfe}
\end{figure}

For the hemisphere-on-plane (HSP) model, Fig. \ref{figmdsstfe} illustrates how the calculated apex-PFEF value $\gamma_{\rm{Pa}}$ depends on the simulation domain dimensions. FEM simulations need to use local densities of mesh elements sufficient to ensure that related numerical errors are smaller than errors due to the finite domain dimensions.  Greater densities of FEM elements need to be used where the electrostatic fields are larger. Near the HSP emitter apex the use of element sizes smaller than $r_{\rm{a}}/30$ is usually enough to evaluate electrostatic fields with the desired precision (see inset of the Fig. \ref{figmdsstfe}); this condition (but written  as ``size  $< r_{\rm{a}}/30$")  also applies to the apexes of other emitter shapes.

The total percentage error (T\%E) $\epsilon_{\rm{t}}$ in the numerical apex-PFEF value $(\gamma_{\rm{Pa}}^{\rm{num}}$), relative to the analytical apex-PFEF value ($\gamma_{\rm{Pa}}^{\rm{anal}}$), can be defined by the formula
\begin{equation}
\epsilon_{\rm{t}} \equiv
\frac{ (\gamma_{\rm{Pa}}^{\rm{anal}} - \gamma_{\rm{Pa}}^{\rm{num}}) }
{\gamma_{\rm{Pa}}^{\rm{anal}}} \times 100 \ \%.
\label{errorreview}
\end{equation}
In this review, errors in numerical FEF values will always be defined in this general way.

\begin{figure}
\includegraphics [scale=0.30] {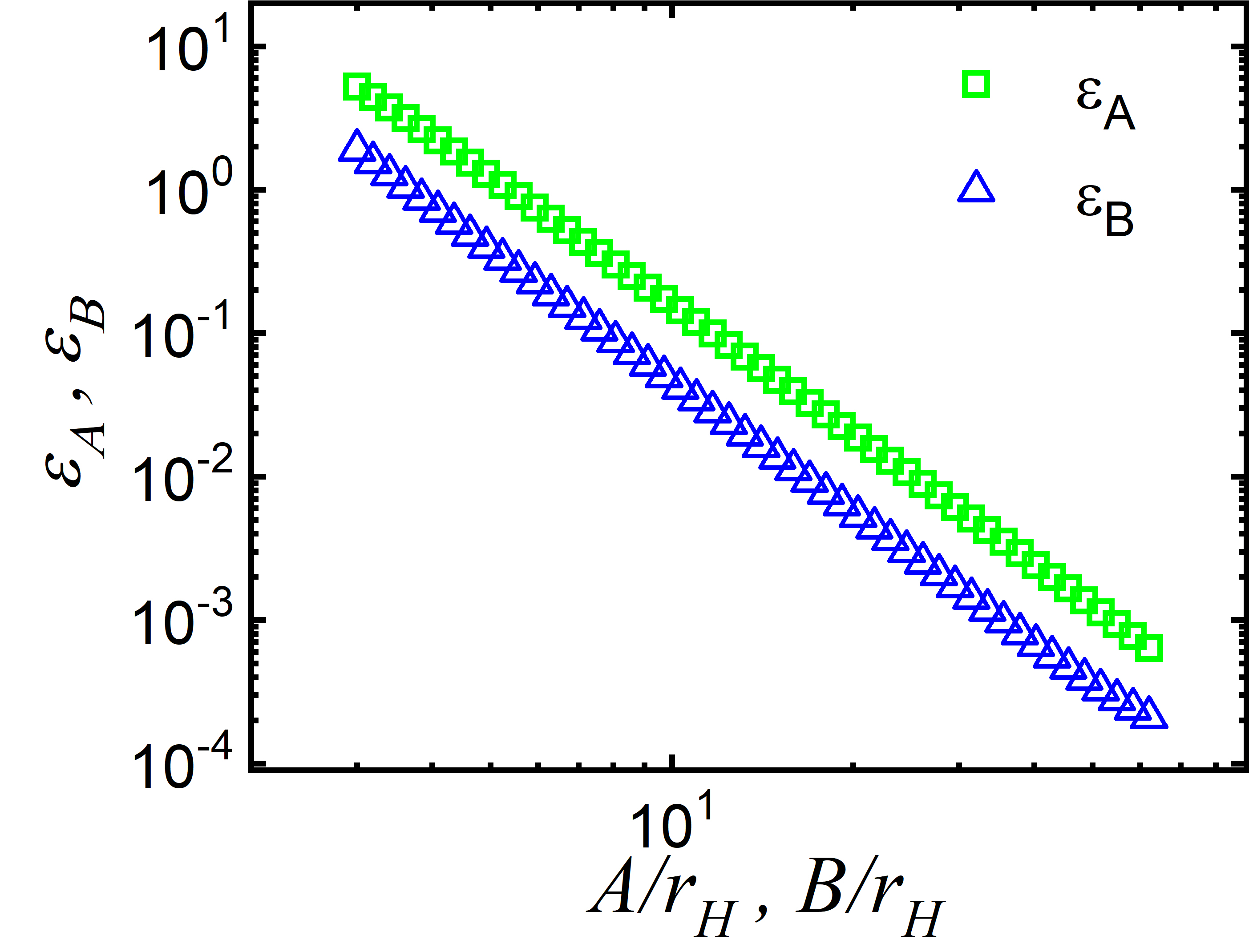}
\caption{(color online) Error plotted against one normalised domain dimension, for much larger values of the other normalised domain dimension. The diagram shows that similar (inverse cubic) behaviour occurs for both domain dimensions (radius $A$ and height $B$).}
\label{errors}
\end{figure}

For the HSP model the apex-FEF is exactly 3 (see Section \ref{AFEFHOP}). It is found that both the radius $A$ and the height $B$ of the simulation domain influence the total percentage error in the simulated apex-FEF value, with the T\%E decreasing significantly as the box dimensions increase. If one of the box dimensions (say the height $B$) is taken as large, and thus does not contribute significantly to the T\%E, then it is found that for moderate values of the other box dimension (in this case the radius $A$), the contribution $\epsilon_A$ of this dimension (as defined by the value of $\epsilon_{\rm{t}}$ in these circumstances) to the T\%E falls off with a power law close to 3. The same effect is found if the roles of box height and radius are reversed. These results are illustrated in Fig. \ref{errors}, where the plots are made against the (dimensionless) normalised quantities $A/r_{\rm{H}}$ and $B/r_{\rm{H}}$, where $r_{\rm{H}}$ is the hemisphere radius.

For the lateral boundary, the origin of the power-law exponent 3 can be understood by considering a small angular element of the circular mirror-dipole. This acts as in the same way as the distant dipole in the two-post-like-emitter situations analysed in Refs \cite{arxiv2018,DallAgnol2018}. The apex-field reduction for the chosen emitter shape is related to the value of the depolarizing field at point ``b" due to the mirror-dipole; for sufficiently large values of box-radius $A$, this depolarising dipole-field falls off as $(A^{-3})$. Each element of the circular mirror-dipole behaves in this way, so the total depolarizing field behaves in this way. This argument is very general, so all post-like shapes are expected to have $(A^{-3})$-type limiting behaviour.

\begin{figure}
\includegraphics [scale=0.30] {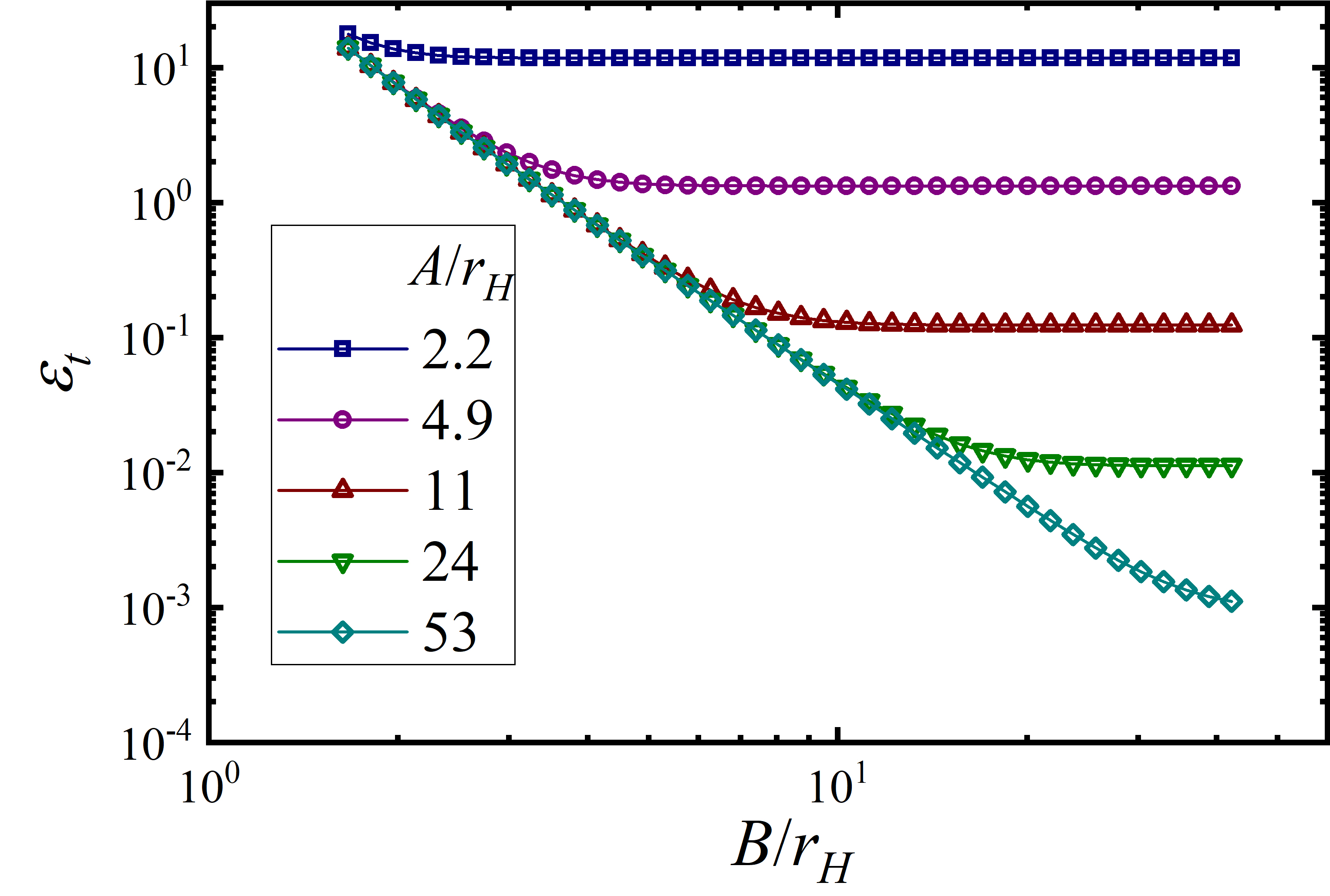}
\caption{(color online) Error due to the finite size of the simulation domain, for a hemisphere of radius $r_{\rm{H}}$.} 
\label{figmdsstfe22}
\end{figure}

For the top boundary, the error behaviour is also an effect associated with a dipole inverse-cube law, but is of a different kind. At sufficient large distances in any direction, the additional field due to a post-like emitter (above that due to the uniform applied field in which it is situated, in PPP geometry) can be modelled as due to a point dipole at ``b". (With a hemisphere on a plane, the modelling is exact.) Hence, at the top boundary, the error in field specification created by imposing a uniform field at the boundary equal to the applied macroscopic field is approximately given by the field distribution on the boundary due to a dipole at ``b".

It appears that in the simulations this error translates into an error in the apex field value that falls off with the inverse cube of box height. The exact electrostatic mechanism by which this simulation error is created when solving Laplace's equation is not yet clear, but the observed inverse-cube dependence on box height looks entirely plausible physically. Again, the argument is very general, so all post-like shapes are expected to have $(B^{-3})$-type limiting behaviour.

\begin{figure}
\includegraphics [scale=0.3] {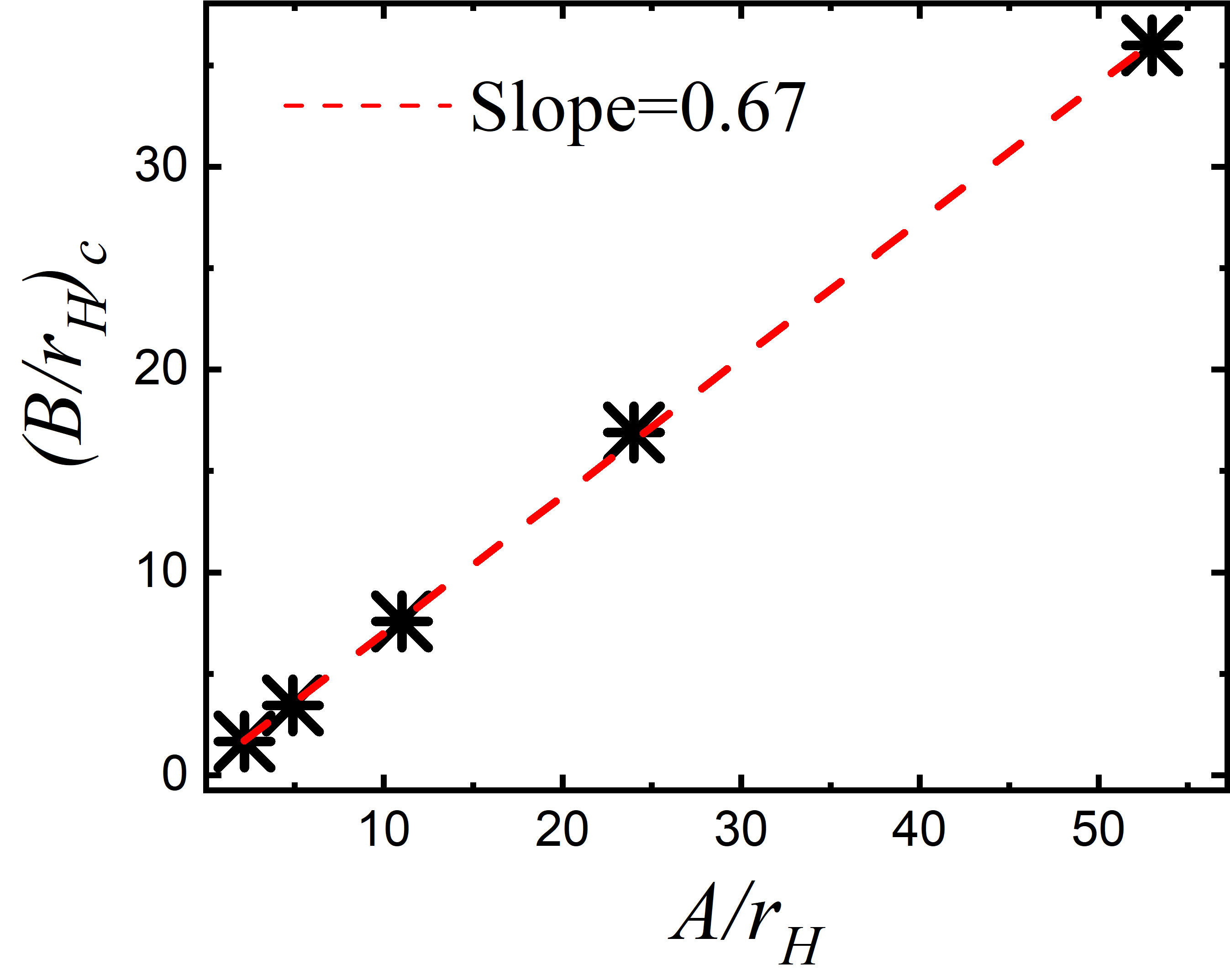}
\caption{(color online) Critical value $(B/r_{\rm{H}})_{\rm{c}}$, as a function of $A/r_{\rm{H}}$. Dashed line has slope $0.67$.} 
\label{crossover}
\end{figure}

The data in Fig. \ref{errors} can usefully be presented in another way, which better illustrates the combined effect of the two sources of error. This is done in Fig. \ref{figmdsstfe22}. For several values of $A/r_{\rm{H}}$, the T\%E $\epsilon_{\rm{t}}$ is plotted as a function of $B/r_{\rm{H}}$. The left-hand (sloping) part of these curves shows the behaviour of the partial percentage error $\epsilon_B$, as already plotted in Fig. \ref{errors}. However, when this error $\epsilon_B$ falls below the partial percentage error $\epsilon_A$ for the chosen value of $A/r_{\rm{H}}$, the T\%E $\epsilon_{\rm{t}}$ ``stabilises" at the relevant value of $\epsilon_A$. The critical value $(B/r_{\rm{H}})_{\rm{c}}$ at which this occurs (defined as the intersection of the extrapolated horizontal line with the slope) is shown in Fig. \ref{crossover} as a function of $A/r_{\rm{H}}$, and is well modelled by the relation $B/r_{\rm{H}} = 0.67 A/r_{\rm{H}}$. This suggests that this critical value corresponds to a critical domain height/radius ratio of $(B/A)_{\rm{c}} = 0.67$, valid for all values of $A$ and $B$ likely to be used in practice.  Equivalent plots could be drawn, and equivalent arguments could be made, with the roles of domain height and radius interchanged. In the regime where $\epsilon_A$ and $\epsilon_B$ are approximately equal, the two errors combine to give the total error $\epsilon_{\rm{t}}$, but $\epsilon_{\rm{t}}$ is not the simple sum of $\epsilon_A$ and $\epsilon_B$.

\begin{figure}
\includegraphics [scale=0.3] {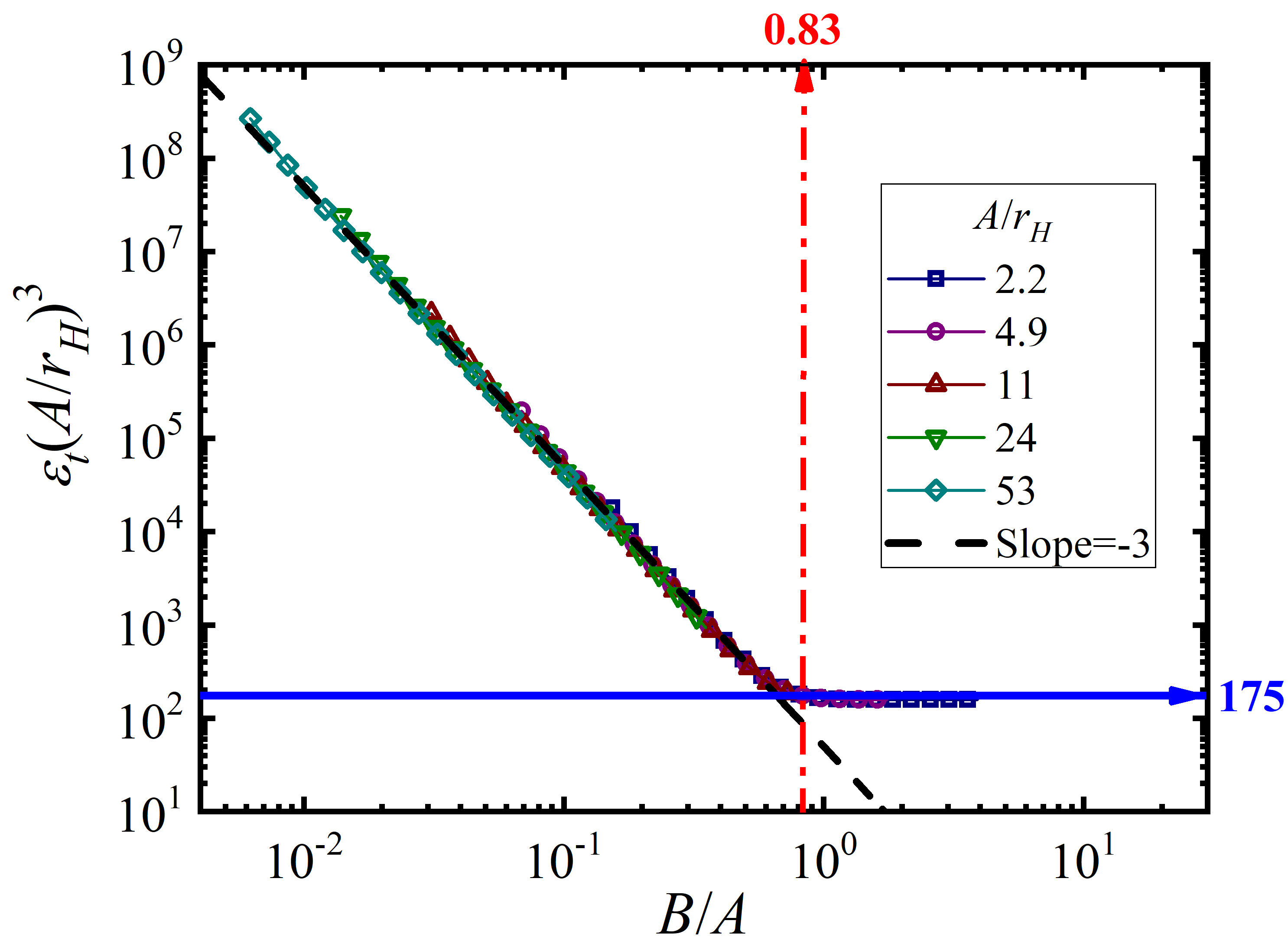}
\caption{(color online) Collapse of all curves shown in Fig. \ref{figmdsstfe22}, by replacing the variable $B/r_{\rm{H}}$ by $B/A$ and the variable $\epsilon_{\rm{t}}$ by $\epsilon_{\rm{t}} \times (A/r_{\rm{H}})^3$. As discussed in the text, well-defined minimum domain dimension criteria can be obtained by using the leftmost point in the plateau of the curves.} 
\label{figCollapse1}
\end{figure}

For the purposes of efficient simulation, the need is to decide a maximum tolerable T\%E, denoted here by $\epsilon_{\rm{tol}}$, and to determine corresponding minimum values of $A/r_{\rm{H}}$ and $B/r_{\rm{H}}$. Fig. \ref{figmdsstfe22}  shows that the curve for $A/r_{\rm{H}}=11$ correspond to $\epsilon_{\rm{t}} = 0.12\%$, for $B/r_{\rm{H}}$ greater than about 10. However, for $\epsilon_{\rm{t}}$-values that do not correspond to the values of $A/r_{\rm{H}}$ used in Fig. \ref{figmdsstfe22}, a more general approach is needed. This is achieved as follows. It is a reasonable hypothesis that the point at which $\epsilon_A$ and $\epsilon_B$ are equal is determined by a particular value of $B/A$ somewhere around unity; it is also known that for the ``stabilised" part of the curves in Fig. \ref{figmdsstfe22} the percentage error $\epsilon_A$ falls off as $(A/r_{\rm{H}})^{-3}$. These things suggest using the data from Fig. \ref{figmdsstfe22} to make a trial plot of $\epsilon_{\rm{t}} \times (A/r_{\rm{H}})^3$ versus $B/A$. As Fig. \ref{figCollapse1} shows, it is found that all the curves in Fig. \ref{figmdsstfe22} then collapse into a single ``slightly broadened curve" applicable to any value of $(A/r_{\rm{H}})$.

The highest of the horizontal sections (or ``plateaus") of these re-plotted curves corresponds to the value $\epsilon_{\rm{t}} \times (A/r_{\rm{H}})^3 = 175$. Hence, in a first approach one can use the formula that the minimum value
$(A/r_{\rm{H}})_{\rm{min}} =  (175/\epsilon_{\rm{tol}})^{1/3} \approx  5.59/(\epsilon_{\rm{tol}})^{1/3}$. The minimum value of $B/A$ is determined by a point at the left-hand end of the plateau, and Fig. \ref{crossover} suggests that a first estimate of this should be taken as $(B/A)_{\rm{min}}= 0.67$ (or perhaps a little higher). In fact, this value corresponds to the intersection of the sloping line and the extrapolated horizontal lines; in order to be on the truly horizontal part of the broadened curve, a slightly higher value needs to be taken. In practice, we use the value $(B/A)_{\rm{min}}= 0.83$ derived in \cite{JVSTB2019}, by a slightly different argument.

It follows that the minimum value $(B/A)_{\rm{min}}$ would be given by $(B/A)_{\rm{min}} \approx 4.63/(\epsilon_{\rm{tol}})^{1/3}$. It has been checked by numerical simulations that when the values of $A$ and $B$ are chosen according to these criteria, then the actual T\%E $\epsilon_{\rm{t}}$ is less than the chosen value of $\epsilon_{\rm{tol}}$. However, the view is taken here that using three significant figures in these formulae is ``overkill", so we have rounded them up to the nearest integer, and (for the hemisphere-on-plane model) recommend here the minimum domain dimensions (MDD) formulae

\begin{equation}
\left(\frac{A}{r_{\rm{H}}}\right)_{\rm{MDD-HSP}} = \frac{6}{\sqrt[3]{\epsilon_{\rm{tol}}}},
\label{mdd-hp-a/rh}
\end{equation}

\begin{equation}
\left(\frac{B}{r_{\rm{H}}}\right)_{\rm{MDD-HSP}} = \frac{5}{\sqrt[3]{\epsilon_{tol}}}.
\label{mdd-hp-b/rh}
\end{equation}

\begin{figure}
\includegraphics [scale=0.3] {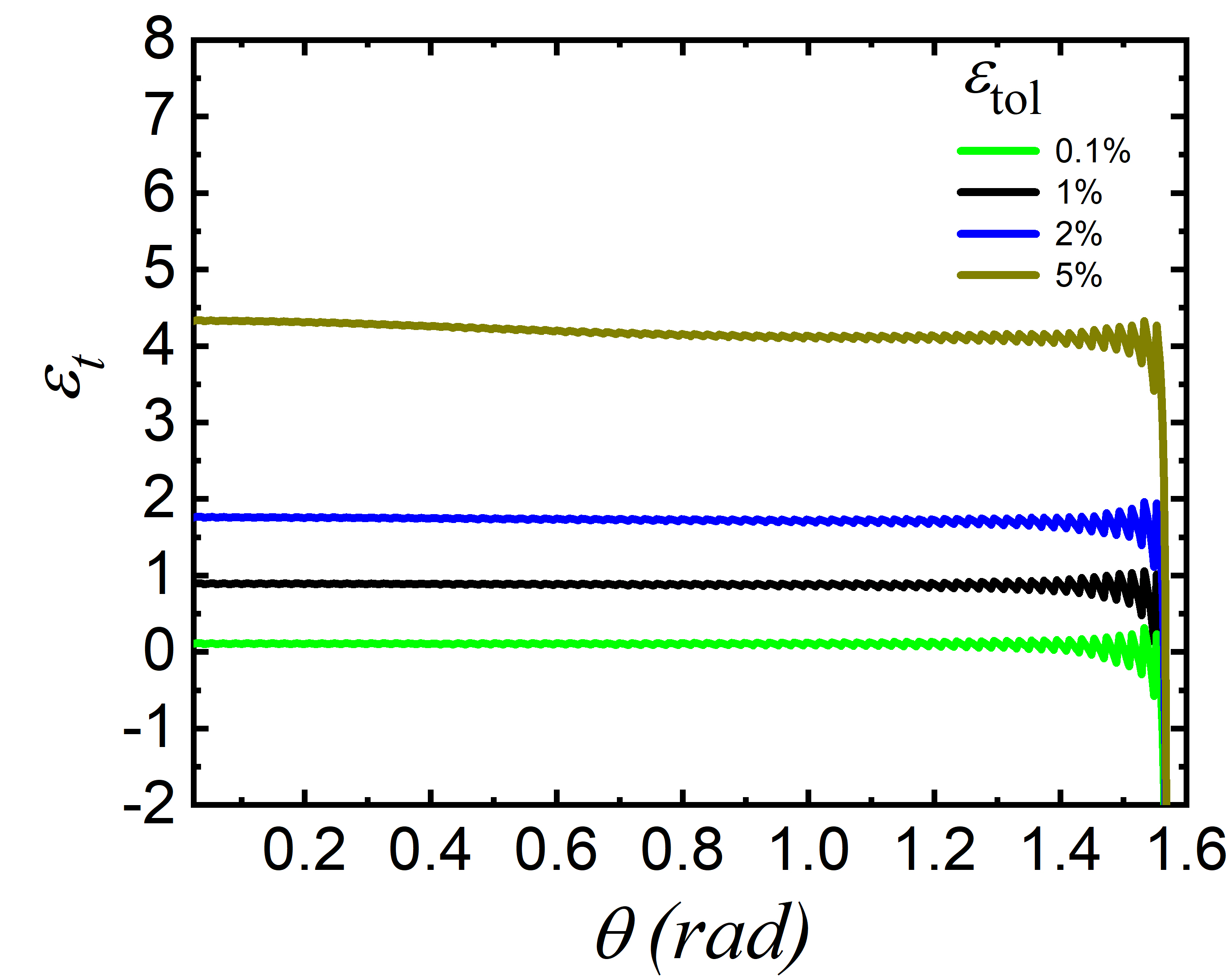}
\caption{(color online) Numerically calculated total percentage error $\epsilon_{\rm{t}}$ as a function of polar angle $\theta$, for various values of $\epsilon_{\rm{tol}}$.} 
\label{figerrorlocal}
\end{figure}

The above treatment is based on calculating apex-PFEFs. As illustrated in Fig. \ref{figerrorlocal}, which shows the numerically calculated total percentage error $\epsilon_{\rm{t}}$ as a function of polar angle $\theta$, we have demonstrated that these MDD values will also guarantee that PFEF-values at positions down the sides of the emitter meet the tolerance used in the equations. [Except that poorer precision is achieved for angles close to $\pi/2$ rad, where the PFEF-value becomes close to zero.] 

\subsection{Hemiellipsoid-on-plane (HEP) emitters}
\label{HEPMDS}
In a hemisphere-on-cylindrical-post model, the apex radius of curvature is equal to the radius of the cylinder and thus to the radius of the protrusion base. This may be a good model for some emitters, but when the emitter shape is such that the apex radius is significantly less than the base radius, then this model does not work well. In this case, the hemiellipsoid-on-plane (HEP) emitter model discussed in this Section may be a better choice. 

\begin{figure}
\includegraphics [scale=0.25] {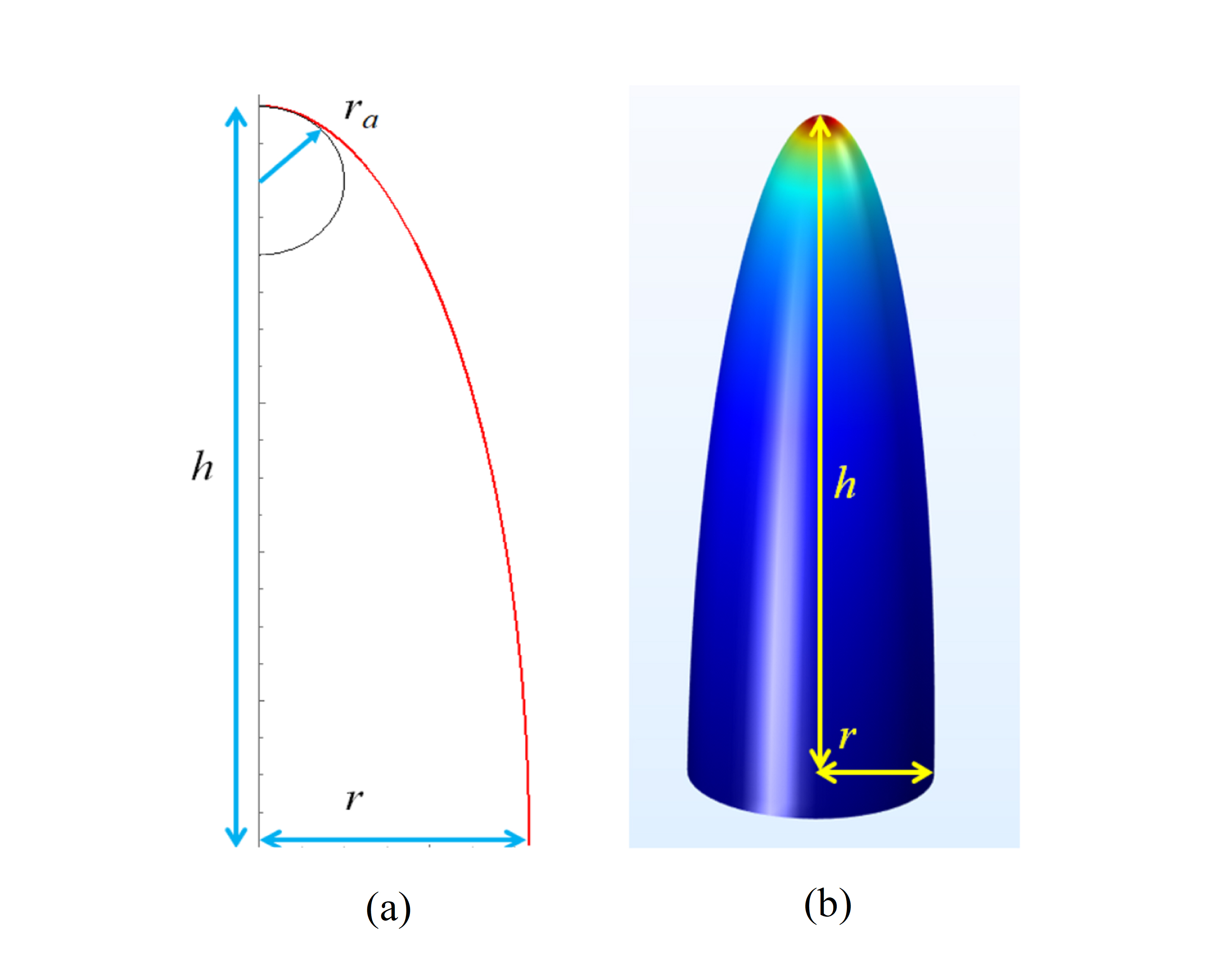}
\caption{(color online) (a) A quarter of an ellipse used to generate an hemiellipsoidal emitter. The base ($\rho$) and apex ($r_{\rm{a}}$) radii, and the total height ($h$), of the emitter are also shown. (b) Hemiellipsoidal emitter obtained by rotating the a quarter of an ellipse shown in (a) around its larger axis. In the color map, a red (blue) color indicates a higher (lower) local electrostatic field.} 
\label{fighemirev}
\end{figure}

As illustrated in Fig. \ref{fighemirev}(a), the HEP-model geometry is obtained by rotating a quarter of an ellipse around its longer axis, to yield the shape shown in Fig. \ref{fighemirev}(b). This shape is characterised by the half-axis lengths of the ellipse, which become the emitter base-radius $\rho$ and total height $h$. For MDD discussions, this shape is usefully characterised by its \textit{aspect ratio} $\nu \equiv h/\rho$ (see Section \ref{AFEFHEP}). Equation (\ref{gama1ellip}) earlier provides an expression for the corresponding apex PFEF.

\begin{figure}
\includegraphics [scale=0.30] {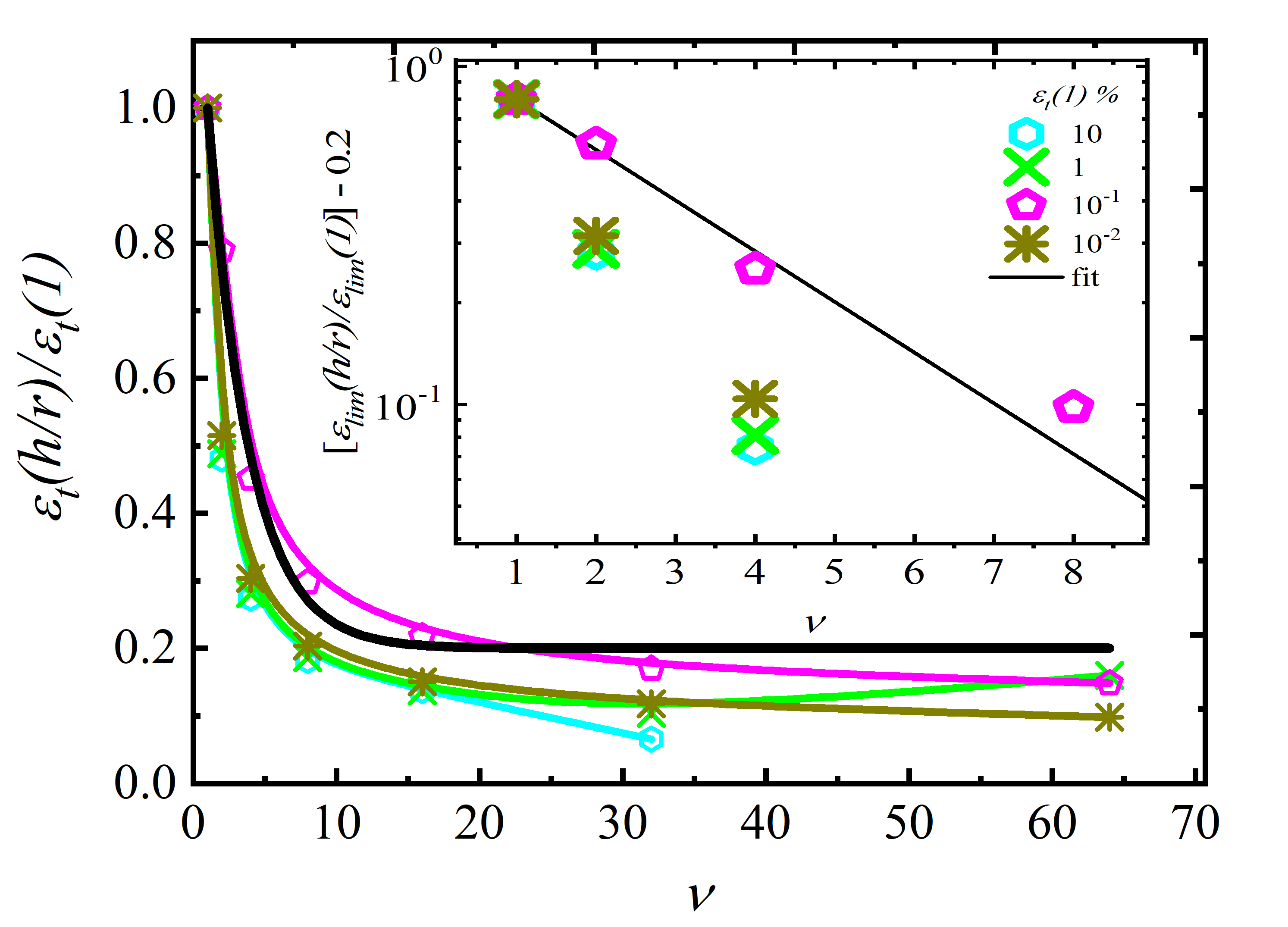}
\caption{(color online) Normalised total percentage errors for apex-PFEF values for a hemiellipsoid-on-plane emitter. The main graph shows the dependence of the ratio  $\epsilon_{\rm{t}}(\nu)/\epsilon_{\rm{t}}(\nu=1)$ on the aspect ratio $\nu$. Note that the total percentage error for large aspect ratios is significantly smaller than the error for $\nu=1$. For the rapidly decreasing part of the main graph, the inset shows $[\epsilon_{\rm{lim}}\left(\nu\right)/ \epsilon_{\rm{lim}} \left(\nu=1\right) - 0.2]$ as a function of $\nu$.  Figure adapted from Ref. \cite{JVSTB2019}.}  
\label{figmdsstfennn2k}
\end{figure}

For the HEP emitter, formulae for minimum domain dimensions have been found by empirical numerical methods, as follows. Simulation-box dimensions are now described by the revised normalised variables $A/h$ and $B/h$. That is, the hemiellipsoid total height $h$ has replaced the hemisphere radius $r_{\rm{H}}$ as the normalising length. Obviously, these new quantities reduce to the old ones in the case that $\nu=1$. In initial trials, for a set of given values of tolerated error $\epsilon_{\rm{tol}}$, minimum dimensions were derived from equations (\ref{mdd-hp-a/rh}) and (\ref{mdd-hp-b/rh}), with the new normalised variables replacing those in the equations. Figure \ref{figmdsstfennn2k} shows how the total error $\epsilon_{\rm{t}}$ decreases as the aspect ratio $\nu$ increases. As $\nu$ increases from 1, the decrease is initially approximately exponential (see inset), but tends to an asymptotic value $\epsilon_{\rm{lim}}$ for $\nu \gtrsim 20$ \cite{JVSTB2019}.

It was shown in \cite{JVSTB2019} that fitting an equation to these results yields the formula 
\begin{equation}
\frac{\epsilon_{\rm{t}}(\nu)}{\epsilon_{\rm{t}}(\nu=1)} \approx 0.2 + 0.8\exp{\left[-0.345 \left( \nu - 1\right)\right]},
\label{mdsbbnnew}
\end{equation}

This result enables us to give the following empirical formulae for the minimum domain dimensions for the hemiellipsoid-on-plane (HEP) emitter: 
\begin{equation}
\left(\frac{A}{h}\right)_{\rm{MDD-HEP}} = 6 \times \sqrt[3]{\frac{0.2 + 0.8\exp{\left[-0.345 \left(\nu - 1\right) \right] }}{\epsilon_{\rm{tol}}}} 
\label{mdsanm}
\end{equation}

\begin{equation}
\left(\frac{B}{h}\right)_{\rm{MDD-HEP}} = 5 \times \sqrt[3]{\frac{0.2 + 0.8 \exp{\left[-0.345 \left(\nu - 1\right)\right]}}{\epsilon_{\rm{tol}}}}.
\label{mdsbnm}
\end{equation}

In practice, it is often useful (as in the following Section) to have the MDD values expressed as a function of the apex sharpness ratio $\sigma_{\rm{a}} = h/r_{\rm{a}}$. From earlier, we have (for the HEP model) $\sigma_{\rm{a}}=\nu^2$, so we can define a function $f(\sigma_{\rm{a}})$ by
\begin{equation}
f(\sigma_{\rm{a}}) {= 0.2 + 0.8 \exp\left[-0.345 \left(\sigma_{\rm{a}}^{1/2} - 1\right)\right]},
\label{f}
\end{equation}

and the above equations become replaced by 

\begin{equation}
\left(\frac{A}{h}\right)_{\rm{MDD-HEP}} = 
6 \times \sqrt[3]{\frac{f(\sigma_{\rm{a}})}{\epsilon_{\rm{tol}}}} 
\label{mdd-hep2-A/h}
\end{equation}

\begin{equation}
\left(\frac{B}{h}\right)_{\rm{MDD-HEP}} = 
5 \times \sqrt[3]{\frac{f(\sigma_{\rm{a}})}{\epsilon_{\rm{tol}}}} 
\label{mdd-hep2-B/h}
\end{equation}

\begin{figure}
\includegraphics [scale=0.4] {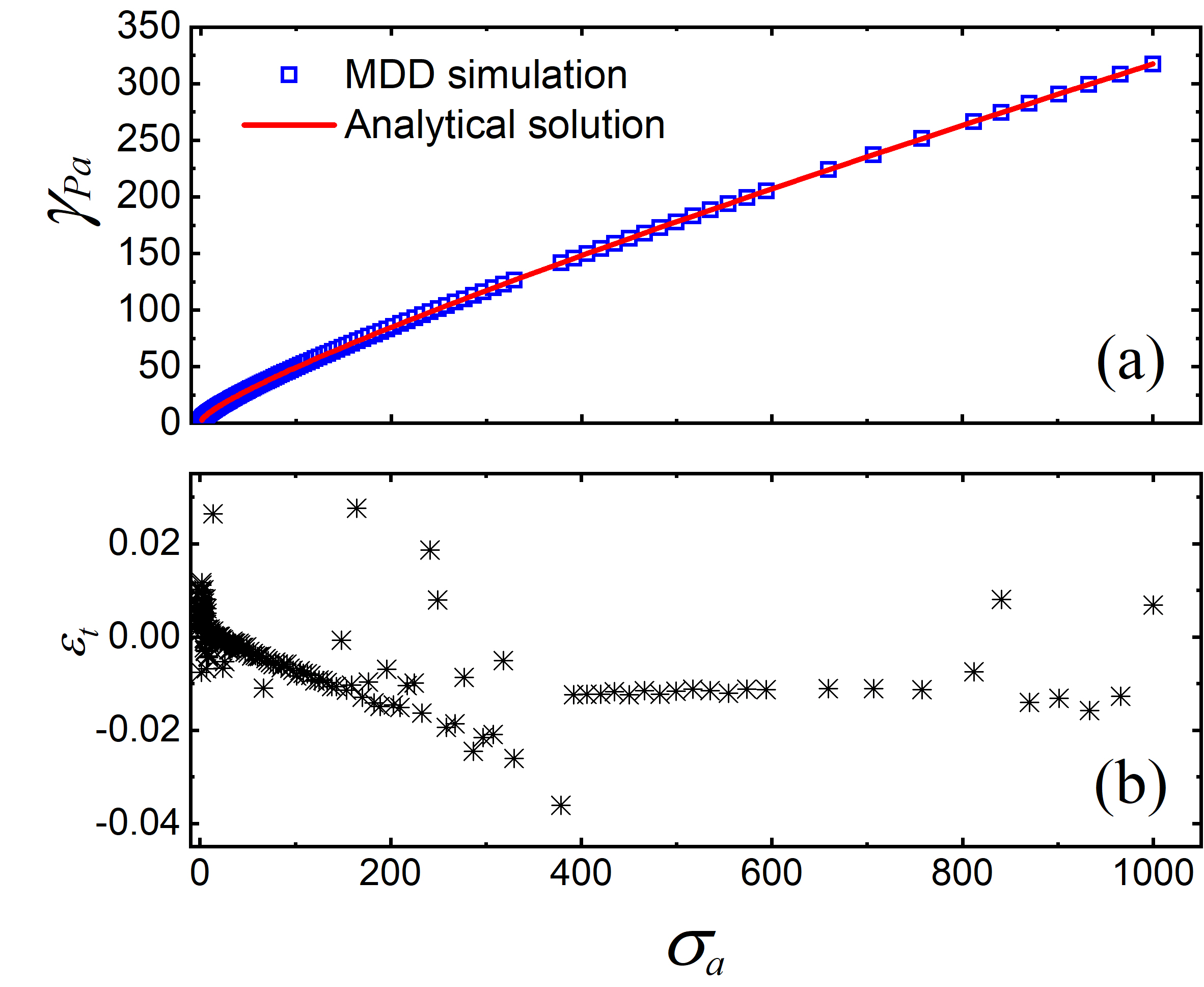}
\caption{(color online) Results for hemiellipsoid-on-plane (HEP) emitter. (a) Comparison between apex-PFEF values derived from analytical formula and from numerical analysis, using MDD values given by eqns (\ref{f}) to (\ref{mdd-hep2-B/h}), with $\epsilon_{\rm{tol}} = 0.1\%$. The results cover the range $ 1.01 \leq \sigma_{\rm{a}} \leq 10^3$, where $\sigma_{\rm{a}}$ [$=h/r_{\rm{a}}$] is the apex sharpness ratio. (b) Corresponding numerically derived  total percentage error. Figure adapted from Ref. \cite{FEF2021}.}
\label{comparison2021}
\end{figure}

We now report the results of using these formulae. Figure \ref{comparison2021}(a) shows a comparison between the exact analytical apex-FEF-value $\gamma_{\rm{a}}^{\rm{anal}}$ and numerical values $\gamma_{\rm{a}}^{\rm{num}}$ found using equations (\ref{f}) to (\ref{mdd-hep2-B/h}), for a required $\epsilon_{\rm{tol}}$-value of $0.1$\%.  Figure \ref{comparison2021}(b) shows the corresponding total percentage errors found in practice. The numerical reasons for the unexpected irregularities in the error graph are not clear, but the T\%E is less than 0.1\% in all cases in the range of $1.01 \lesssim \sigma_{\rm{a}} \lesssim 1000$. This will be our target precision for other shapes as well.

\subsection{Other emitter-model shapes}

\begin{figure}
\includegraphics [scale=0.35] {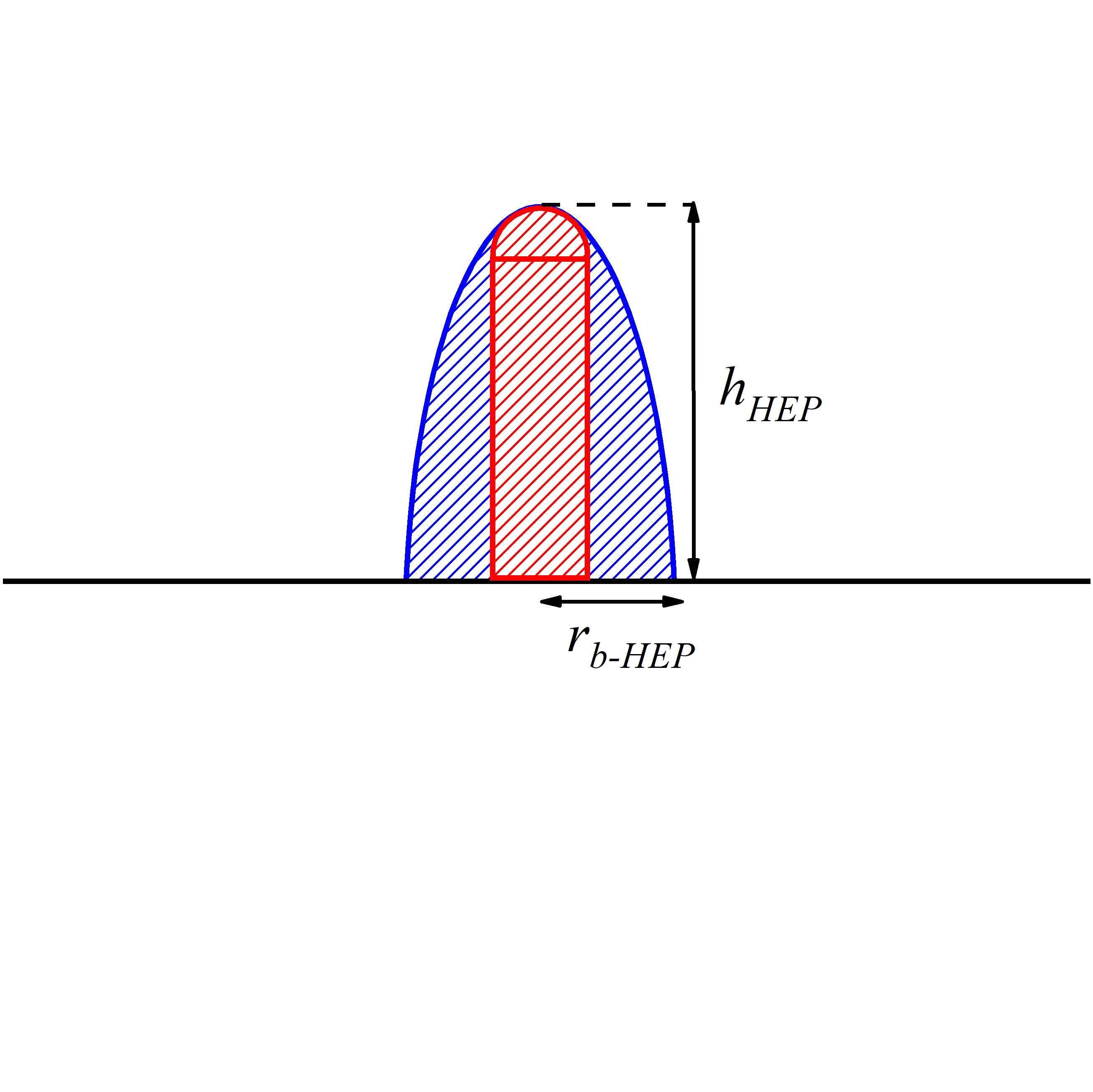}
\vspace{-1.8cm}
\caption{(color online) Representation of an hemiellipsoid that circumscribes the HCP-model emitter. The height $h_{\rm{HEP}}$ and apex radius of the hemiellipsoid are taken equal to those of the HCP model emitter; the hemiellipsoid base-radius $r_{\rm{b-HEP}}$ is also indicated.} 
\label{figmds2hemell}
\end{figure}

We now consider other single-protrusion shapes. If shapes of this kind, other than the HP and HEP models, are to be simulated, then a method is needed to define minimum domain dimensions. Our chosen procedure is as follows. As shown in Fig. \ref{figmds2hemell}, a hemiellipsoidal shape is defined that circumscribes the intended emitter shape. The height and apex sharpness ratio of this circumscribing hemiellipsoid are then used in eqns (\ref{f}) to (\ref{mdd-hep2-B/h}) in order to specify minimum domain dimensions. The thinking behind this approach is that the closer any part of the emitter shape is to the simulation box boundaries, then the worse will be the errors. Hence, since the shape inside the circumscribing hemiellipsoid is everywhere less distant from the lateral simulation-box boundary, the errors are expected to be less than they would be for the hemiellipsoid.

It may be that, for some shapes, different procedures could be specified that would lead to slightly smaller MDD values; however, our view is that more detailed investigations of this kind can wait until there is a demonstrable need for them.

As an example, Fig. \ref{figmds2hemell} shows the much studied hemisphere-on-cylindical-post (HCP) emitter, with height $h_{\rm{HCP}}$ and apex-radius $r_{\rm{HCP}}$, located within the circumscribing hemiellipsoid. In the case shown, the height $h_{\rm{HEP}}$ and apex radius of the hemiellipsoid are taken equal to those of the HCP model emitter. But, in fact, the parameters of any hemiellipsoid that totally encloses the emitter of interest will guarantee the condition $\epsilon_{\rm{t}} < \epsilon_{\rm{tol}}$.

With emitter systems that are not rotationally symmetric (for example, small emitter clusters) a similar procedure can be applied. One can draw a hemiellipsoid that encompasses the whole emitter system system, and use the MDD-HEP formulae to decide working values of $A/h_{\rm{HEP}}$ and $B/h_{\rm{HEP}}$ that will implement a given tolerance percentage error. In some cases it may be possible to draw hemiellipsoids of this kind in more than one way, so these working values (although fit for purpose) may not represent true minimum MDD values.

\subsection{The extrapolation method} 
\label{extrap}

As part of the work described above we found \cite{JVSTB2019}, and later used \cite{FEF2021}, an even more accurate numerical method of finding PFEF values---the so-called \textit{extrapolation method}. This Section describes the method and validates it by using the known analytical results for the hemiellipsoid-on-plane (HEP) model, stated earlier.

\begin{figure}[h!]
\includegraphics [scale=0.28] {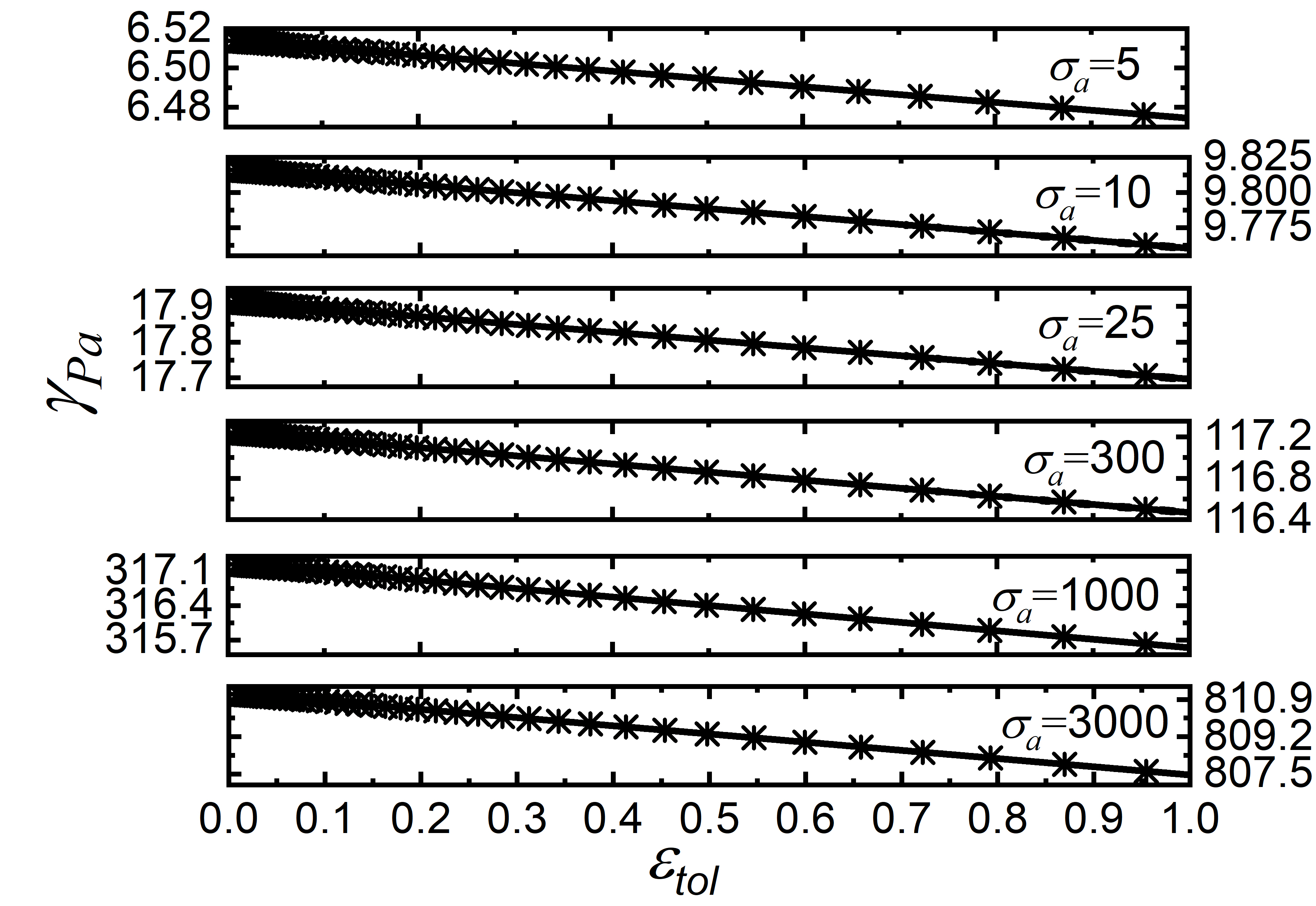}
\caption{Apex PFEF values ($\gamma_{\rm{Pa}}$) for the hemiellipsoid-on-plane (HEP) model, calculated using finite-element methods and minimum domain dimensions. For a fixed apex radius of 50 nm, and chosen values of apex sharpness ratio $\sigma_{\rm{a}}$, the apex PFEF $\gamma_{\rm{Pa}}$ is plotted as a function of the chosen tolerated error $\epsilon_{\rm{tol}}$. The $\gamma_{\rm{Pa}}$-values found by extrapolating these curves to zero error are shown in Table \ref{tabhep}. For other details see text.} \color{red} 

\label{extrapMDSHemi}
\end{figure}

For a fixed value (50 nm) of apex-radius $r_{\rm{a}}$, and a given value of apex sharpness ratio $\sigma_{\rm{a}}$, eqns (\ref{mdd-hep2-A/h}) and (\ref{mdd-hep2-B/h}) can be used to determine minimum domain dimensions (MDD) for a chosen value of the tolerated error $\epsilon_{\rm{tol}}$. Using these domain dimensions then generates a specific prediction for the HEP-model apex-PFEF $\gamma_{\rm{Pa}}$. The results of carrying out this procedure, for several $\sigma_{\rm{a}}$-values, and for a large number of $\epsilon_{\rm{tol}}$-values for each $\sigma_{\rm{a}}$-value, are shown in Fig. \ref{extrapMDSHemi}. For any chosen $\sigma_{\rm{a}}$-value, a regression line can be fitted to the related points and the ``extrapolated" PFEF-value that corresponds to $\epsilon_{\rm{tol}}=0$ determined.

It turns out that this extrapolated numerical value of the apex-PFEF $\gamma_{\rm{Pa}}$ is highly accurate, as is illustrated in Table \ref{tabhep}. All percentage differences from the known analytical HEP-model results are less than 0.001\%, over a wide range of values of apex sharpness ratio $\sigma_{\rm{a}}$. This finding suggests strongly that this extrapolation methodology should produce reasonably accurate PFEF-values for post-like shapes for which there is no known exact analytical result.

We note that this is a procedure that has been established \textit{in PPP geometry}. It may be that analogous procedures exist for other geometries, but this is not yet established.

\begin{table}
\centering
\renewcommand{\arraystretch}{1.5}

 \caption{
 Comparisons of apex-PFEF ($\gamma_{\rm{Pa}}$) values for the hemiellipsoid-on-plane (HEP) model. For specific values of the apex sharpness ratio $\sigma_{\rm{a}}$, the numerical result from the extrapolation method is compared with the known analytical result given by eqns (\ref{nu}) to (\ref{arrarb}).}

 \begin{tabular}{|c|c|c|c|}
 \hline
 $\sigma_{\rm{a}}$ & $\gamma_{\rm{Pa}}$ - extrapolated & $\gamma_{\rm{Pa}}$ - analytical & \% difference \\

 \hline
 \hline

$5$  & $6.51429$ & $6.51430$ & $-0.00015\%$    \\ \hline
$10$  & $9.81657$ & $9.81664$ & $-0.00071\%$     \\ \hline
$25$  & $17.91436$ & $17.91441$ & $-0.00028\%$    \\ \hline
$300$   & $117.24949$ & $117.24912$ & $+0.00032\%$    \\ \hline\
$1000$ & $317.25944$ & $317.25871$ & $+0.00023\%$   \\ \hline
$3000$ & $811.1931$ & $811.1916$ & $+0.00018\%$    \\ \hline

\end{tabular}
\label{tabhep}
\end{table}

\subsection{General comments}%

Obviously, the treatments here have employed a cylindrical simulation box. The other main option (which may seem more obvious to some readers) is to use a cuboidal simulation box with a square base. For some problems, in particular the analysis of depolarisation effects in regular square arrays, a cuboidal box of this kind is needed. But for analysis of single protrusions and small clusters the cylindrical box is markedly superior. This is for two main reasons. First, the higher symmetry of the cylindrical box makes meshing problems easier, and very significantly reduces the amount of computing time required to carry out a particular analysis. Second, at least with the finite-element package that we are using, we find in practice that the numerical accuracy and consistency of our results is better when using cylindrical simulation boxes.

As has been shown clearly in this review and in earlier work, the accuracy of electrostatic finite-element analysis can depend significantly on the relative sizes of the object being simulated and the simulation box. For this reason, we recommend that ideally the reported results of electrostatic finite-element simulations should always be accompanied by details of (a) the shape and dimensions of the simulation box, (b) the target accuracy of the simulations, and (c) the methods used to decide on the box dimensions.

\section{Application to other single emitters}
\label{OtherSingle}

\subsection{Apex PFEF-values for the HCP model}
\label{MDDHCP}

Following the pre-2003 work discussed earlier, the HCP model was investigated numerically by Read and Bowring (RB) \cite{RBowring}, using a Boundary Element Method, and by several research groups  \cite{ZhuRef,Podenok06,ZENG2009,Roveri16,JVSTB2019,FEF2021}, using Finite Elements Methods. Resulting fitted formulae for the apex PFEF were presented either as a power series in $(h/r_{\rm{a}})$ or as an expression of the form
\begin{equation}
\gamma_{\rm{Pa}} \approx C \left( D + \frac{h}{r_{\rm{a}}} \right)^{\kappa}.
\label{EdgRead}
\end{equation}

Edgcombe and Valdr\`e (EV) \cite{Edgcombe2001,Forbes2003}, using FEM, had reported that, in the range $4\leq h/r_{\rm{a}} \leq 3000$, their numerical results for $\gamma_{\rm{Ma}}$ were adequately fitted, to within 3\%, by taking $C=1.2$, $D=2.15$ and $\kappa = 0.90$. RB \cite{RBowring}, using BEM, claimed that their numerical results would be fitted, with accuracy better than 0.4\%, by taking $C=1.0782$, $D=4.7$ and $\kappa = 0.9152$, at least for $31 \leq h/r_{\rm{a}} \leq 3001$.

Using our MDD extrapolation method, as described in Section \ref{extrap}, it has been possible to find $\gamma_{\rm{Pa}}$ for the HCP model with high accuracy. A least-squares fitting procedure enables a four-parameter fitting formula, which---following \cite{FEF2021}---we prefer to write in the form
\begin{equation}
\gamma_{\rm{Pa}} = a\left[b + \left( \frac{h}{r_a} \right)^{c}\right]^d,
\label{Newfpf}
\end{equation}
where $a = 0.86391$, $b=0.97756$, $c=0.52989$ and $d = 1.77667$.  In the range $4\leq \sigma_{\rm{a}} \leq 3000$, this formula reproduces the extrapolation method result with accuracy better than 0.41 \%.

\begin{figure}[h!]
\includegraphics [scale=0.30] {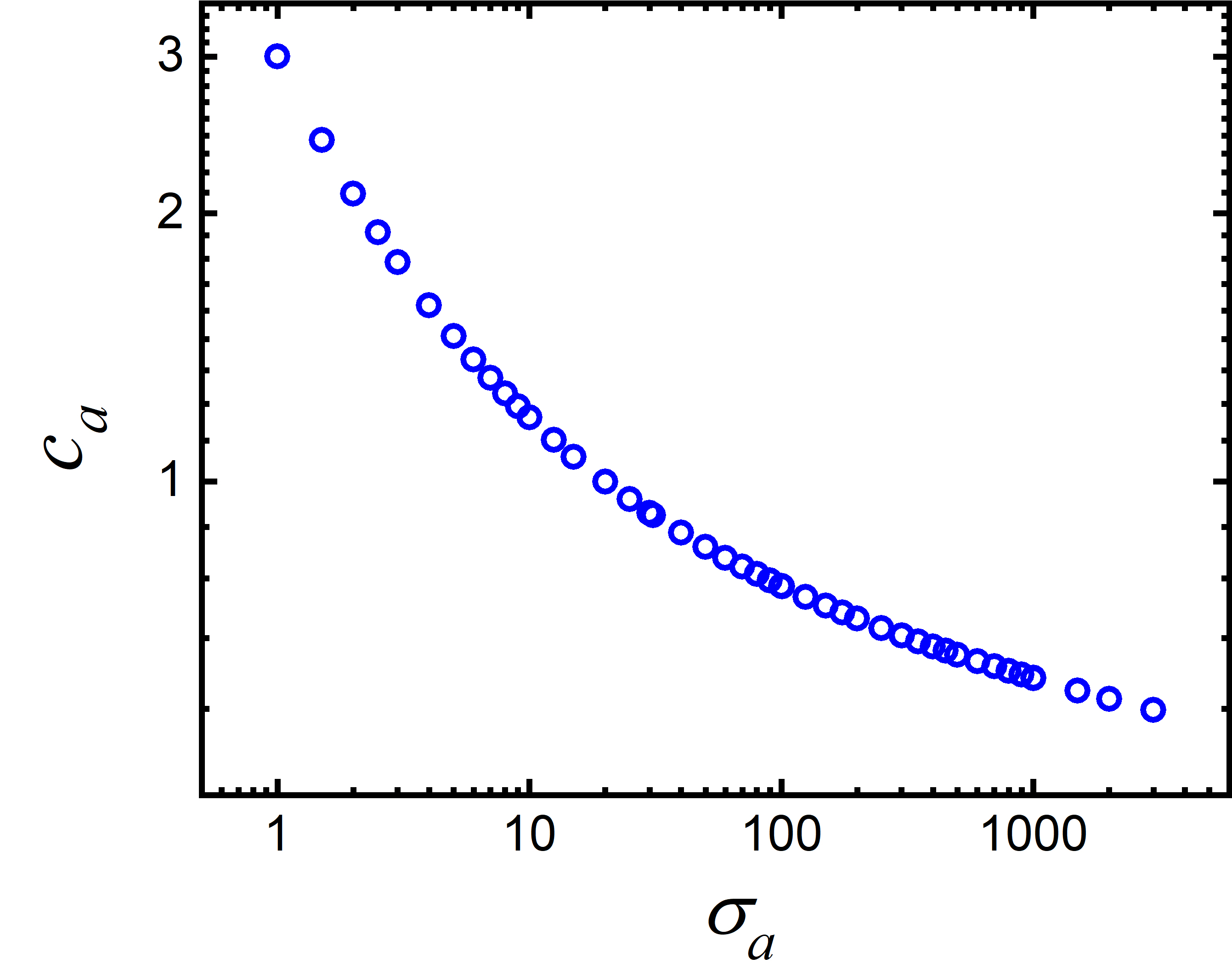}
\caption{In relation to the ``conducting post formula", eq. (\ref{gam-Fcondpost}), to show how the correction factor $c_{\rm{a}}$ depends on the apex sharpness ratio $\sigma_{\rm{a}}$.} 
\label{cafig}
\end{figure}

Table \ref{tab} compares these various numerical and formula-based apex-PFEF estimates, and also provides a value for the correction factor $c_{\rm{a}}$ in eq. (\ref{gam-Fcondpost}). Our conclusion is that our values and the RB values are better than the EV values, and that in nearly all cases formula (\ref{Newfpf}) is slightly better than formula (\ref{EdgRead}). The general consistency of our results and the RB results gives additional confidence that these results for the HCP model are numerically accurate. \color{black}
Figure \ref{cafig} illustrates how the correction factor $c_{\rm{a}}$ depends on the apex sharpness ratio $\sigma_{\rm{a}}$.

\begin{table*}
 \centering
 \renewcommand{\arraystretch}{1.5}
 \caption{Comparison of apex-PFEF ($\gamma_{\rm{Pa}}$) estimates for the hemisphere-on-cylindrical-post (HCP) emitter model, for selected values of the apex sharpness ratio $\sigma_{\rm{a}}$.  Estimates 1 and 4 are taken from Edgcome and Valdr\`{e} (EV) \cite{Edgcombe2001};  estimates 2 and 5 from Read and Bowring (RB) \cite{RBowring}; estimates 3 and 6 from our work. Estimates 1,2 and 3 were derived numerically; estimates 4, 5 and 6 were derived from algebraic formulae fitted to the numerical results. Values for $c_{\rm{a}}$ in formula (\ref{gam-Fcondpost}) have been derived from our extrapolated results.} 

 \begin{tabular}{|c|c|c|c|c|c|c|c|}

  \hline
 
  $\sigma_{\rm{a}}$ & $\gamma_{\rm{Pa}}$(1) & $\gamma_{\rm{Pa}}$(2) & $\gamma_{\rm{Pa}}$(3) & $\gamma_{\rm{Pa}}$(4) & $\gamma_{\rm{Pa}}$(5) & $\gamma_{\rm{Pa}}$(6) & $c_{\rm{a}}$  \\ \hline 
  & Num.(EV) \cite{Edgcombe2001} & Num.(RB) \cite{RBowring} & extrapolated & Eq. (\ref{EdgRead}) (EV) & Eq. (\ref{EdgRead}) (RB) & Eq. (\ref{Newfpf}) & from $\gamma_{\rm{Pa}}$(3)   \\

 \hline
 \hline

$1$ &  $2.97$ & $-$ & $3.0002$ & $3.3702$ & $5.3024$ & $2.9013$ & $3.00$   \\ \hline
$31$ & $28.2$ & $28.41$ & $28.4078$ & $28.0296$ & $28.4251$ & $28.4435$ & $0.91$  \\ \hline
$101$ & $76.4$ & $76.99$ & $76.9869$ & $77.8581$ & $76.7596$ & $76.9442$ & $0.76$   \\ \hline
$301$  & $200$ & $202.2$ & $202.078$ & $205.434$ & $202.881$ & $202.164$ & $0.67$   \\ \hline
$1001$ & $598$  & $602.8$ & $602.008$ & $603.130$ & $603.335$ & $603.044$ & $0.60$  \\ \hline
$3001$ & $-$  & $1645$ & $1661.75$ & $1618.08$ & $1643.29$ & $1662.91$ & $0.55$   \\ \hline

\end{tabular}
\label{tab}
\end{table*}

\subsection{MDD results for the apex-PFEF for other post-like emitters}

\begin{figure}[h!]
\includegraphics [scale=0.35] {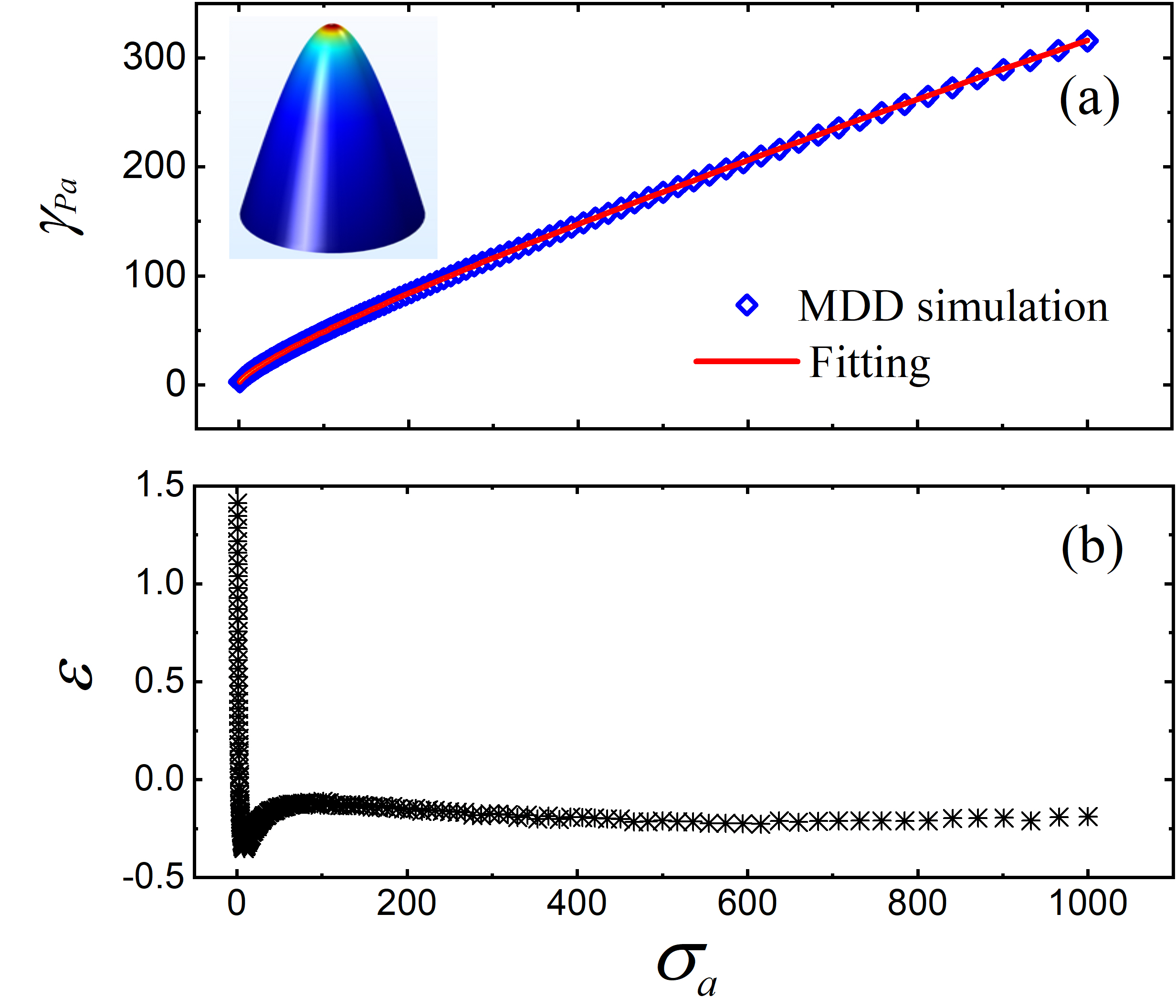}
\caption{(a) Numerical ($\gamma_{\rm{Pa}}^{\rm{num}}$) and ``fitted" apex-PFEF ($\gamma_{\rm{Pa}}^{\rm{fit}}$) values for a paraboloidal emitter, as a function of apex sharpness ratio $\sigma_a$. The inset color map indicates the local field distribution, with red (blue) indicating higher (lower) local field values. In (a), the blue diamonds represent values derived by the MDD numerical approach and the red line is a least-squares fitting to these points using eq. (\ref{Newfpf}). (b) Related error defined as $\epsilon \equiv \left[ (\gamma_{\rm{Pa}}^{\rm{num}} - \gamma_{\rm{Pa}}^{\rm{fit}}) / 
\gamma_{\rm{Pa}}^{\rm{num}} \right] \times 100 \;  \%$. Figure adapted from Ref. \cite{FEF2021}.}
\label{Parresult}
\end{figure}

The good fitting properties of eq.(\ref{Newfpf}) suggests that it could perhaps be applied to other emitter shapes. As an example, we consider a paraboloidal emitter. Figure \ref{Parresult}(a) shows numerical and fitted apex-PFEF values, as functions of apex sharpness ratio $\sigma_{\rm{a}}$. Least-squares fitting using Eq. (\ref{Newfpf}) yields $a=0.571$, $b=1.125$, $c=0.42$ and $d = 2.133$. Figure \ref{Parresult}(b) plots the total percentage error $\epsilon_{\rm{t}}$, showing that the four-parameter form (\ref{Newfpf}) is a very good fitting function in this case too.

\begin{figure}[h!]
\includegraphics [scale=0.32] {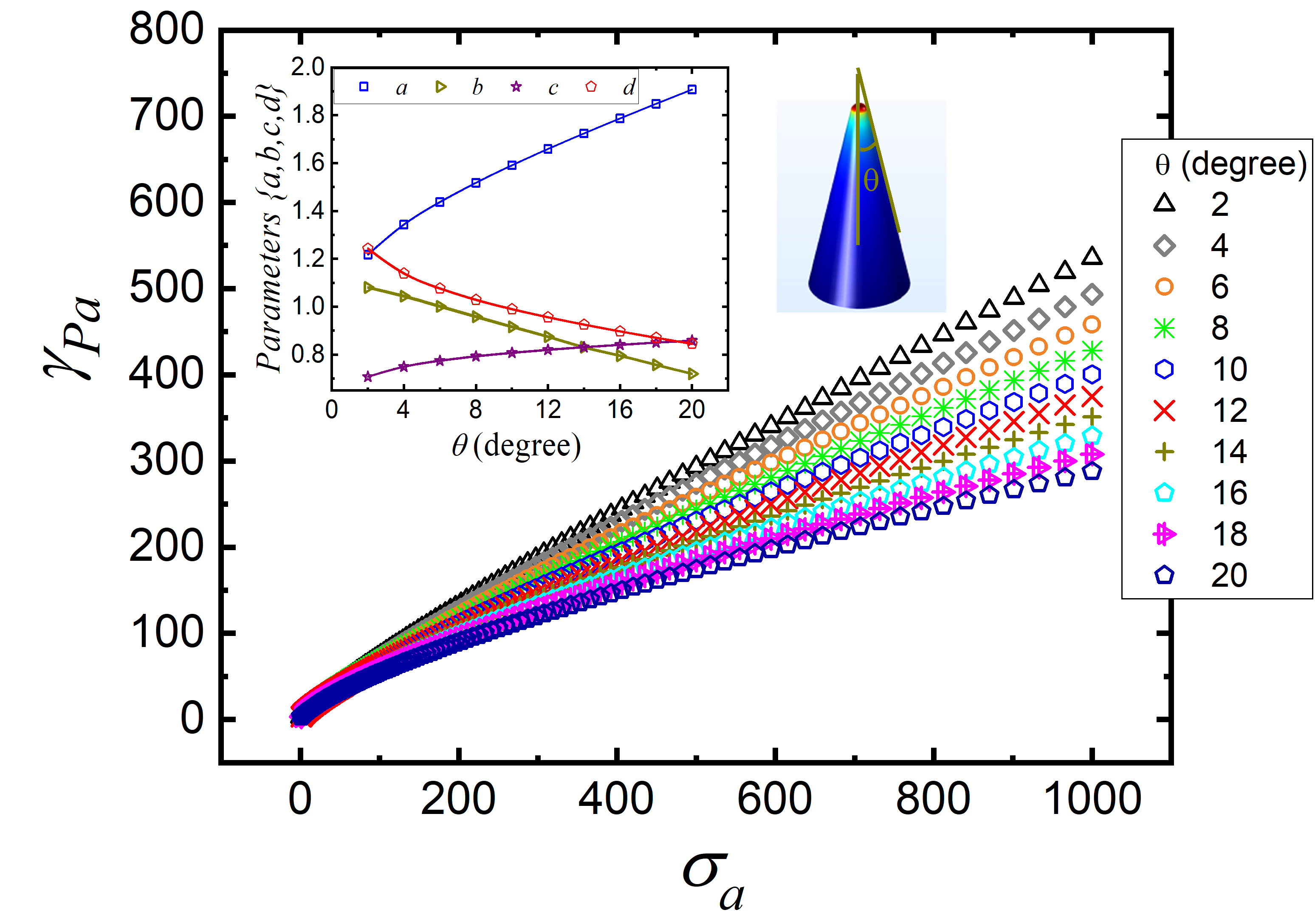}
\caption{Apex-PFEF values for the spherically rounded (truncated) cone (SRC) emitter model, as a function of $\sigma_a$, for various values of the cone apex half-angle $\theta$. The inset shows how the fitting parameters $a,b,c,d$, derived using eq. (\ref{fitparhcpn}), vary with $\theta$. Figure adapted from Ref. \cite{FEF2021}.}
\label{SRC}
\end{figure}

A second example is the \textit{spherically rounded (truncated) cone (SRC)} emitter model, originally discussed in \cite{FEF2021}. In this case, the vertex half-angle $\theta$ is an additional independent variable, and the parameters in our four-parameter fitting formula become functions of $\theta$, as shown in the inset to Fig \ref{SRC}.  For ``a" we assume a dependence of the form
\begin{equation}
a = \omega_{\rm{a}} + \lambda_{\rm{a}} \theta^{\mu_{\rm{a}}},
\label{fitparhcpn}
\end{equation}
where $\omega_{\rm{a}} $, $\lambda_{\rm{a}}$ and $\mu_{\rm{a}}$ are fitting parameters. Similar expressions are obtained for ``b", ``c" and ``d". For fitting over the range $2^{o} \le \theta \le 20^{o}$, values of the resulting fitting parameters are shown in Table \ref{tab2}.

\begin{table*}
   \centering
 \renewcommand{\arraystretch}{1.5}

 \caption{Values of the fitting parameters $\omega$, $\lambda$ and $\mu$ used in modelling the SRC emitter, for each  of the parameters $a,b,c,d$ [see Eq.(\ref{fitparhcpn})].} 
 
 \begin{tabular}{|c|c|c|c|}

  \hline
 $Parameter$ & $\omega$ & $\lambda$ & $\mu$ \\

 \hline
 \hline

$a$ &  $0.997$  & $0.1475$ & $0.6061$  \\ \hline
$b$ & $1.131$ & $-0.0241$ & $0.9488$ \\ \hline
$c$ & $0.277$ & $0.394$ & $0.1296$   \\ \hline
$d$  & $0.1821$ & $-0.497$ & $0.2242$   \\ \hline

\end{tabular}
\label{tab2}
\end{table*}

Other significant cylindrical-symmetry emitter shapes are considered in the literature but are not discussed here. In particular, these include open-cap nanotubes \cite{Tanaka_2004,XanthakisOpen} and cusp-like emitter shapes. We expect our minimum domain dimensions (MDD) arguments to apply to these emitters. but no dedicated studies have yet been carried out.

\section{Advanced single-emitter topics}
\label{AdvSingle}

\subsection{Effects related to ``external" series resistance}
\label{SerResSect}

When there is significant electrical resistance in the current path between the high voltage generator (HVG) and the emitter apex, this can cause a Fowler-Nordheim (or related) data plot to bend downwards at the left-hand (high voltage, etc.) side. This behaviour is sometimes called ``saturation", and can affect the plot interpretation.

Series resistance effects of this general kind have been discussed theoretically in many papers, including \cite{Miranda04,Hong06,Chen10,Kang12,ForbesJordan,Bachmann17,Vdiff,ForbesPreConvert}.

The effect occurs because an electronically ideal FE system has an intrinsic \textit{emission resistance} $R_{\rm{e}}$ that is associated with the emission process itself and is voltage and current dependent.

To illustrate the effect theoretically, we use Extended Murphy-Good FE theory and write an expression for the ``emission current" $I_{\rm{e}}$ in terms of the ``emission voltage" $V_{\rm{e}}$ (i.e., the voltage applied between the emitting surface and a counter-electrode) in the following simplified way
\begin{equation}
I_{\rm{e}} = A_{\rm{fC}}^{\rm{SN}} C V_{\rm{e}}^2 \exp[-B/V_{\rm{e}}],
\label{emres1}
\end{equation}
where $A_{\rm{fC}}^{\rm{SN}}$, $B$ and $C$ will normally be weakly voltage-dependent quantities. The \textit{emission resistance} $R_{\rm{e}}$ is then given by
\begin{equation}
 R_{\rm{e}}  =  V_{\rm{e}}/I_{\rm{e}} =  (V_{\rm{e}} A_{\rm{fC}}^{\rm{SN}} C)^{-1} \exp[B/V_{\rm{e}}] ,
\label{emres2}
\end{equation}

At lower emission voltages the dominating term in this equation will be the exponent. When $V_{\rm{e}}$ is very low, then the exponent and the emission resistance will be very large. As the emission voltage increases, the emission resistance steadily decreases. At some point the decreasing emission resistance may become comparable with the series resistance, and the FN plot ceases to be linear.

The illustrative discussion below is based on PPP geometry, but analogous effects occur in other geometries.

Resistance in series with the emission resistance can arise in several ways. Resistances can be associated with:  (a) the post itself; and/or (b) a contact resistance at the point where the post joins the substrate. These are ``internal" resistances (i.e., between the plates) and can affect the between-plates field distribution, as discussed in Section \ref{voltloss}. 

Resistances can also be associated with (for example): (c) a resistive (non-metallic) substrate on which the post-like emitter stands; (d) a series transistor fabricated for current-limiting purposes; (e) poor electrical contacts (and/or the use of resistive materials) in the mechanical emitter support system; and/or (f) a high-value ``safety resistor" (if any voltage drop across this is not subtracted off before processing experimental results). All these constitute \textit{external} series resistances.

For illustration, we now consider the case of an emission resistance in series with a \textit{constant} series resistance $R_{\rm{sr}}$. This will illustrate the principles, but the analysis of real systems could be vastly more complicated.
\begin{figure}[h!]
\includegraphics [scale=0.9] {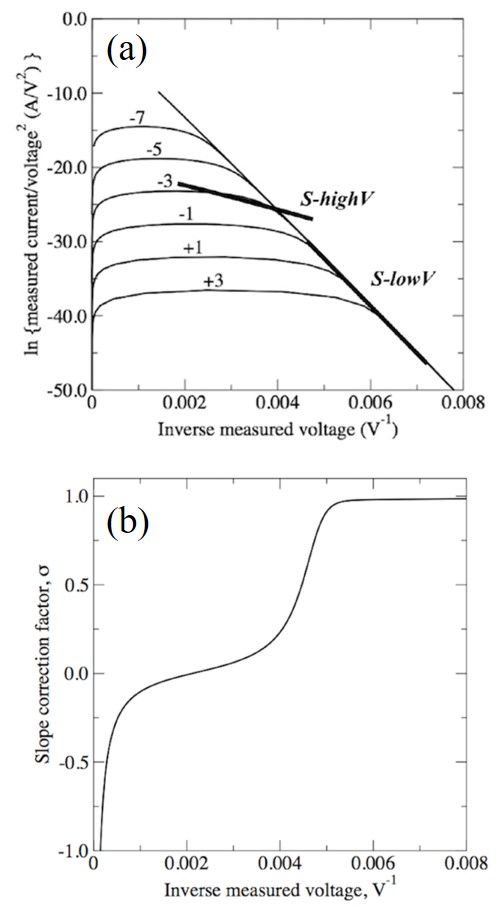}
\caption{Simulated circuit performance when a constant high resistance $R_{\rm{sr}}$ is placed in series with a field electron emitter with formal emission area $A_{\rm{fC}}^{\rm{SN}}$, modelled using EMG FE theory. (a) Fowler-Nordheim-type plot using measured current and voltage. A curve marked ``$N$" corresponds to a value ($A_{\rm{fC}}^{\rm{SN}} R_{\rm{sr}}) = 10^N$ $\Omega$ m$^2$. (b) To show how the slope correction factor $\sigma_{\rm{sr}}$ depends on $1/V_{\rm{m}}$, for $N = -1$. Adapted from Ref. \cite{ForbesJordan}.} 
\label{SerResFig}
\end{figure}

Even for the simple system just indicated, the algebra is slightly messy, so is not presented here. The results of illustrative simulations are shown in Fig. \ref{SerResFig}, as a FN plot using the \textit{measured} current and voltage. It is clear that the plot slope is constant in the low-voltage region, but that its magnitude then decreases as the measured voltage (and hence the emission voltage) increase. At some point the decreasing emission resistance may become comparable with the series resistance, so the FN plot ceases to be linear.

For an electronically ideal FE system, or in a low-voltage range of a more complicated system that is acting in an ideal fashion in this low-voltage range, the usual MG/EMG formula for extracting a value of the true characteristic plate-FEF $\gamma_{\rm{PC}}$ is
\begin{equation}
\gamma_{\rm{PC}}^{\rm{extr}}({\rm{true}})  =  - {\rm{s}_{\rm{t}}} {b {\phi}^{3/2} d_{\rm{sep}}} / S_{\rm{Vm}}^{\rm{fit}}({\rm{low}}V_{\rm{m}}),
\label{serres1}
\end{equation}
where $S_{\rm{Vm}}^{\rm{fit}}({\rm{low}}V_{\rm{m}})$ is the slope of the line fitted to the low-voltage part of the FN plot, and the other symbols have their usual meanings.

Clearly, if the slope of a line fitted to a ``high-voltage" part of the plot  were used is this formula, then the formula would yield a false value of the actual PFEF.

In formal principle, one could get a correct formula for interpreting the high-voltage part of the plot by replacing  eq.(\ref{serres1}) by
\begin{equation}
\gamma_{\rm{PC}}^{\rm{extr}}({\rm{true}})  =  - {\sigma_{\rm{sr,t}}} {b {\phi}^{3/2} d_{\rm{sep}}} / S_{\rm{Vm}}^{\rm{fit}},
\label{serres2}
\end{equation}
where $\sigma_{\rm{sr,t}}$ is an appropriately defined/determined fitting value of a generalised slope correction factor.

In reality, $\sigma_{\rm{sr,t}}$ is expected to be a relatively sensitive function of the precise range of $V_{\rm{m}}$-values over which the measurements have been taken, and accurate prediction of the effective correction factor to be used in any particular case looks to be exceptionally difficult.

Also note that experimentally observed ``saturation effects"  are not necessarily due to series-resistance effects. As discussed in \cite{Vdiff}, the ``voltage-difference" effects discussed in the next Section may also be a possibility, and may in some or many cases provide a more plausible explanation of ``saturation" than series-resistance effects.

\subsection{PFEF reduction due to voltage difference along an emitter}
\label{voltloss}

\textit {Background}. When an emitting protrusion is sufficiently long/high and sufficiently resistive, a non-negligible voltage $V_{\rm{d}}$ $[\equiv V_{\rm{a}} - V_{\rm{b}}]$ can exist between its apex and its base. This voltage $V_{\rm{d}}$ is needed to drive the emission current along the protrusion. To ensure language suitable for both FE and FI, $V_{\rm{d}}$ is called here the \textit{voltage difference}; other names have been used in past FE literature.

As shown below, this voltage difference leads to a physical reduction in the magnitude of the apex field and hence in the operating value $\gamma_{\rm{Pa}}^{\rm{op}}$ of the apex PFEF. The related field-enhancement theory involves \textit{finite-current electrostatics}, rather than the usual zero-current electrostatics found in textbooks. 

Voltage-difference effects cause $\gamma_{\rm{Pa}}^{\rm{op}}$ to become dependent on emission current. Thus, voltage-difference effects are one possible ``complication" that can cause non-linearity in Fowler-Nordheim and similar data-analysis plots. As noted above, it is not impossible that voltage-difference effects could be a more common cause of observed non-linearity in such plots than the FE community currently realises.

A voltage difference along an emitting protrusion (or along a needle-type emitter) causes a shift (or ``energy deficit") in the emitted \textit{total energy distribution} (TED). This shift can be detected experimentally by retarding-potential energy analysis. A voltage difference can also be detected by a direct-contact technique \cite{Minoux2005}.

In FE, $V_{\rm{d}}$-effects were apparently first detected in 1962 by Russell \cite{Russ62b}, who found that the critical bias necessary to collect electrons from semiconductor emitters was a function of the measured voltage (and hence the emission current). Clear early experimental reports of TED shifts (albeit on needle-geometry emitters) include the work of Braun, Forbes, et al. \cite{Braun75, ForGaP96} on GaP, and Mousa \cite{Mousa96} on a carbon emitter. Fursei and Egorov \cite{FurEg69} were probably the first to suggest that $V_{\rm{d}}$-effects would change PFEFs and might explain non-linearity in FN plots. Fuller discussions are given in \cite{Vdiff,ForVL-IVNC17} (see https://doi.org/10.13140/RG.2.2.20609.97120).

In field ion microscopy (FIM), experimentally observed voltage-difference effects were reported in 1970 by Morgan \cite{Morgan70}. Uranium dioxide ($\rm{UO_2}$) appears to undergo a rapid change in resistivity (roughly equivalent to a semiconductor-metal transition) at a temperature around 100 K. A needle-like $\rm{UO_2}$ FIM specimen can be imaged by FIM on both sides of the transition, but there is a change by a factor of 3 to 4 in the best image voltage. This implies changes by a factor of 3 to 4 in the best image field (and hence in the apex shape factor $k_{\rm{a}})$ for the needle-like ion emitter.

These things were (and are) thought to be associated with a change from a metal-like situation (where there is little voltage difference between the emitter apex and the support structure to which it is attached) to a resistive situation (where there is a significant voltage difference). In the latter case, it is expected that usually most of the voltage difference will occur near the quasi-conical tip of the needle.

Much later, in 2013, in the context of the atom-probe tomography of resistive semiconductors, an apex-field reduction effect (by a factor of around 2) was found theoretically by Vurpillot and colleagues \cite{Vur13}. In their simulations they replaced the tip section (last $1\ \mathrm{}{\mu} \rm{m}$) of a metallic needle by a dielectric of the same shape. For further discussion, see ref. \cite{Vur13} and p.178 in \cite{MilFor14}.

The detailed discussion of voltage-difference-related effects in needle-geometry semiconductor emitters is highly complex, for both emission polarities, and is outside the scope of this review.

\bigskip

\textit{Basic PPP-geometry theory}. Discussion in PPP geometry is somewhat simpler. For a resistive post in PPP geometry, the simple formula (\ref{VL5}) below was first obtained by Minoux et al. \cite{Minoux2005}, by fitting to the results of numerical simulations on a model of a carbon nanotube. Later, Forbes \cite{Vdiff} presented a simple physical derivation of the formula, and argued that the physical effect was a general phenomenon in finite-current electrostatics. The present treatment is a slightly improved version of the Ref. \cite{Vdiff} treatment.   

When a voltage difference exists between ``a" and ``b", eq. (\ref{Phi}) is replaced by 
\begin{equation}
(\Delta \mathit{\Phi})_{\rm{model}} + (\Delta \mathit{\Phi})_{\rm{XD}} =  (\Delta \mathit{\Phi})_{\rm{d}}  =  V_{\rm{d}} ,
\label{VL1}
\end{equation}
where $(\Delta \mathit{\Phi})_{\rm{d}}$ [$= \mathit{\Phi}_{\rm{a}} - \mathit{\Phi}_{\rm{b}}$] is the electrostatic potential difference (ESPD) corresponding to $V_{\rm{d}}$, with constant work function over the system surfaces being assumed (for simplicity), as before.

For the case considered here of a single emitting protrusion, $(\Delta \mathit{\Phi})_{\rm{XD}} = - E_{\rm{P}} h$, as before. Further, if the secondary contribution that $E_{\rm{P}}$ makes to the total apex ES field (here denoted simply by $E_{\rm{a}}$) is disregarded, as when deriving eq (\ref{Field}), then---for the basic FSEPP model in use---we have ${(\Delta \mathit{\Phi})_{\rm{model}} = \Delta \mathit{\Phi}(\rm{FSEPP)}} {\approx E_{\rm{a}} r_{\rm{a}}}$. Equation (\ref{VL1}) thus reduces to 
\begin{equation}
E_{\rm{a}} r_{\rm{a}} \approx E_{\rm{P}} h + V_{\rm{d}} .
\label{VL2}
\end{equation}
Algebraic manipulation then yields an expression for the operating apex PFEF: 
\begin{equation}
\gamma_{\rm{Pa}}^{\rm{op}} \equiv E_{\rm{a}}/E_{\rm{P}} \approx (h/r_{\rm{a}}) (1 + V_{\rm{d}} / E_{\rm{P}} h)    
\label{VL3}
\end{equation}
\begin{equation}
 \approx  \gamma_{\rm{Pa}}^{\rm{zc}} \; (1 + V_{\rm{d}} / E_{\rm{P}} h),    
\label{VL4}
\end{equation}
where eq. (\ref{gam-FSEPP}) has been used to replace $(h/r_{\rm{a}})$ by the \textit {zero-current apex PFEF} $\gamma_{\rm{Pa}}^{\rm{zc}}$. (This has also been called the ``small-current apex FEF", but it is now thought that``zero-current" is a  better description.)

When this equation is applied to FE, then $E_{\rm{P}}$ is negative and $V_{\rm{d}}$ is positive. For field ion emitters the signs are reversed. To get a formula that applies to both polarities, eq.(\ref{VL3}) is re-written as
\begin{equation}
\gamma_{\rm{Pa}}^{\rm{op}} \approx  (1 - |V_{\rm{d}}| / |E_{\rm{P}}| h) \;  \gamma_{\rm{Pa}}^{\rm{zc}}.    
\label{VL5}
\end{equation}
This equation shows clearly that significant apex-PFEF reduction occurs when the voltage change $|V_{\rm{d}}|$ along the emitter becomes a significant fraction of the magnitude $|E_{\rm{P}} h|$ of the driving PD (which is provided, in the single-emitter case, by the applied (inter-)plate field alone).

We emphasise that, although the basic theory given here is based on the behaviour of a single post-like protrusion in PPP geometry, this FEF-reduction effect is a general phenomenon in finite-current electrostatics, and should occur with more-or-less any form of resistive emitter with a surface that possesses sharp features.
\\

\textit{Comments.} (1) Minoux et al. \cite{Minoux2005} have shown (by means of numerical simulations) that a formula similar to (\ref{VL5}), but with a numerical correction factor, is obtained when there is additional resistance at the base of the protrusion, due to the presence of a resistive ``contact layer" between the protrusion and the substrate.

(2) Current-dependent changes in apex-VCL and apex-FEF values can also occur at the tips of semiconductor field emitters, in particular those with quasi-conical tips, due to semiconductor-physics effects. This has been pointed out by Groening et al.  \cite{Groen99}, amongst others. The theory of such effects can be highly complex, and is not yet fully explored.

\subsection{Effect of field-emitted vacuum space charge}
\label{SectionFEVSC}

Another effect that can reduce apex-field values (and increase apex-VCL values), for a given voltage difference between the emitter apex and the counter-electrode, is the presence of \textit{field-emitted vacuum space charge (FEVSC)}. In the past, this has been discussed using PPP geometry, with one plate taken as the emitter. Obviously, this is a one-dimensional (1D) model and does not strictly apply to real needle-shaped or post-shaped emitters, but it has merit as an indication of the maximum size of FEVSC effects. In the FEVSC model discussed here, the voltage between the plates is denoted by $V_{\rm{e}}$ and termed the \textit{emission voltage}. 

The original work on vacuum space-charge was done in the context of thermally emitted positive ions, and is associated with the names of Child \cite{Child1911} and Langmuir \cite{Langmuir1913,Langmuir1923}. However, in Child-Langmuir theory the boundary condition at the emitter surface is that the electrostatic field be zero. Thus, strictly, Child-Langmuir theory does not apply to field emitters: the need is for theories of FEVSC in which the boundary condition at the emitter surface is a field value sufficient to sustain the emission (ion or electron emission) of interest.

For electrons, the first theory of this kind was developed by Stern et al. \cite{Stern} in 1929, and a more comprehensive study was reported by Barbour et al. \cite{Barbour53} in 1953. An overview of then-existing work on 1-D models, and a rationalization of 1-D theory, was presented by one of us in 2008 \cite{Forbes08-FEVSC}. The main results of this 2008 analysis are presented here (but, to avoid clashes, different notation is used).

A convenient language to discuss FEVSC effects is the following. For a given emission voltage, the \textit{Laplace field} $E_{\rm{Lp}}$ is the surface field in the absence of FEVSC and the \textit{Poisson field} $E_{\rm{Ps}}$ is the field when FEVSC is present. A field reduction factor $\mathit{\Theta}_{\rm{sc}}$ due to space charge is then defined by
\begin{equation}
\mathit{\Theta}_{\rm{sc}} = E_{\rm{Ps}} / E_{\rm{Lp}} .    
\label{ThetaSC}
\end{equation}

The role of FEVSC theories is to determine mathematical expressions for $\mathit{\Theta}_{\rm{sc}}$; with 1D theories the fields involved are interpreted as emitter apex fields. For simplicity in what follows we omit the subscript ``SC".

A convenient theoretical form is found by defining a constant here denoted by $\kappa_{\rm{sc}}$ and a dimensionless field-dependent parameter here denoted by $\xi$ (but by $\zeta$ in \cite{Forbes08-FEVSC}), using the equations 
\begin{equation}
\kappa_{\rm{sc}} \equiv {\epsilon_0}^{-1} (m/2|ne|)^{1/2}.    
\label{FEVSC1}
\end{equation}
where $m$ is the mass and $ne$ is the charge of the emitted electron or ion, and
\begin{equation}
\xi \equiv \kappa_{\rm{sc}} {|V_{\rm{e}}}|^{1/2} |J_{\rm{Ps}}| / |{E_{\rm{Ps}}|}^2 , 
\label{FEVSC2}
\end{equation}
where $|J_{\rm{Ps}}|$ is the magnitude of the related local emission current density, evaluated for field magnitude $|E_{\rm{Ps}}|$. The LECD can either be an assumed experimental value, or a value predicted by using some specified expression for LECD. The parameter $\xi$ has been called the \textit{space-charge strength}.

An analysis of FEVSC electrostatics analogous to that made by Barbour et al \cite{Barbour53} yields the equation
\begin{equation}
9 \xi^2 {\mathit{\Theta}}^2 - 3{\mathit{\Theta}} - 4\xi + 3 = 0   
\label{FEVSC3}
\end{equation}
called in \cite{Forbes08-FEVSC} the \textit{dimensionless planar FEVSC equation}.

This has a general solution that can be implemented on a spreadsheet, but the mathematics is messy because there is a branch point at $\xi= 1/2$ (see \cite{Forbes08-FEVSC} for details). Several regimes can be identified where simplified equations apply. Those most of interest are as follows.

In the \textit{low-$\xi$ regime}, the general solution reduces to the Stern et al. 1929 result (see \cite{Stern}, eq. (13))
\begin{equation}
\mathit{\Theta} = 1 - (4/3) \xi + 3{\xi}^2 - 8{\xi}^3 \approx 1 - (4/3) \xi,   
\label{FEVSC4}
\end{equation}
where the second approximation is valid if $\xi$ is sufficiently small.

In the \textit{very-high-$\xi$ regime} the general solution reduces to
\begin{equation}
\mathit{\Theta} = (2/3){\xi}^{-1/2} {[1-3/(4\xi)]}^{1/2}.   
\label{FEVSC5}
\end{equation}
This \textit{very-high-$\xi$} result is the FEVSC  \textit{strong space-charge limit}. The formula has the form [$(2/3){\xi}^{-1/2}$] (see \cite{Forbes08-FEVSC}) of the strong space-charge limit for thermally emitted vacuum space-charge (Child's Law), multiplied by the factor in square brackets, which gets closer to unity as $\xi$ increases.

The theory in Ref. \cite{Forbes08-FEVSC} can be used make an estimate of when FEVSC effects are likely to become significant. From eq.(18) in \cite{Forbes08-FEVSC} it can be found that $\mathit{\Theta}=0.90$ corresponds to $\xi \approx 0.0896$. Bearing in mind that, in a 1-D model
\begin{equation}
|V_{\rm{e}}| = |E_{\rm{Lp}}| d_{\rm{gap}} =  |E_{\rm{Ps}}| d_{\rm{gap}} / \mathit{\Theta} ,   
\label{FEVSC6}
\end{equation}
where the separation between the planar electrodes is here denoted by  $d_{\rm{gap}}$, eq. (\ref{FEVSC2}) can be re-arranged into the form
\begin{equation}
 |J_{\rm{Ps}}| / |{F_{\rm{Ps}}|}^{3/2}  =  (\xi {\mathit{\Theta}}^{1/2} \kappa_{\rm{sc}}^{-1}) /d_{\rm{gap}}^{1/2}. 
\label{FEVSC7}
\end{equation}
\begin{equation}
{\approx (3.737 \times 10^{-7} \; \rm{m}^{1/2}}) / {d_{\rm{gap}}^{1/2}}. 
\label{FEVSC8}
\end{equation}
where we have used the fact that the value\footnote{In the original paper \cite{Forbes08-FEVSC}, the units of $\kappa_{\rm{sc}}$ are given incorrectly.}  of $\kappa_{\rm{sc}}$ for an electron is $1.904 \times 10^5$ A$^{-1}$ V$^{3/2}$.

For example for $d_{\rm{sep}} =$ 1 mm, using the 1956 Murphy-Good FE equation and $\phi =$ 4.50 eV, this yields the result that there is a FEVSC-induced 10 \% reduction in LECD at a field around 6 V/nm, at which point the LECD is about $4 \times 10^{10}$ A/m$^2$.

\subsection{Blade-type geometries}

Most LAFEs involve irregular arrays of point-like emission site, but there has been some past interest (e.g., \cite{SpindtWedge}) in arrays of emitters that have the shape of ``walls", ``wedges", ``knife-edges" or ``blades". We refer to these generally as \textit{blade emitters}. Practical blade emitters have finite length, and may have rounded ends, but theoretical modelling starts by considering blade emitters of infinite length.

Recently, the growing interest in the possibility of FE from the edges of graphite flakes has been recognised in the publication of a new handbook \cite{Sai22}, containing much useful material, in particular an overview of emission from graphene edges by Purcell and colleagues \cite{PurcellInSaito}.

Emission theory for blade emitters differs from standard forms of emission theory, because the effects of quantum confinement in the direction ``across the blade" need to be taken into account. Early investigations were carried out by ZhiBing Li and colleagues \cite{Qin11,ZBL11}; more recent theoretical developments include \cite{JC17} and are discussed in \cite{Sai22}, in particular in \cite{ZBLInSaito}.

Early treatments of the electrostatics of single-blade emitters include \cite{Miler2007} and \cite{Read-GR}. L. K. Ang investigated blade emitters with a cusp-like cross-section \cite{AngLor}. Shiffler and colleagues investigated depolarization effects related to two parallel blade emitters \cite{Tang2011}. Filippov and senior colleagues have investigated straight and rounded blade emitters with several different cross-sectional profiles \cite{Filippov19,Filippov22}.  There remains scope for further work on the electrostatics of blade emitters, and for a more detailed review in due course.

\subsection{Line-charge models}

In principle, line-charge arguments can use either a linear array of point charges (and/or dipoles), or a continuous linear distribution of charge. The aim is to choose a charge distribution such that one of the equipotentials models the emitter-shape of interest. Potentials and fields at other positions are then determined by the chosen charge distribution. If an exact emitter shape is required, choosing the related charge distribution can be non-straightforward.

Line-charge arguments can be applied in both field electron and field ion emission, but have found more use in FE. The theory of applying line-charge distributions to needle-shaped emitters and to field emission has been discussed, for example, by Harris et al. \cite{Jensen2015} and Zheng et al. \cite{Zheng20}.

The first people to have used line-charge arguments in FE seem to have been Vibrans \cite{Vibrans1964a,Vibrans1964} and Egorov and co-workers \cite{Egorov99}. They have been used extensively in the work of Harris, Jensen and colleagues (e.g., \cite{JensenAPL2015,Jensen2015,JensenP}), and also by Biswas and colleagues (e.g., \cite{Biswas2018c3,BiswasAIP2019}).

Holographic-type electron microscopy experiments have been carried out in order to verify the form of line charges, and find \cite{Zheng20} that an extra charge near a rod/needle apex is required: this finding is consistent with the basic version of the FSEPP model, where only a single point charge is used to model the emitter.

In general terms, the accuracy of the line-charge models lies between the FSEPP model and the FEM simulations.  For a short period, before the recent development of highy accurate FEM simulations described earlier, line-charge models were useful in exploring the two-emitter interaction. Their main usefulness now looks likely to be in niche applications where FEM simulations are difficult to implement. A particular niche may be the theory of small finite arrays of randomly placed emitters. A line-charge approach has also recently been used to model a complicated dielectric structure \cite{Egorov2021}.

\subsection{Atomistic electrostatic modelling}

At present, most electrostatic modelling in field electron emission is ``above the atomic level", but atomic-level modelling has long been needed in field ion emission to deal with: (a) the interpretation of field ion images of individual atoms (e.g., \cite{Forbes85,Katnagallu18}; (b) the physics of field adsorption (e.g. \cite{ForbesWafi}; (c) the issue of ``which (partially charged) surface atom evaporates next, in the process of field evaporation (e.g., \cite{Rolland15})"; and (d) the locations of the specimen electrical surface (e.g., \cite{ForbesES}) and hence of the critical surface inside which ordinary field ionization cannot occur. The electrical-surface location also affects the location of the FE tunnelling barrier \cite{JAP2019RG}.

Most of the earlier modelling has been classical in nature, using point charges and dipoles where relevant, but this is now largely being replaced by modern methods based on density functional theory (DFT).

Detailed review of atomistic electrostatic modelling is outside the scope of this article, but we think it useful to indicate the FE contexts in which it is already known that it will be needed in future, as part of the problem of predicting local emission current densities.

(*1) Understanding the details of FE from carbon nanotubes, including the probably important role of patch fields and the possible role of field penetration into the emitter apex. Some particularly relevant discussions are \cite{PRL2004Zhi,Qin,WangIPE,ZBL15rev,JAPFEF2019}. An important conclusion of work so far is the strong indication that better consistency between experiment, classical models and quantum-mechanical models is found if modern DFT calcuations are used to derive the \textit{induced} atomic-scale charge, field and potential distributions, by subtracting off the zero-field results from the finite-field results. 

(*2) Understanding the details of FE from flat atomically structured surfaces. There is already some relevant modern work (e.g. \cite{Bruno2016,Bruno2019}), but more is needed. This work needs to be linked with related work on positively charged surfaces (e.g., \cite{Sanchez04}).

(*3) Investigating the effect of single-atom adsorption and single-atom-defects on otherwise flat surfaces (e.g., \cite{DjuraDefects}).

(*4) Investigating the electrostatics and  emission properties of atomically sharp emitters, such as those developed by Fink \cite{Fink88}, and of the 3-atom tips used in gas field ion sources for scanning ion microscopy (e.g., \cite{HeSIM}). 

(*5) In both FE and FI emission, investigating sub-atomic level and bond-level emission properties, where these can be now be observed experimentally (e.g., \cite{HataFEM07,Tautz18,Ohta20,Yanag22}).

The exploration of atomic-level and sub-atomic-level FE and FI emission physics seems likely to become a linked growth area in these theoretical specialities. There also seems a need for a ``properly joined up" theory of charged surfaces that relates the classical and quantum mechanical approaches.

\subsection{Two-Stage Field Enhancement \break and ``Schottky's Conjecture"}
\label{Schottky}

A question of basic interest is what is the total field enhancement generated by a two-stage structure, in which one protrusion stands on the apex of another. Experimentally, it is known (e.g., \cite{Huang2005,APL2020,Biswas2021}) that such arrangements can generate a high total apex FEF. The basic physics can be understood qualitatively by considering PPP geometry and using FSEPP models for two HCP posts.

A ``large" post, of height $h_{\rm{b}}$ and apex radius $r_{\rm{b}}$, stands directly on the emitter plate; a ``small" post, of height $h_{\rm{a}}$ and apex radius $r_{\rm{a}}$, stands on the apex of the large post. Let $E_{\rm{B}}$ denote the (inter-)plate field (i.e., the electrostatic field at the emitter plate surface in the absence of both posts).  Let $E_{\rm{b}}$ denote the apex field for the large post, in the absence of the small post, and let the corresponding PFEF $\gamma_{\rm{Pb}} \equiv  E_{\rm{b}} / E_{\rm{B}}$.

If the small post were standing directly on the emitter plate (in the absence of the large post), then let its apex field be $E_{\rm{a}}^{(1)}$ and the related apex FEF be $\gamma_{\rm{a}}^{(1)} \equiv E_{\rm{a}}^{(1)} /E_{\rm{B}}$. When the small post is standing on the apex of the large post, then let its apex field be  $E_{\rm{a}}^{(2)}$ and the related single-stage FEF be $\gamma_{\rm{a}}^{(2)} \equiv E_{\rm{a}}^{(2)} /E_{\rm{b}}$. 

By definition, the total apex PFEF for the two-stage structure is $\gamma_{\rm{Pa,tot}} \equiv E_{\rm{a}}^{(2)}/E_{\rm{B}}$. Expansion yields 
\begin{equation}
\gamma_{\rm{Pa,tot}} \equiv E_{\rm{a}}^{(2)}/E_{\rm{B}} =  (E_{\rm{b}}/E_{\rm{B}}) (E_{\rm{a}}^{(2)}/E_{\rm{b}})  =  \gamma_{\rm{Pb}} \times \gamma_{\rm{a}}^{(2)} ,
\label{2stage}
\end{equation}
Thus, when the fields $E_{\rm{B}}$, $E_{\rm{b}}$ and $E_{\rm{a}}^{(2)}$, and related apex FEFs, are defined as above, then it follows from these definitions alone that the total apex FEF is found by multiplying the separate FEFs for the two stages. we refer to this as \textit{two-stage field enhancement}. Analogous multi-stage formulae can also be given.

In the limiting special case where $h_{\rm{a}}<<r_{\rm{b}}$, and the situation can be treated as if the small post were standing on a planar surface, this relation reduces to
\begin{equation}
\gamma_{\rm{Pa,tot}}  =  {(\gamma_{\rm{Pa,tot}})}^{\rm{SchC}} \approx  \gamma_{\rm{Pb}} \times \gamma_{\rm{a}}^{(1)}  =  \gamma_{\rm{Pb}} \times \gamma_{\rm{Pa}}    .
\label{2stage-2}
\end{equation}

It is thought that Schottky \cite{Schottky1923} was the first person to consider the theory of two-stage electrostatic effects of this general kind, and eq. (\ref{2stage-2}) is often now called \textit{Schottky's Conjecture}. Schottky's mathematical analysis was, in fact, somewhat different, and there seems to be a discrepancy in his published equation (see p.78 of \cite{Schottky1923}).

In reality, Stern et al. in 1929 \cite{Stern} seem to have been the first in FE to use the physics of formula (\ref{2stage-2}), when they argued that putting a small hemisphere on top of a much larger hemisphere would generate a total apex FEF of 9.

A matter of some interest is how to establish a condition for breakdown of eq. (\ref{2stage-2}). The most reliable approach is via finite element methods (e.g., \cite{deAssisSchConj,deAssisJAP}). However, the issues involved can be explored qualitatively by using the idea of differences in ``driving potential difference" (as discussed earlier) to compare the behaviours of a ``small" HCP-model post (of height $h_{\rm{a}}$ and apex radius $r_{\rm{a}}$)  when: (1) it is located in an uniform applied field; and (2) it stands on the apex of a ``large" HCP model post (of height $h_{\rm{b}}$ and apex radius $r_{\rm{b}}$).

In case (1), the driving potential difference between the top and bottom of the small post is $\Delta {\mathit{\Phi}}_1 = -E_{\rm{b}}h_{\rm{a}}$, where the post is treated as in an uniform field equal to the field $E_{\rm{b}}$ at the apex of the large post.

In case (2) the FSEPP model is used to represent the electrostatic effects of the large post, by placing a charge $Q$ at the sphere centre. The driving potential difference
$\Delta {\mathit{\Phi}}_2$ for the small post is then estimated by
\begin{equation}
\begin{split}
\Delta {\mathit{\Phi}}_2 \approx  \left[ Q/(4\pi\epsilon_{0}) \right]  \left[ 1/(r_{\rm{b}} + h_{\rm{a}}) - 1/r_{\rm{b}} \right]  \\
= E_{\rm{b}} r_{\rm{b}} \left[ 1/( 1+ h_{\rm{a}}/r_{\rm{b}}) - 1 \right]  \\
\approx E_{\rm{b}} \left[ - h_{\rm{a}} + (1/2)(h_{\rm{a}}^2/r_{\rm{b}}) \right] 
\label{2stage2}
\end{split}
\end{equation}
Hence the fractional difference $N$ between the cases is
\begin{equation}
N \equiv (\Delta {\mathit{\Phi}}_2 - \Delta {\mathit{\Phi}}_1) / \Delta {\mathit{\Phi}}_1
\approx  -h_{\rm{a}}/2r_{\rm{b}}.
\label{2stage3}
\end{equation}
If, say, we wish the error to be less than 10 \% ($|N|<0.1$), then this model suggests the criterion
\begin{equation}
h_{\rm{a}} < \; \approx \; 0.2 \;r_{\rm{b}}.
\label{2stage4}
\end{equation}
\begin{figure}[h!]
\includegraphics [scale=0.28] {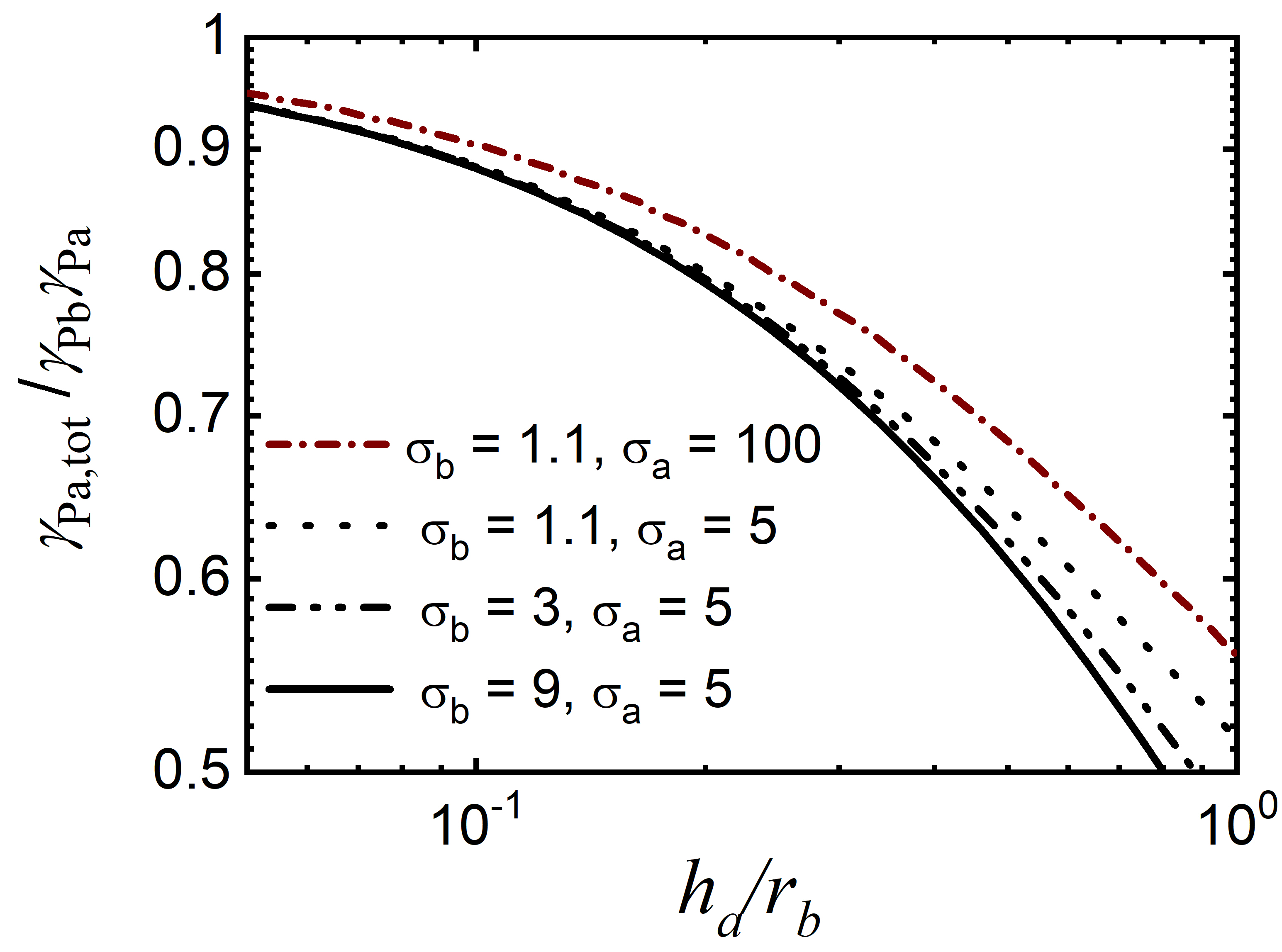}
\caption{To show, for a two-HCP-stage model in PPP geometry, how Schottky's Conjecture loses accuracy with increase in the ratio ``second-stage-post-height $(h_{\rm{a}})$" divided by ``first-stage-post-apex-radius $(r_{\rm{b}})$". The vertical axis shows the ratio defined by: the total apex FEF for the two-stage geometry, divided by the product of the single-stage apex FEFs for the situations where the HCP-model emitters stand individually on the planar substrate. All FEFs are calculated numerically, using FEM simulations. Plots are shown for different combinations of the apex sharpness ratios for the large (first-stage) post ($\sigma_b$) and the small (second-stage) post ($\sigma_a$). It is seen that fall-off of 10 \% is reached for $h_{\rm{a}}  \approx 0.1 \; r_{\rm{b}}$.}     
\label{SchConjFig2}
\end{figure}

Figure \ref{SchConjFig2} shows the results of a numerical calculation \cite{deAssisSchConj} for this two-HCP-stage model. This shows that, in this geometry, significant breakdown of the conjecture in fact occurs for $h_{\rm{a}} > \; \approx \; 0.1 \;  r_{\rm{b}}$. The difference from eq. (\ref{2stage4}) is due to the approximations in FSEPP-type models.

Broadly similar conclusions were reached by Jensen, Harris and colleagues, using line-charge models \cite{Harris20192}.

An equivalent condition for a blade-type emitter was derived by Ryan  Miller et al. \cite{Miller2009}.

Variants of the analytical theory given above can be developed for cases where the small protrusion stands, not at the apex of the large protrusion, but at some other point on the protrusion. 

Various other methods of investigating two-stage field-enhancement effects and/or the limitations of eq. (\ref{2stage-2}) have been explored in the literature, for example \cite{APL2001,Huang2005,Miler2007,Miller2009,JensenAIP2016,deAssisJAP,JVSTB2017,Harris2019,Harris20192,Edgar2019,JensenJAP19,Biswas2020,Biswas2021}.

More generally, it is thought that two-stage field enhancement plays an important role in many physical situations, particularly in some electrical-breakdown situations, and can contribute to the development of highly effective technological field emitters. A particular example of this is the ``looped fibre + fibrils'' configuration discussed in Ref. \cite{APL2020}, where experimental data-analysis that passes an Orthodoxy Test validity check appears to suggest that a two-stage field enhancement factor of order 10$^5$ has been achieved.

\subsection{Collector adjacency effects}
\label{adjacency}

\textit{Collector adjacency effects} occur when the counter-electrode (here called the \textit{collector}) is sufficiently close to the emitter that this affects the values of the apex local field and related field enhancement factors. Adjacency effects have been widely discussed in FE literature (e.g., \cite{Miller1,NevYar82,Miller2,BonardPRL2002,XQWang04,Surrey05a,Silva2005,Hii06,ZXu06,Febre06,ZENG2009,Biswas2021}), usually for PPP geometry and HCP model emitters, though also for other models. 

In PPP geometry, adjacency effects have been discussed for both the relevant plate-FEF and the relevant gap-FEF. As discussed below, these two FEF-types have qualitatively different behaviour. 

It is simplest to consider an electronically ideal emitter of constant height $h$ and constant apex sharpness ratio $\sigma_{\rm{a}} = h/r_{\rm{a}}$. One needs to establish the effect of decreasing the gap length $d_{\rm{gap}}$ between the emitter apex and the collector, but there are options for discussing this.

Our initial approach is to reduce the interplate separation $d_{\rm{sep}}$ whilst keeping the interplate field magnitude $|E_{\rm{P}}|$ $[=|V_{\rm{P}}|/d_{\rm{sep}}]$ constant, by appropriately reducing the interplate voltage magnitude $|V_{\rm{P}}|$. In these circumstances the magnitude $|E_{\rm{a}}|$ of the apex local field would be constant, independent of $d_{\rm{sep}}$, if there were no adjacency effects. In practice, as $d_{\rm{sep}}$ is reduced from a large value, a point is reached where $|E_{\rm{a}}|$ begins to \textit{increase}, and the situation moves towards a \textit{narrow-gap regime} where $d_{\rm{gap}} << r_{\rm{a}}$ and
\begin{equation}
|E_{\rm{a}}| = |V_{\rm{G}}|/d_{\rm{gap}} = |V_{\rm{P}}|/d_{\rm{gap}} .
\label{narrowgap}
\end{equation}
Here, $|V_{\rm{G}}|$ is the magnitude of the voltage between the emitter apex and the adjacent collector; for an electronically ideal system this equals the magnitude $|V_{\rm{P}}|$ of the interplate voltage.

Since the apex plate-FEF $\gamma_{\rm{Pa}} = |E_{\rm{a}}|/|E_{\rm{P}}|$, it follows that $\gamma_{\rm{Pa}}$ has the same behaviour as that just described for $|E_{\rm{a}}|$.

If, instead, one holds the interplate voltage constant, then the behaviour of the plate-FEF is the same as just described, but the apex local field magnitude $|E_{\rm{a}}|$ now has a more complicated dependence on the plate separation $d_{\rm{sep}}$.

In all the above cases, the point at which a ``significant" dependence on $d_{\rm{sep}}$ begins to develop (for likely practical values of $h/r_{\rm{a}}$) is usually stated to be around $d_{\rm{sep}} < h/2$ to $d_{\rm{sep}} < h/3$. However, it is expected that the details will depend somewhat on the shape of the emitter apex, and in particular on the apex radius. If the issue is important, then specific FEM-based calculations are recommended.

In our view, the approach just described, in terms of plate-FEFs, is often preferable. However, there is an alternative approach that starts from the properties of the gap. In the narrow-gap regime, the apex field is given by eq. (\ref{narrowgap}). As the gap length $d_{\rm{gap}}$ increases, this formula needs (for an electronically ideal emitter) to be replaced by
\begin{equation}
|E_{\rm{a}}| = |V_{\rm{G}}|/\zeta_{\rm{Ga}} = |V_{\rm{P}}|/\zeta_{\rm{Ga}} .
\label{GVCL}
\end{equation}
where the (apex value $\zeta_{\rm{Ga}}$ of the) \textit{gap-voltage conversion length} is defined by this equation.

The magnitude of the (mean) \textit{gap-field} $E_{\rm{G}}$ is defined by
\begin{equation}
|E_{\rm{G}}| \equiv |V_{\rm{G}}|/d_{\rm{gap}} = |V_{\rm{P}}|/d_{\rm{gap}} .
\label{GField}
\end{equation}
Hence the apex value $\gamma_{\rm{Ga}}$ of a \textit{gap-field enhancement factor (GFEF)} can be defined by
\begin{equation}
\gamma_{\rm{Ga}} \equiv |E_{\rm{a}}|/|E_{\rm{G}}| = d_{\rm{gap}}/\zeta_{\rm{Ga}} .
\label{GFEF2}
\end{equation}
As $d_{\rm{gap}}$ increases from a very low value, the gap-FEF transitions from being unity to being effectively equal to the large-gap plate-FEF.

In our view, it may be more straightforward to deal directly with the gap VCL, which transitions from being equal to the gap length to being equal to the usual apex local VCL, but formulation in terms of gap-FEFs is more usual.

Various numerical discussions of the transition process, for both plate-FEFs and gap-FEFs, have been given in the citations earlier, but in our view have not yet reached the stage where  a quantitatively definitive summary can be presented. 

The general effect reported here, namely transition from a ``narrow-gap regime'' to a ``wide-gap regime" as the gap length increases, also occurs in point-plane and point-point electrostatic gemoetries, but detailed discussion is outside the scope of this review. 

\color{black}

\section{Electrostatic Depolarization}
\label{DepolSect}

\subsection{Introduction}

When two or more post-like emitters stand on a conducting plane the emitters interact electrostatically. For simplicity, most existing  numerical discussions consider only the special case of identical, cylindrically symmetrical, post-like emitters. This review also considers only this special case. The ratio $c/h$ of the \textit{emitter separation} $c$ to the common height $h$ is then of particular relevance. The term ``emitter separation" refers to the distance between the post axes. The ratio itself is called here the \textit{separation ratio}.

As already discussed, each post, together with its electrical image in the conducting plane, forms a finite electrostatic dipole.

The apex-PFEFs for two or more posts are mutually decreased if the post separation is decreased. Our preferred name for this effect is \textit{electrostatic depolarization}, since the physics of what is happening is a decrease in the magnitude of the finite electrostatic dipole. Depolarization effects associated with \textit{point} electrostatic dipoles have long been part of the theory of charged surfaces \cite{ForNegPhi78,ForbesWafi}. In FE literature, the finite-dipole effect is also known as ``screening" and ``shielding". 

In the two-emitter situation, the magnitude of the depolarization can be quantified by the ($+ve$) \textit{fractional reduction} ($-\delta$) in the apex-PFEF, as given by 
 \begin{equation}
 -\delta \equiv \frac{(\gamma_{\rm{Pa}})_1 - (\gamma_{\rm{Pa}})_2}{(\gamma_{\rm{Pa}})_1}.
 \label{depol}
 \end{equation}
 where the subscripts ``1" and ``2" label the one-emitter and two-emitter situations, respectively. The value of ($-\delta$) depends on the separation $c$ of the emitters.
 
 Depolarization effects also influence field and FEF values at emitter surface positions other than the apex, but the numerics of this has not yet been systematically explored, and such effects are outside the scope of this review.

As discussed by Forbes \cite{RFJAP2016}, two separate effects contribute to the reduced apex-PFEF value. The primary cause is so-called \textit{charge-blunting}, which results from the ``driving potential difference" term $(\Delta \mathit{\Phi})_{\rm{XD}}$ in eq. (\ref{Phi}). The name ``charge blunting" is used because the effect of $(\Delta \mathit{\Phi})_{\rm{XD}}$ is to force charge back from the emitter tip onto the conducting plane.

A much smaller contribution is made by the so-called \textit{neighbour field} effect, in which the field due to the depolarising dipole forms part of the``external field contribution" $E_{\rm{Xa}}$ in eq. (\ref{Field}). Both contributions are discussed in Ref. \cite{RFJAP2016}. For simplicity, the analytical discussions here mostly disregard the neighbour field effect; however, it is automatically included in the FEM electrostatic simulations.   

If ($c/h$) is relatively small ($\approx 1$), the depolarization related to charge blunting must be described directly in terms of contributions to the driving potential difference. This approach can be used at all separations, as in \cite{RFJAP2016}. But, if the emitters are reasonably well separated, then the effect is conveniently described as due to a \textit{depolarizing field} $E_{\rm{D}}$. This opposes the inter-plate field and makes, to the driving potential difference, a contribution $-E_{\rm{D}} h$ that reduces the total magnitude of $(\Delta \mathit{\Phi})_{\rm{XD}}$.

If the emitters are very well separated, then the theory effectively reduces to the mutual depolarizing effect of point dipoles: this is standard electrostatics and is used (for example) in the classical theories of charged surfaces  (e.g. \cite{ForbesArray98}) and of field adsorption, (e.g. \cite{ForbesWafi}), as employed in the context of field ion microscopy.

When more than two emitters are present, the depolarization effects are basically additive. For any particular emitter under analysis, all emitter pairs need to be considered, but with contribution magnitudes that depend on the pair separation. With small clusters or finite arrays, emitters near the centre of the cluster experience greater total depolarization than those near the edges of the cluster or array, due to the geometry of the situation. Thus, field enhancement factors are higher near the edges of the cluster or array.

This effect can be seen in the ``7-emitter cluster" figure below, and has also been found in earlier numerical simulations, for example those of Murata et al. in 2001 \cite{MurataArray01}, Bocharov and Eletskii in 2005 \cite{BochElet} and Smith and Silva in 2009 \cite{Silva2009}, and in the line-charge  models used by Harris, Jensen and colleagues (e.g., \cite{JensenP}).  Its existence is well established experimentally [private discussion at conferences], though sparsely described in FE literature. This edge effect can be a hindrance in the design of practical devices. Sometimes, steps to mitigate its consequences have been explored (e.g., \cite{Harris2016,HarrisJAP2016,JensenP}). 

An implication of this theoretical additivity is that the two-emitter interaction is the basic building block for cluster theory and array theory and requires careful attention. This is carried out in Sections \ref{PhysExpl} to \ref{CPEE}. Depolarization theory relating to larger clusters and arrays of post-like emitters is not yet fully established, and is too large a topic to be included in this review; Section \ref{Multi} provides some introductory remarks.

Numerical explorations of the two-emitter situation were initially made by Harris, Jensen, and colleagues \cite{JensenAPL2015,Jensen2015}, using line-charge emitter models, and led to diagrams broadly similar to Fig. \ref{HCPCPEE}. Following earlier empirical thinking (e.g., \cite{joS}), they interpreted their work in a manner that implies an \textit{exponential} fall off in ($- \delta$) with separation, at large separations. They also found \cite{JensenAPL2015} that, at very small separations, where the value of $c$ was only a few  post radii and was reducing, the apex-PFEF went through a minimum and began to increase again.   

The later exploration by Forbes \cite{RFJAP2016}, using two FSEPP model emitters, suggested that (apart from the regime at very small separations) there were at least two ($c/h$) regimes. At large ($c/h$) values, ($-\delta$) diminished with the inverse cube of distance. However, as ($c/h$) moved towards unity, there was a change-over to a regime where ($c/h$) changed more slowly. This is illustrated below. A second discussion [arXiv:1803.03167], which was later confirmed in further papers \cite{JPCM2018,DallAgnol2018}, firmly associated the inverse-cubic regime with the electrostatics of finite dipoles.

The two regimes were originally called the ``closely separated" and ``widely separated" regimes. We consider it better to call the first of these regimes (where $c/h$ is roughly between 0.1 and 1) the \textit{moderately separated regime}, and to suppose the existence of a \textit{narrowly separated regime} where $c/h < \sim 0.1 $ (but $c/r_{\rm{a}} >> 1)$.   

At very small values of $(c/h)$, de Assis and Dall'Agnol \cite{CPEE2017} identified a regime  where the position of maximum field on the emitter tip moves away from the  axis. This regime has been called the \textit{close proximity regime} and is discussed in Section \ref{CPEE}.  This regime presumably either overlaps with or is identical with the narrowly separated regime found with the line-charge models, but details of the relationship have not yet been explored.

In the ideal world one needs, for each of the four \textit{separation regimes}, physically derived analytical formulae for $\gamma$ and/or $- \delta$. In practice these formulae will not be accurate, so one also needs both a numerical treatment and a fitting formula. In practice, at present, none of the analytical formulae or fitting formulae cover all four regimes, so one needs to specify ranges of applicability for formulae.

As shown below, a further complication is that values of $\gamma$ and $- \delta$ will also depend on the apex sharpness ratio, and simulation reports need to take this into account.

All these things will be different in detail for posts of different shapes. We therefore concentrate on the behaviour of the well-studied hemisphere-on-cylindrical post (HCP) emitters, on the assumption that behaviours for other post shapes will be qualitatively similar.

However, as things stand at present, even for HCP model emitters our quantitative understanding of the two-post interaction (although well advanced) is not yet complete. Section X reports the present situation. As a preliminary, Section X B presents a physical explanation of electrostatic depolarization.

\subsection{Reduction of the apex-FEF for two identical emitters: A physical explanation}
\label{PhysExpl}

\begin{figure}[h!]
\includegraphics [scale=0.35] {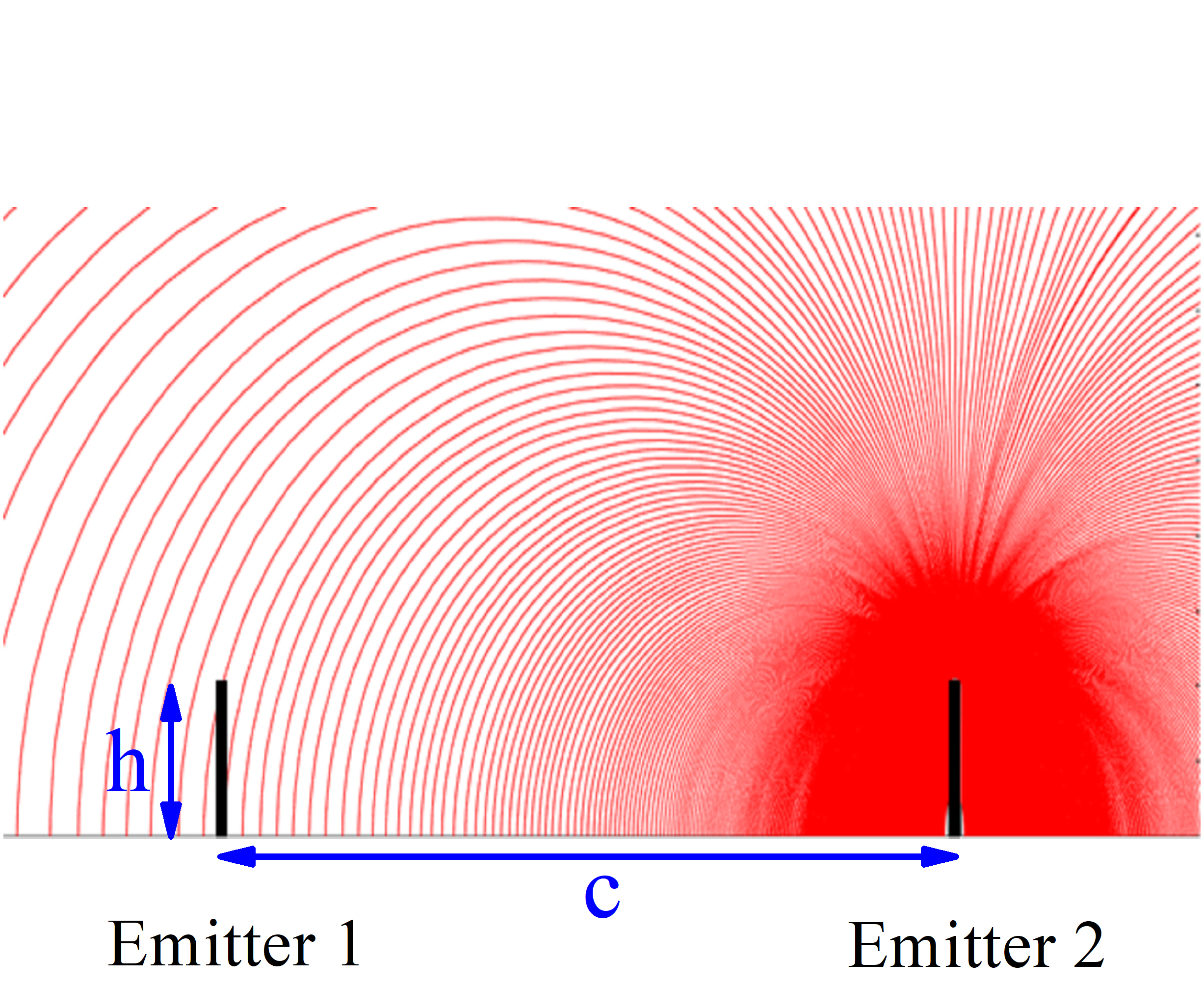}
\caption{Illustration of the depolarizing field related to two identical emitters of height $h$, separated by a distance $c$. In this case, the emitters 1 and 2 are far from each other and (together with their substrate images) form finite dipoles. Electrostatic field lines are shown in red. Figure adapted from Ref. \cite{DallAgnol2018}.} \label{DEP}
\end{figure}

Electrostatic depolarization is a physical effect that appears to be unfamiliar to many people. A physical explanation may be helpful.

Consider two identical conducting posts that stand on a grounded horizontal flat conducting plane and are situated in a vertically uniform, macroscopic field $E_{\rm{P}}$. Each post develops a charge-distribution near its tip, and there is a corresponding image distribution on the opposite side of the grounded plane.

In the lowest order of approximation, these two distributions can each be represented by a finite electrostatic dipole of moment $p$. Thus, from the electrostatic point of view, we are considering two parallel dipoles that tend to mutually depolarize each other, as illustrated in Fig. \ref{DEP}. In the lowest order of approximation (which is adequately valid if $(c/h)$ is sufficiently large, each dipole generates, at the base of the other post, a depolarizing field $E_{\rm{D}}$ given by:
\begin{equation}
E_{\rm{D}} = - \frac{p}{4 \pi \epsilon_0 c^3}.
\label{depE}
\end{equation}
There is an issue of what are the best units to use for measuring $p$. This is addressed in the Appendix.

If $c$ is sufficiently large, then the depolarizing field strength varies relatively little along the length of the post, and we can define an effective electrostatic field $E_{\rm{eff}}$ acting on the post by

\begin{equation}
E_{\rm{eff}} = E_{\rm{P}} + E_{\rm{D}}.
\label{E_eff}
\end{equation}

The polarization of each emitter is actually induced by $E_{\rm{eff}}$ rather than $E_{\rm{P}}$. Hence,

\begin{equation}
p=\alpha E_{\rm{eff}},
\label{polariz1}
\end{equation}
where $\alpha$ is the effective polarizability of each emitter. As a result, combining Eqs. (\ref{depE}), (\ref{E_eff}) and (\ref{polariz1}) yields
\begin{equation}
E_{\rm{eff}} = {\mathit{\Theta}}_{\rm{D}} E_{\rm{P}}  \equiv \frac{E_{\rm{P}}}{1 + \alpha/4\pi \epsilon_0 c^3}.
\label{polariz2}
\end{equation}
where the correction factor $\mathit{\Theta_{\rm{D}}}$ is defined via eq. (\ref{polariz2}).

Thus, in the theory in Section \ref{CDPD}, in the context of the $(\Delta \mathit{\Phi})_{\rm{XD}}$ term, the expression $-E_{\rm{eff}} h$ [$= - {\mathit{\Theta}}_{\rm{D}} E_{\rm{M}} h] $ takes the place of $-E_{\rm{M}} h$, and the outcome is the following formula for the fractional reduction $(- \delta)$ in apex-FEF:
\begin{equation}
(-\delta) = 1 - {\mathit{\Theta}}_{\rm{D}}  =  \frac{\alpha}{\alpha + 4\pi \epsilon_0 c^3}.
\label{polariznn}
\end{equation}

In the limit of large separation, this becomes an inverse-cubic dependence on separation. This prediction has been verified by FEM simulations,
as illustrated in following discussion.

Clearly, result (\ref{polariznn}) is not strictly valid if the depolarizing field varies along the post. In such circumstances, it is better to treat the depolarizing dipole as a pair of negative and positive charges, as is done in the Forbes FSEPP-model approach \cite{RFJAP2016}, and directly develop an expression for the driving potential difference $(\Delta \mathit{\Phi})_{\rm{XD}}$ in eq. (\ref{Phi}).

This FSEPP-model approach yields a general expression (for the charge-blunting contribution) of the form
\begin{equation}
(- \delta)  = \frac{C_1}{C_1 + C_0}  \approx C_1 ,
\label{Forbes-delta}
\end{equation}
where $C_0$ and $C_1$ are geometrical parameters defined algebraically in \cite{RFJAP2016}. In the limit of large separation (i.e., in the ``widely separated regime") $-\delta$ is given adequately by
\begin{equation}
(- \delta)  \approx 2 r_{\rm{a}} {\ell}^2 / c^3  \approx 2 \times \left( \frac{h}{r_{\rm{a}}} \right)^{-1} \times \left(\frac{c}{h} \right)^{-3} .
\label{Forbes-delta2}
\end{equation}
where $r_{\rm{a}}$ denotes the radius of the floating sphere, and $\ell = h-r_{\rm{a}}$ is the height of the centre of the floating sphere above the ground plane. It can be shown that this expression is exactly equivalent to $\alpha/4\pi \epsilon_0 c^3$.

\begin{figure}[h!]
\includegraphics [scale=0.10] {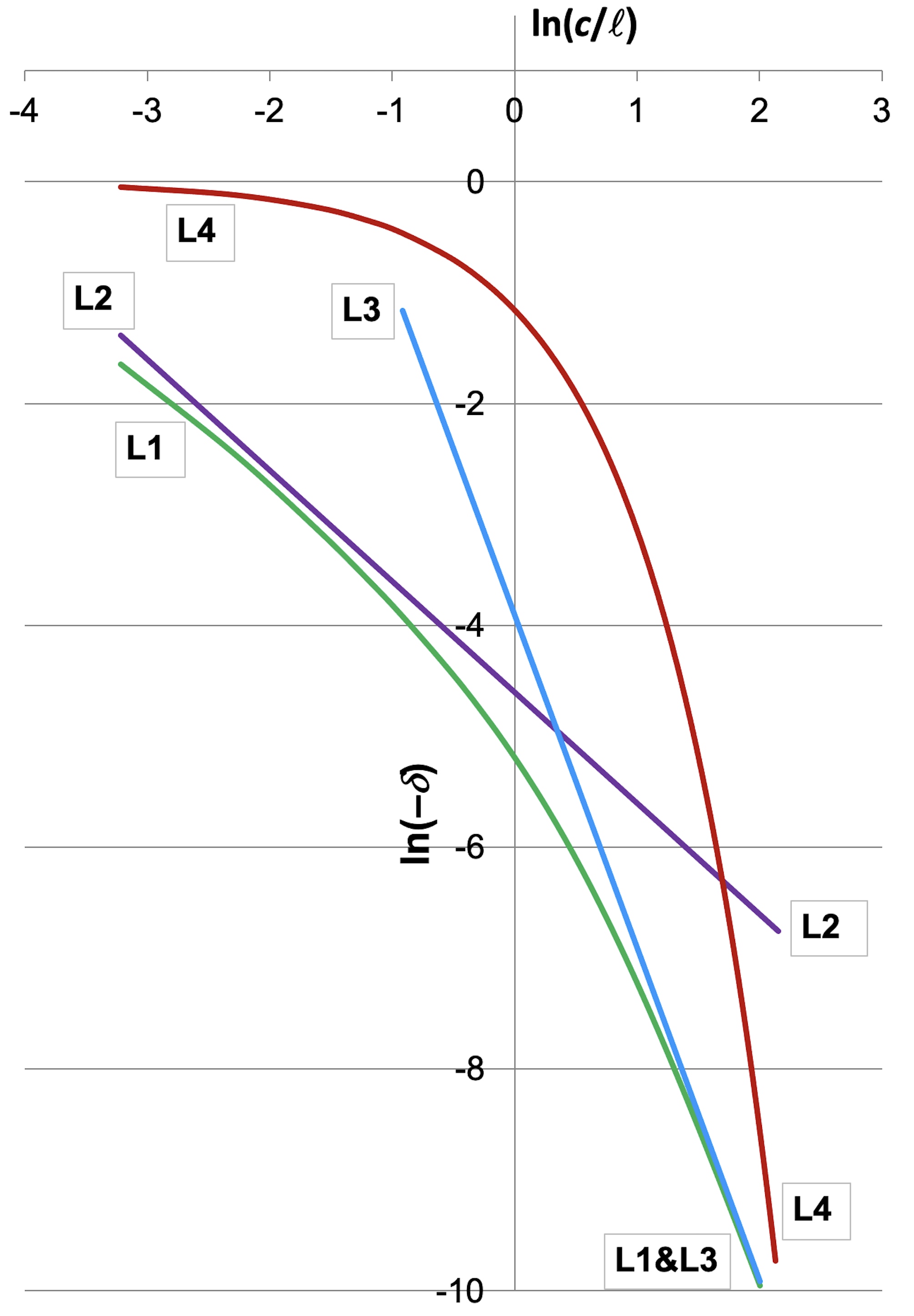}
\caption{Older depolarization predictions for two HCP-model emitters, showing how the fractional reduction ($- \delta$) in apex PFEF depends on the ratio of the emitter separation $c$ to the length $\ell$ of the cylindrical post. An illustrative value of $\ell/r_{\rm{a}} = 100$ has been assumed, where $r_{\rm{a}}$ is the post and apex radius. Line L1 is the precise result from the FSEPP model, as used in \cite{RFJAP2016}. L2 is the simple approximation for behaviour at moderate separation ratios $c /\ell$. L3 is the limiting behaviour at large separation ratios. L4 is the prediction derived by Bonard et al. \cite{bonard} by fitting a formula to the results of finite-element analysis of (what seems to have been) a small linear array. (But note that this Bonard prediction is not strictly applicable to the two-emitter situation). Reproduced from Ref. \cite{Vdiff}.}  
\label{FSEPPnew2}
\end{figure}

\begin{figure}[h!]
\includegraphics [scale=0.35] {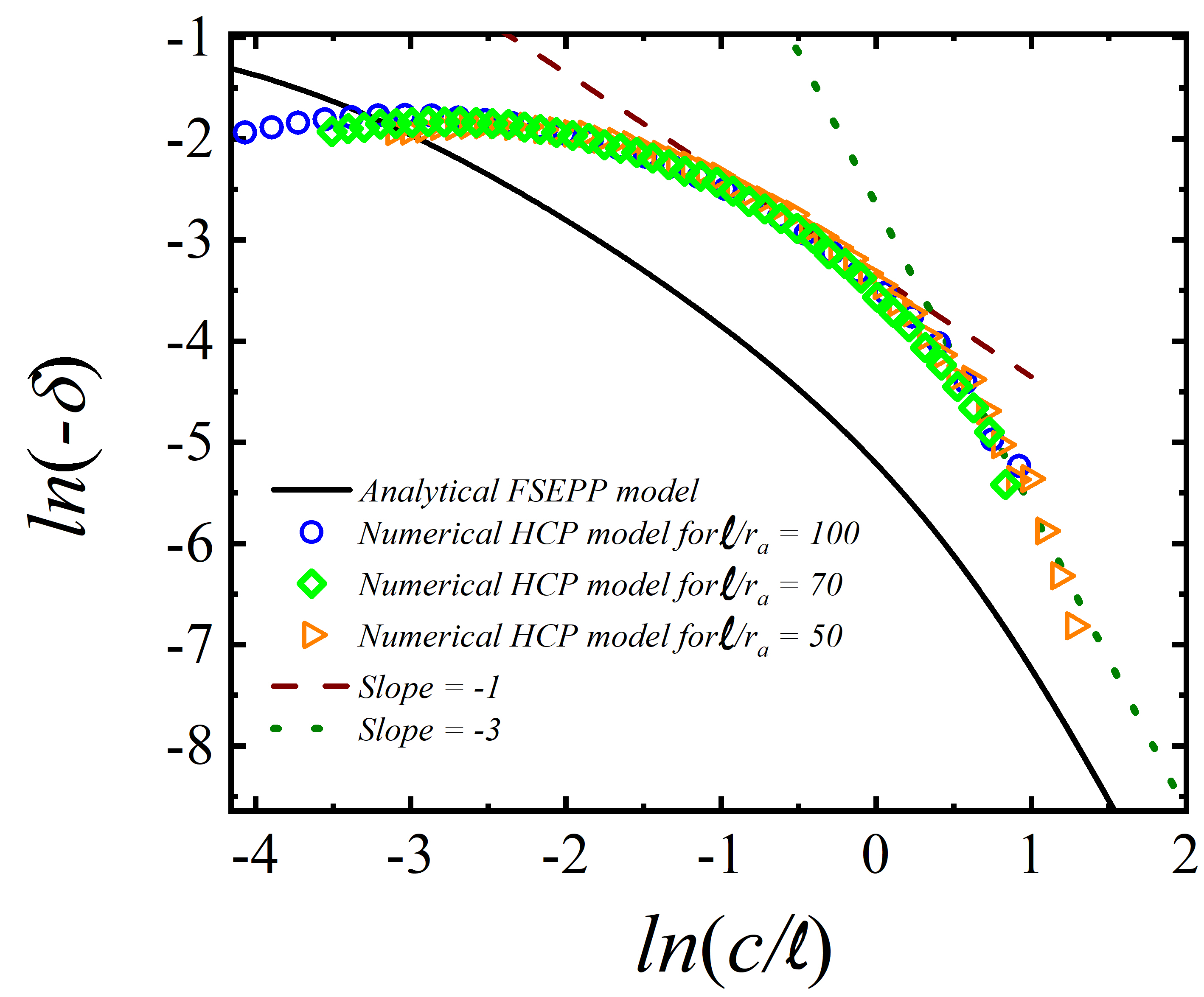}
\caption{Comparison of the predictions of the FSEPP model and FEM simulations, for mutual depolarization of two HCP-model posts with $\ell/r_{\rm{a}} =100$. FEM-simulation results for two other values of $\ell/r_{\rm{a}}$ are also shown.} 
\label{FSEPP_HCP}
\end{figure}

At somewhat smaller separations, but in conditions where $c \gg r_{\rm{a}}$ (i.e., in the ``moderately separated regime") $-\delta$ is given adequately \cite{RFJAP2016} by 
\begin{equation}
(- \delta)  \approx  r_{\rm{a}} / c  \approx  \left( \frac{h}{r_{\rm{a}}} \right)^{-1} \times \left(\frac{c}{h} \right)^{-1}.
\label{Forbes-delta3}
\end{equation}
These regimes, together with a plot of the exact variation of $C_1$ are shown as a ln-ln plot in Fig. \ref{FSEPPnew2},
for an illustrative emitter with $\ell/r_{\rm{a}} = 100$.
The analytical models display the underlying physics of the two-emitter situation, but for accurate numerics it is best to use FEM electrostatic simulations. Figure \ref{FSEPP_HCP} shows FEM-simulation results for geometries that include that considered in Fig. \ref{FSEPPnew2}. The inverse-cubic fall-off in ($ - \delta$) at large $c/h$ values is well confirmed, but the ``moderately separated" regime is poorly defined and has a limited range, and in the ``narrowly separated regime" $( - \delta)$ goes through a maximum as $c/h$ decreases.

\subsection{Effect of the apex sharpness ratio on the fractional reduction of the apex-FEF}

Equations (\ref{Forbes-delta2}) and (\ref{Forbes-delta3})  predict the physical effect that $(-\delta)$ will depend on the apex sharpness ratio $h/r_{\rm{a}}$, in both the ``moderately separated" and ``widely separated" regimes. Since this applies to the two-emitter situation, it will also apply to clusters and arrays. However (as noted in our earlier paper \cite{DallAgnol2018}), this factor (or dependence) does not clearly appear in most earlier work on electrostatic depolarization, either in analytical discussions of multi-emitter situations or in line-charge models of such situations.

\begin{figure}[h!]
\includegraphics [scale=0.35] {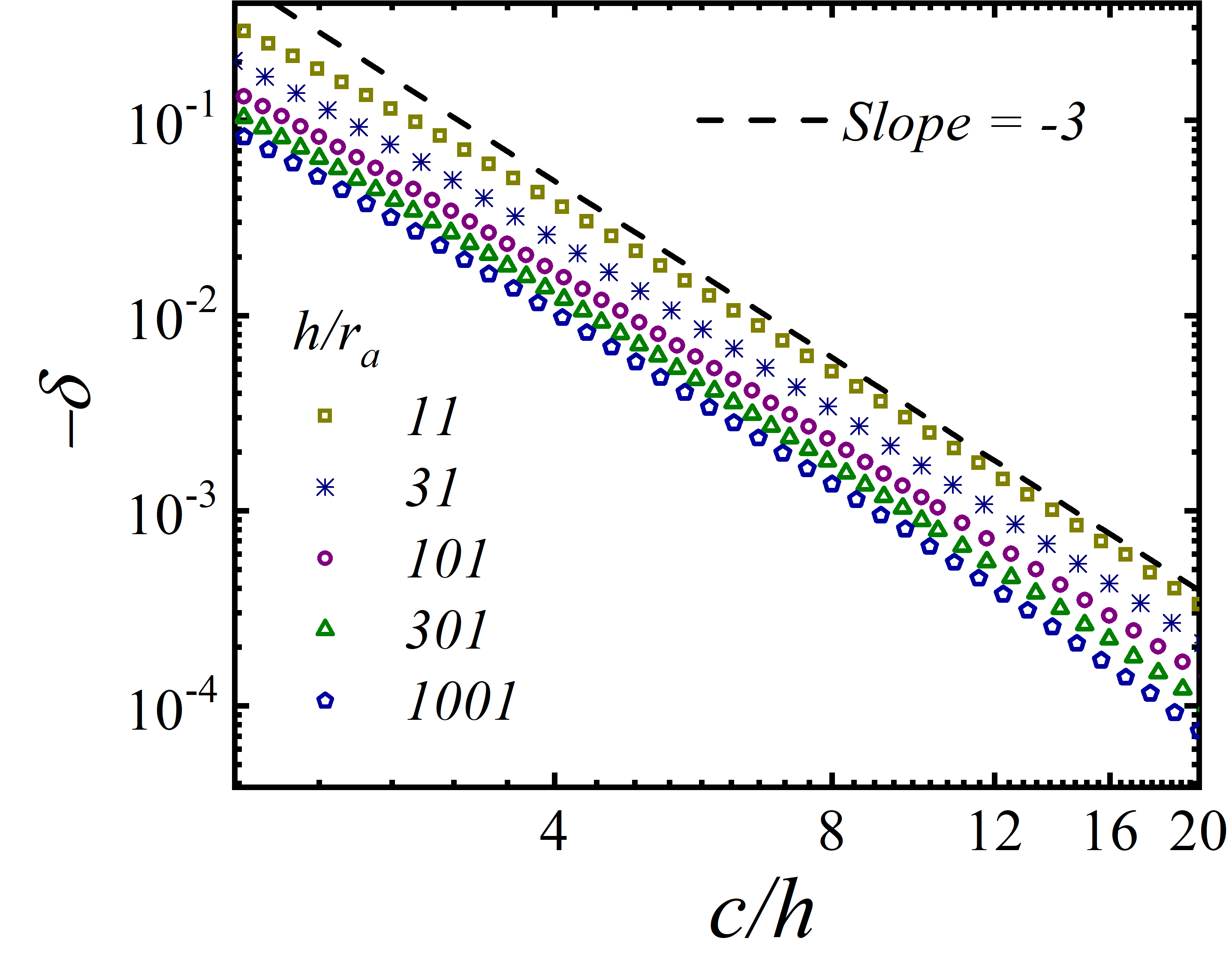}
\caption{To show, for HCP-model emitter and its ring-like image with respect to the rotated right-hand boundary, how the dependence of $(- \delta)$ on the separation ratio $c/h$ is parameterically dependent on the value of the apex sharpness ratio $h/r_{\rm{a}}$. Figure adapted from Ref. \cite{Adson2020}.}
\label{HCPPrefactor}
\end{figure}

In the circumstances of Fig. \ref{FSEPP_HCP} the effect is relatively small. For HCP-model emitter and its ring-like image with respect to the rotated right-hand boundary Fig. \ref{HCPPrefactor} shows that it appears more obviously in the widely-separately regime when a greater range of values of $h/r_{\rm{a}}$ is considered. 

The implication of this confirmation is that all earlier analytical treatments of multi-emitter interactions (except, perhaps, those in Ref. \cite{RFJAP2016}), should be treated with moderate reservations, and that all earlier line-charge treatments and FEM simulations should be treated as applying only to the apex sharpness ratio assumed in the research in question.

\subsection{Close proximity electrostatic effect}
\label{CPEE}

In Fig. \ref{FSEPP_HCP}, the maxima in simulated values of ln$(- \delta)$ imply that the reduced apex-PFEF value goes through a minimum and then begin to increase again, as $c/h$ decreases. As already noted, the existence of this minimum was originally shown by Harris, Jensen and colleagues in 2015 \cite{JensenAPL2015}, using line-charge modelling. They also found, as would be expected, that for an ``internal" emitter in a cluster, the depth of the minimum depends on the number of emitters in the cluster.

\begin{figure}[h!]
\includegraphics [scale=0.35] {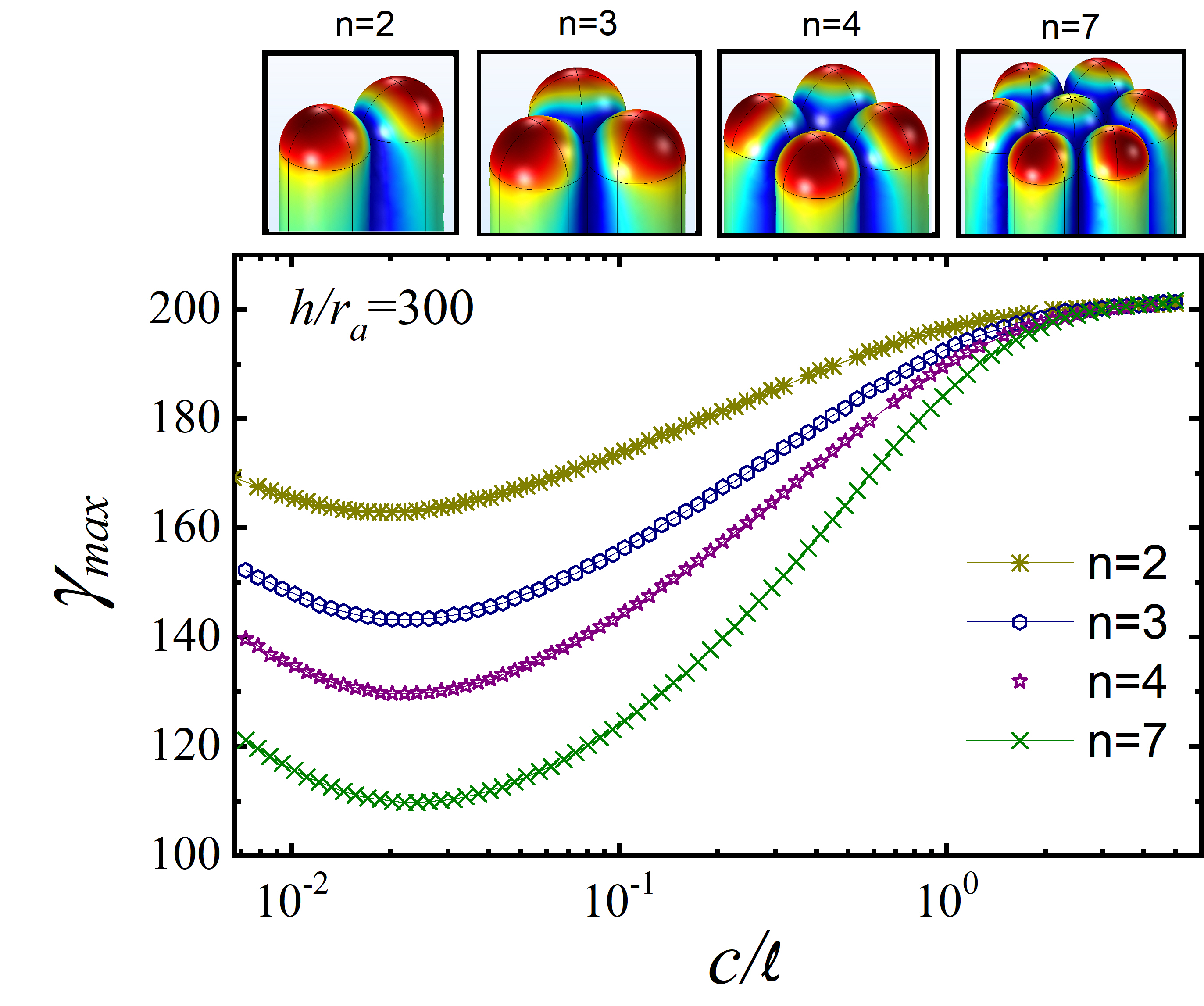}
\caption{To show how the maximum local PFEF, $\gamma_{Pmaxm}$, varies as a function of the ratio $c/\ell$ (effectively $c/h$), for clusters with different numbers $n$ of HCP-model emitters. 
The data are presented for $\ell/r_a=299$ ($h/r_a=300$).  The upper panels show the local-PFEF distribution [red (blue) indicates higher (lower) local PFEF], for the situation where all emitters are in contact with their neighbours. Figure adapted from Ref. \cite{CPEE2017}.}
\label{HCPCPEE}
\end{figure}

Careful analysis \cite{CPEE2017} has shown that (in the two-emitter situation) a second effect also occurs. This is that the position ``max" of maximum field enhancement moves away from the emitter apex, in a direction away from the other emitter. This effect is demonstrated in Fig. \ref{HCPCPEE}, for a illustrative apex sharpness ratio of 300, and (as already noted) has been called the \textit{close proximity effect}.

This figure also shows that this effect occurs with closely spaced clusters of emitters, with the displacement of the maximum PFEF $\gamma_{\rm{Pmax}}$ always being outwards from the centre of the cluster, and with (obviously) the extent of the reduction in $\gamma_{\rm{Pmax}}$ increasing with the number of emitters in the cluster. With large clusters or large finite arrays of closely spaced emitters, this ``displacement outwards" effect is expected to mainly affect the emitters at the edges of the cluster/array.

The magnitudes of all these effects are significantly influenced by the apex sharpness ratio, although qualitative features remain the same. For details see \cite{CPEE2017}.

\subsection{Fitting formulae for simulation results}

Ideally we need to have fitting formulae for all the FEM-simulations results, but this issue remains to be systematically addressed. For the widely separated regime, a formula has been suggested \cite{DallAgnol2018} that we write here in the form 
\begin{equation}
(-\delta)_{\rm{fit}} = \frac{K_1}{K_2 + (c/h)^3}.
\label{widefit}
\end{equation}

\begin{figure}[h!]
\includegraphics [scale=0.30] {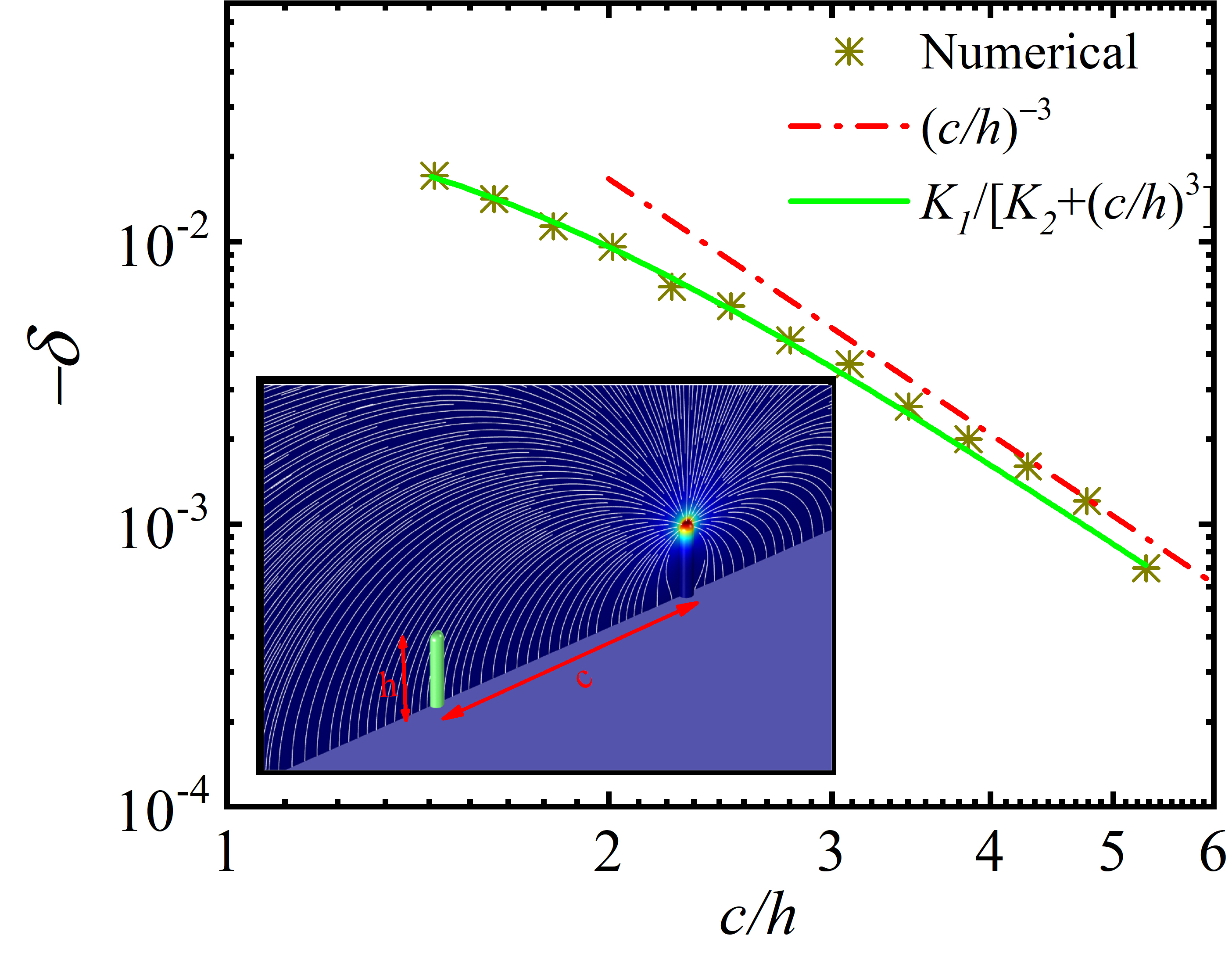}
\caption{Comparison, for two HCP-model emitters with apex sharpness ratio 51, between FEM numerical results, inverse-cubic decay with respect to the separation ratio $c/h$, and the fitting function given by Eq. (\ref{widefit}), with $K_1 = (0.109 \pm 0.004)$ and $K_2 = (3.3 \pm 0.3)$. The inset illustrates the depolarization effect being modelled, for $c/h = 4$: for the right-hand-side emitter it shows the PFEF distribtion near its apex [red (blue) indicates higher (lower) local PFEF] and the field distribution due to this emitter. Figure adapted from Ref. \cite{DallAgnol2018}.} 
\label{TwoHCP}
\end{figure}

For two HCP-model emitters, with $h/r_{\rm{a}} = 51$, Fig.\ref{TwoHCP} illustrates the use of this formula. In the range considered ($c/h > 1.5)$ the fit appears to be good. 

The values of $K_1$ and $K_2$ will depend on the apex sharpness ratio, but the good fitting properties of eq.(\ref{widefit}) in the ``widely separated" depolarization regime  have now been verified for many different physical systems.

For the hemisphere-on-plane (HSP) emitter model, the best fit is given by $K_1 = (0.9433 \pm 0.0006)$, $K_2 = (12.00 \pm 0.02)$. This result has relevance to discussion of the effect of ``bumps" on a macroscopically flat emitter surface.

\subsection{Multi-emitter depolarization effects: clusters and arrays}
\label{Multi}

In principle, four types of multi-emitter situations are of theoretical interest: larger regularly spaced clusters (both linear and non-linear), finite regular arrays, infinite regular arrays, and random clusters and arrays. Although there is a quite a large body of work on regularly spaced finite clusters and arrays, much of it is in need of reassessment, and we have felt it imprudent to provide extensive citations here. An overview of some of this work has been provided by Harris et al. \cite{Harris2015AIP} and a useful FSEPP-model treatment has been given by Zhbanov et al. \cite{Screen2}. What is really needed is a separate critical review that builds on the detailed discussion of the two-emitter situation in this review, and on older work involving arrays of point dipoles, and uses derived insights to re-assess and extend the existing FE ``cluster" and ``array" literature.

Exceptions to this thinking are a few papers (some already cited) that have carried out numerical simulations on finite regular arrays, and some papers on random arrays that explore some of the specific issues involved. For the random arrays, we note in particular the numerical analysis of Read and Bowring \cite{RBowring}, the analysis of Biswas and Rudra \cite{Biswas2018c3} and the conceptual discussion by Bieker et al. \cite{Bieker2019}.

\section{Summary and future tasks}
\label{Sum&Tasks}

\subsection{Summary}

In this review, which we think is the first of its kind, we have employed systematic terminology and definitions and have described the basic principles and formulae of field emitter electrostatics. The theory has been formulated in a ``polarity independent" manner that applies to both field electron and field ion emitters.

The distinction between electronically ideal and electronically non-ideal emitters has been noted, as has the fact that the ``electrostatics" of electronically ideal emitters needs to be seen as part of the theories of the chemical thermodynamics and statistical mechanics of electrons.

For ideal emitters, when the emitter is in static electrical equilibrium, the Fermi level is constant throughout the emitter. Making the simplifying assumption that the local work function is uniform across all surfaces in the system leads to the requirement that the classical electrostatic potential is constant across (and ``just outside") the surface of the emitter and its support electrode. This constant-potential surface provides the boundary condition both for classical analytical analyses and for electrostatic simulations using finite-element methods.

In this review we have concentrated on two application areas: (i) the electrostatics of the needle-type emitters used in the projection technologies (field electron and ion sources, field electron and ion microscopies and spectroscopies, and atom-probe microscopy); and (ii) the electrostatics of the post-like emitters used in basic models of the arrays of emitters used in large-area field electron emitters.

In view of the continuing research interest in FE from carbon nanotubes and related structures, we have paid special attention to the electrostatics of the hemisphere-on-cylindrical-post (HCP) emitter model and to the interaction between two posts of this type, and have identified four separation regimes. We have provided a careful review of our ``minimum domain size" finite-element procedure for finding highly accurate values for plate-type field enhancement factors; this procedure is applicable both to the HCP model and to other post-type emitter models.

We have also provided brief accounts of, and some key citations for, some other effects that can influence FEF values.

We also note that we expect the ``conducting post formula" to find many applications in electrostatics in general.

\subsection{Future tasks}

The writing of this review has drawn our attention to aspects of field emitter electrostatics where further basic research and/or a mini-review of existing knowledge could be useful. We think the following topics would be of current interest.

(*1) A more careful treatment and discussion of the correction factor $c_{\rm{a}}$ in the conducting post formula, eq.(\ref{gam-Fcondpost}).

(*2) Review and extension of knowledge about how FEFs vary with location on the emitter surface, particularly near the emitter apex.

(*3) Review of two-stage field enhancement and further FEM investigations into the limits of validity of Schottky's conjecture. 

(*4) Further detailed analysis of the two-identical-post situation: (i) to explore the narrowly-spaced and close-proximity separation regimes more carefully; (ii) to establish the influence of the apex sharpness ratio; and (iii), if possible, to establish a fitting formula or formulae that cover all regimes.

(*5) Systematic treatment and review of the theory of arrays, initially infinite and finite regular arrays, later random arrays and clusters.

(*6) More detailed analysis and review of collector adjacency effects.

(*7) A review of the electrostatics of blade geometries.

(*8) Review of the electrostatics of point-plane situations (including ``ball-probe" geometries).

(*9) Development of the electrostatics of three-electrode devices, such as nanoscale vacuum channel transistors.

(*10) More careful exploration of the effects of variations in local work function, and of resulting patch fields.

(*11) Exploration of depolarization effects amongst emitters with different apex sharpness ratios.

All the above need to be done for metal systems. There is also a need for equivalent results where non-metals are involved, in particular for the electrostatics of semiconductor posts, and for point-plane analyses where the plane is the front surface of a semiconductor.

All the above involve the theory of classical conductors and semiconductors. As already noted, there is also a need for atomistic-level electrostatics. Much attention has already been given to carbon nanotubes and some to other carbon-related materials, but further research is needed, in order to make improved connection between FE theory and experiment. The electrostatics of atomically structured  metal surfaces also needs more detailed research, but probably the immediate need is for a brief review that brings together existing results from both field electron and field ion emission.

Beyond this there is likely to develop a need for sub-atomic-level electrostatics, in order to help interpret related experiments, but this is largely unexplored research territory.

In short, there is a vast amount of future research that needs to be done in the context of field electron and ion emitter electrostatics. This review has been able to cover only a limited part of the topic. But our hope is that our work can provide both a stimulation and a partial basis for future field emitter electrostatics research. We also hope that our work can contribute more generally to electrostatics knowledge.

\section{Acknowledgements}

TAdA is grateful for the financial support from the
Conselho Nacional de Desenvolvimento Cient\'{i}fico e Tecnol\'{o}gico
(CNPq) under Grant No. 310311/2020-9.

\bigskip

\section*{Appendix:  Units of \break Electric Dipole Moment}

In the modern international system of equations (the ``ISQ"), the SI unit for electric dipole moment is the ``coulomb metre" (C m) or equivalent. However, this unit is inconveniently large for discussing atomic-scale phenomena.

Before the 1970s reforms, a common unit of electric dipole moment was the Gaussian-system unit the ``Debye", which was equal to the strength of a dipole consisting of positive and negative charges each of strength one electrostatic unit (e.s.u.) separated by a distance of 1 \AA. The Debye is still widely used in theoretical chemistry, but this unit is inconvenient in many contexts (including field emitter electrostatics) because the Debye is not an SI-related unit and is not coherent with SI units.

The system of customary units often used in field emission is based on the same set of equations (the ISQ system) as SI units, but on atomic-level units. In this customary system, a convenient unit for electric dipole moment is the strength of a dipole consisting of positive and negative changes, each of magnitude equal to the ISQ elementary charge $e$, separated by a distance of one nanometre (nm).

The natural name for such a unit might be (analogous to the electronvolt) the ``electron-nanometre". However, this name is not allowed by the current rules of the international system of measurement: instead, by multiplying ``electron" by ``V" (the symbol for ``volt"), and dividing ``nanometre" by ``V",  it can be written that: 1 electron-nanometre $\equiv 1$ eV$\cdot (\rm{V/nm})^{-1}$. This unit is internationally recognised for continued use alongside SI units, and can be seen as derived from the ISQ formula for the potential energy $U$ of a dipole of moment $p$ aligned along an applied field $E$, namely
\begin{equation}
U = - p E .
\label{dipole-moment1}
\end{equation}
The relationship between the numerical values (in useful units) of ISQ electric dipole moment $p_{\rm{ISQ}}$ and Gaussian electric dipole moment $p_{\rm{s}}$ is [see 
Wikipedia entry on ``Debye"]:
\begin{equation} 
{p_{\rm{ISQ}} / [\rm{eV}\cdot(\rm{V/nm})^{-1}]} \; {=  0.02081943 \cdot p_{\rm{s}} / [\rm{Debye}]} . 
\label{dipole-moment2}
\end{equation}
That is: a Gaussian value of 1 Debye converts into an ISQ (FE customary units) value of 0.02081943 $\rm{eV}\cdot(\rm{V/nm})^{-1}$.


\providecommand{\newblock}{}

\end{document}